\documentclass[journal]{IEEEtran}

\usepackage[pdftex]{graphicx}
\usepackage[cmex10]{amsmath}
\usepackage[]{algorithm2e}
\usepackage{url}
\usepackage{color}
\usepackage{cite}
\usepackage{bm}
\usepackage[font = footnotesize]{caption}
\usepackage{subcaption}
\usepackage{lipsum}
\usepackage{epstopdf}
\usepackage{fancyhdr}
\usepackage[percent]{overpic}
\usepackage{multirow}
\usepackage{multicol}
\usepackage{comment}
\usepackage{graphics}
\usepackage{graphicx}
\usepackage{tikz}
\usepackage{amssymb, amsmath, amsbsy}
\usepackage{upgreek}
\usepackage{cancel}
\usepackage{color}
\usepackage{mathdots}
\usepackage{mathrsfs}
\usepackage{stackrel}
\usepackage[thinlines]{easytable}
\usepackage{physics}
\usepackage{upquote}
\usepackage{mathtools}

\DeclarePairedDelimiter\ceil{\lceil}{\rceil}
\DeclarePairedDelimiter\floor{\lfloor}{\rfloor}
\providecommand{\abs}[1]{\lvert#1\rvert}

\newcommand{\red}[1]{{\color{black}#1}}

\newcommand{\rev}[1]{{\color{black}#1}}

\begin{document}
%\relscale{1}

%\title{Low Complexity Digital Predistortion Using Adaptive Spline-Based Hammerstein Model}
%\title{Reduced-Complexity Digital Predistortion: \\ Adaptive Spline-Interpolated LUT Methods}
\title{Gradient-Adaptive Spline-Interpolated LUT Methods for Low-Complexity Digital Predistortion}
%\title{Low-Complexity Digital Predistortion: \\Gradient-Adaptive Spline-Interpolated LUTs}
%\title{Low-Complexity Digital Predistortion Using Gradient-Adaptive Spline-Interpolated LUTs}

\author{
Pablo~Pascual~Campo,
Alberto~Brihuega,~\IEEEmembership{Student Member,~IEEE,}
Lauri~Anttila,~\IEEEmembership{Member,~IEEE,}
Matias~Turunen,
Dani~Korpi,~\IEEEmembership{Member,~IEEE,}
Markus~All\'en, %~\IEEEmembership{Member,~IEEE,}
and~Mikko~Valkama,~\IEEEmembership{Senior Member,~IEEE}
%\\
%%% Commenting out sponsors now, to save space
%\thanks{This work was partially supported by the Academy of Finland (grants \#288670, \#301820, \#310991, and \#315858), Nokia Bell Labs, and the Doctoral School of Tampere University. The work was also supported by the Finnish Funding Agency for Innovation (``RF Convergence'' project).}
\thanks{P. Pascual Campo, A. Brihuega, L. Anttila, M. Turunen, M. All\'en, and M. Valkama are with the Department
of Electrical Engineering, Tampere University, Tampere,
Finland. e-mail: pablo.pascualcampo@tuni.fi}%
%\thanks{D. Korpi was with the Department of Electrical Engineering, Tampere University, Finland. Currently, he is with Nokia Bell Labs, Espoo, Finland.}%
\thanks{D. Korpi is with Nokia Bell Labs, Espoo, Finland.}
%\thanks{Manuscript received \rev{Month XX}, 2020.}
}%\vspace{-5pt}

% 

% Acronyms
% ACLR (Adjacent channel power ratio)
% AMAM (
% BF (Basis Functions)
% DPD (Digital Predistortion)
% DSP (Digital Signal Processing)
% DUT (Device Under Test)
% FLOPs (Floating Point Operations)
% GMP (Generalized Memory Polynomial)
% LMS (Least Mean Squares)
% LS (Least Squares)
% LUT (Lookup Table)
% mmWave
% MP (Memory Polynomial)
% NMSE (Normalized Mean Square Error)
% PA (Power Amplifier)
% PAPR (Peak-To-Average-Power-Ratio)
% PSD (Power Spectral Density)

% make the title area
\maketitle
%
%\rev{.. lets make all mods with \textbackslash rev\{.\} command}
%\\
%\\
%
\begin{abstract}
In this paper, new digital predistortion (DPD) solutions for power amplifier (PA) linearization are proposed, with particular emphasis on reduced processing complexity in future 5G and beyond wideband radio systems. The first proposed method, referred to as the spline-based Hammerstein (SPH) approach, builds on complex spline-interpolated lookup table (LUT) followed by a linear finite impulse response (FIR) filter. % facilitating basic memory modeling. 
The second proposed method, the spline-based memory polynomial (SMP) approach, contains multiple parallel complex spline-interpolated LUTs together with an input delay line such that more versatile memory modeling can be achieved. For both structures, gradient-based learning algorithms are derived to efficiently estimate the LUT control points and other related DPD parameters. %{Additionally, comprehensive computational complexity analyses are carried out, while also comparing to ordinary gradient-adaptive memory polynomial (MP) DPD}. 
Large set of experimental results are provided, with specific focus on 5G New Radio (NR) systems, showing successful linearization of multiple sub-6~GHz PA samples as well as a 28~GHz active antenna array, incorporating channel bandwidths up to 200~MHz. Explicit performance-complexity comparisons are also reported between the SPH and SMP DPD systems and the widely-applied ordinary memory-polynomial (MP) DPD solution. The results show that the linearization capabilities of the proposed methods are very close to that of the ordinary MP DPD, particularly with the proposed SMP approach, while having substantially lower processing complexity. 
\end{abstract}

\IEEEpeerreviewmaketitle

\section{Introduction}

%  trim={<left> <lower> <right> <upper>}
\begin{figure*}[ht]
    \centering
    \begin{subfigure}[t]{0.49\textwidth }
        \includegraphics[width=\textwidth, trim = 26 8 30 23,clip]{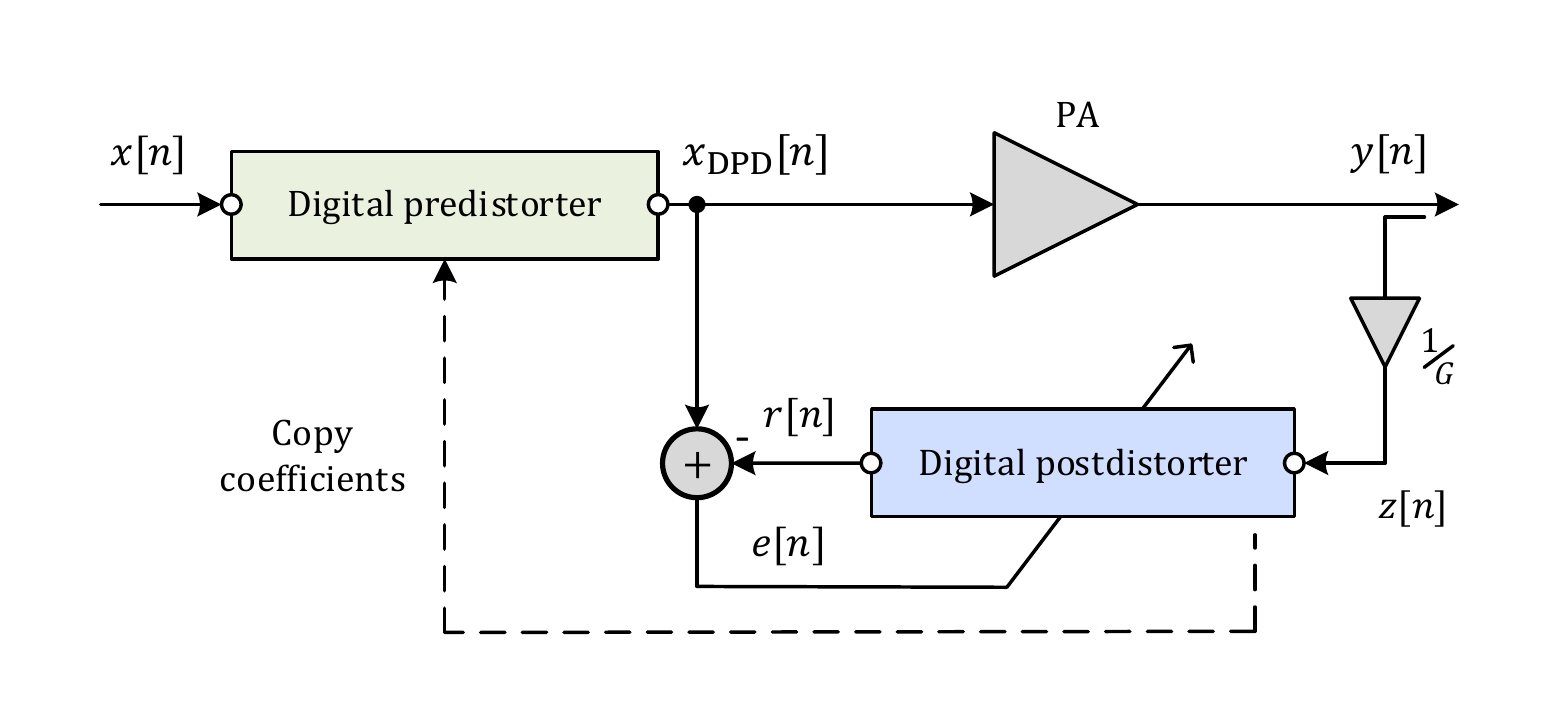}
        \caption{ }
        \label{fig:ILA}
    \end{subfigure}
    ~ %add desired spacing between images, e. g. ~, \quad, \qquad, \hfill etc.
      %(or a blank line to force the subfigure onto a new line)
    \begin{subfigure}[t]{0.43\textwidth}
        \includegraphics[width=\textwidth, trim = 13 22 15 9,clip]{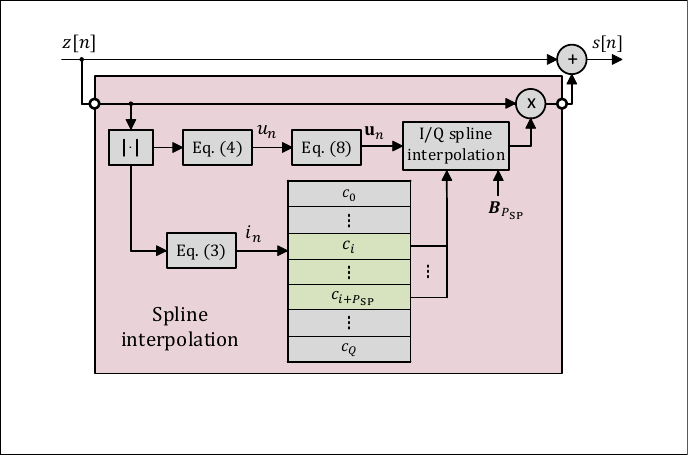}
        \caption{ }
        \label{fig:interp}
    \end{subfigure}
    \vspace{-1.5mm}
    \caption{\quad Illustration of a) the considered DPD system building on indirect learning architecture (ILA), and b) \rev{the injection-based complex spline-interpolated LUT scheme utilized inside the proposed digital predistorter and digital postdistorter entities.}} \vspace{-2.5mm}
    \label{fig:DPD_concept}
    \vspace{-2mm}
\end{figure*}

\IEEEPARstart{M}{odern} radio communication systems, such as the 4G LTE/LTE-Advanced and the emerging 5G New Radio (NR) mobile networks, build on multicarrier modulation, most notably orthogonal frequency division multiplexing (OFDM)\red{\cite{2018DahlmanL5G}}. OFDM waveforms are known to contain high peak-to-average power-ratio (PAPR)\red{\cite{ghannouchi2009behavioral,taxonomy}}, which complicates utilizing highly nonlinear power amplifiers (PAs) in transmitters operating close to saturation {\cite{ghannouchi2009behavioral,6153399,Swedes_review}}. Digital pre-distortion (DPD) is, generally, a well-established approach to control the unwanted emissions and nonlinear distortion stemming from nonlinear PAs, see, e.g.,\red{\cite{ghannouchi2009behavioral
,4497808,abdelaziz2018decorrelation,6153399,6612754, kim2001digital}} and references therein. Especially when combined with appropriate PAPR reduction methods \cite{zhu_2011}, DPD based systems can largely improve the transmitter power efficiency while keeping the unwanted emissions within specified limits.

Some of the most common approaches in PA direct modeling as well as DPD processing are the memory polynomial (MP)\cite{kim2001digital,ghannouchi2009behavioral,tehrani2010comparative} and the generalized memory polynomial (GMP)\cite{morgan2006generalized,mkadem2016multi,ghannouchi2009behavioral,tehrani2010comparative}, both of which can be interpreted to be special cases of the Volterra series\red{\cite{ghannouchi2009behavioral,4161079,1362669,Schetzen2006}}. Such approaches allow for efficient direct and inverse modeling of nonlinear systems with memory, while also supporting straight-forward parameter estimation, through, e.g., linear least-squares (LS), as they are known to be linear-in-parameters models\red{\cite{tehrani2010comparative}}. However, the processing complexity per linearized sample is also relatively high, particularly with GMP and other more complete Volterra series type of approaches, though also some works exist where complexity reduction is pursued \cite{1362669, 6353238, 4350085, 4264109,9018116,8255667}. 
\rev{Specifically, the works in~\cite{4350085,4264109,6963510} present predistorter and PA modeling methods that build on spline-based basis functions -- an approach that is technically considered also in this article, in the form of spline-interpolated lookup tables (LUTs).}
%With such approaches, a coefficient DPD vector is multiplied with each individual spline basis function to predistort the input signal. In the learning stage, this vector is adapted following a LS fitting technique.}

In this paper, we develop and describe \rev{two new DPD solutions} whose linearization capabilities are similar to those of the well-established \rev{polynomial-based solutions}, while at the same time offering a substantially reduced DPD main path processing and parameter learning complexities. The development of such reduced-complexity DPD solutions is mainly motivated by the following four facts or tendencies. First, the channel bandwidths in NR are substantially larger than those in LTE-based systems. Specifically, up to 100 MHz and 400 MHz continuous channel bandwidths are already specified in NR Release-15 at frequency range 1 (FR-1; below 6~GHz bands) and FR-2 (24-40~GHz bands), respectively, 
\cite{3GPPTS38104}, which imply increased DPD processing rates. Second, the actual unwanted emission requirements, particularly in the form of \rev{total radiated power (TRP)} based adjacent channel leakage ratio (ACLR), are largely relaxed in NR FR-2 systems, being only in the order of 26-28~dB \cite{3GPPTS38104}, increasing the feasibility of simplified DPD solutions. Third, the medium range and local area base-stations adopt substantially reduced transmit powers \cite{3GPPTS38104}, compared to classical macro base-stations, hence the available power budget of the DPD solutions is also reduced. Finally, as observed recently in \cite{Swedes_review}, even continuous learning may be needed at FR-2 and other mmWave active array systems, hence developing methods which reduce the parameter learning complexity becomes important. 

\rev{The first new DPD method proposed in this paper, referred to as the spline-based Hammerstein (SPH) approach, builds on complex spline-interpolated LUT followed by a linear finite impulse response (FIR) filter. The interpolation allows to use a small amount of points in the LUT, while the linear filter facilitates basic memory modeling. Gradient-based learning algorithms are also derived, to efficiently estimate the LUT control points as well as the linear filter parameters in a decoupled manner. The second proposed DPD method, referred to as the spline-based memory polynomial (SMP), consists of multiple parallel spline-interpolated LUTs and an input delay line such that more versatile memory modeling can be achieved when summing together the outputs of the parallel LUTs. Through spline interpolation, the size of all parallel LUTs can be kept small, while gradient-adaptive learning rule is again derived to estimate the control points of the involved parallel LUTs.} \rev{For both proposed models -- the SPH DPD and the SMP DPD -- comprehensive computational complexity analyses are provided, while also comparing to ordinary gradient-adaptive canonical MP DPD system}. Then, extensive RF measurement results are provided, covering several different FR-1 PA samples, channel bandwidth cases as well as base-station classes. Additionally, a state-of-the-art 28~GHz active antenna array, specifically Anokiwave AWMF-0129, is successfully linearized with 100~MHz and 200~MHz 5G NR channel bandwidths.

%Additionally, a linear FIR filter accounts for the memory effects present in the real system. The method is designed in a decoupled way, being able to adapt individually its different coefficients.

%To support the proposed model, four independent experiments with four real DPD measurement setups have been performed. In the first two experiments, the DPD model is tested with two different PAs oriented to serve as the amplification stage of Local Aera and Medium Range BS, according to the new 5G NR standards. In the third experiment, it is shown how the proposed DPD scheme can also linearize a conventional Wide Area BS PA, in this case a 100 Watts peak-power PA. Finally, the fourth experiment shows the linearization of a 64-antenna large active array, working in mmWave frequencies, specifically at 28 GHz. The normalized mean square error (NMSE), error vector magnitude (EVM), and adjacent channel power ratio (ACLR) performance metrics are calculated in each case to verify the linearization performance of the setups.

%In brief, the proposed design is shown to be a simple behavioral identification model, being able to outperform the classical predistortion models in terms of computational complexity by adjusting the input range into simpler modeling regions, yet achieving a similar performance in terms of NMSE, EVM, and ACLR.

\rev{
%Regarding the comparison to the existing state-of-the-art solutions, 
In general, it is noted that LUT-based PA linearization is, as such, a well-known approach, see, e.g., \cite{6612754,liang2014quadratic, molina2017digital, jardin2007filter,5497423} and the references therein. However, the PA memory aspects are not considered in \cite{liang2014quadratic}, while fairly sizeable LUTs without interpolation are considered in \cite{6612754,jardin2007filter}. Additionally, a linearly-interpolated LUT-type implementation of a memory polynomial is described in \cite{molina2017digital} while the learning is based on classical LS model fitting.} 
\rev{Furthermore, in~\cite{5497423}, a DPD structure that includes two parallel Hammerstein systems compensating for the PA AM-AM and AM-PM responses, with Catmull-Rom spline interpolation, is presented. The model identification is based on a separable LS technique, specifically using a Levenberg-Marquardt algorithm to identify the DPD coefficients. The main path and training complexities are thus high when compared to the methods presented in this article. %, stemming e.g. from the facts that I/Q-to-polar and polar-to-I/Q conversions are needed in the main DPD path while matrix inversion is needed in DPD learning. 
It is finally also noted that multi-dimensional LUT based solutions exist \cite{4336139,1599577}. However, %their ability to handle substantial memory is fairly limited. 
the LUT size in the nested LUT scheme in \cite{4336139} grows exponentially with the memory depth, thus requiring unfeasible total LUT size when the linearized system exhibits substantial memory. The 2-dimensional LUT technique in \cite{1599577} is, in turn, limited in its memory modeling capability, since it uses a weighted average of past amplitude samples to index the second LUT dimension. %2-dimensional (2D) LUT based DPD solutions exist, however, their ability to handle substantial memory is known to be fairly limited unless unfeasible total LUT sizes are considered.
}
%
%\\
%.. to be checked .. Another parallel approach to linearize an arbitrary PA is the use of LUTs. The obvious advantage of those models is the large reduction in terms of computational complexity, allowing the use of simpler approaches and easier implementations in real time devices. The drawback is, however, a less accurate modeling of the PA, leading to reduced levels of linearization performance and distortion suppression. To smooth this drawback, several LUT models have already been presented in the literature, where authors combine the use of LUT with interpolation schemes~\cite{molina2017digital,liang2014quadratic} or FIR filters~\cite{jardin2007filter} after the LUT to boost the predistortion performance.}
%
%\rev{..it is also noted that the works in~\cite{4350085,4264109} present predistorter methods that build on spline-based basis functions -- an approach that is technically considered also in this article. With such approaches, a coefficient DPD vector is multiplied with each individual spline basis function to predistort the input signal. In the learning stage, this vector is adapted following a LS fitting technique.}

\vspace{-4mm}
In the DPD system context of Fig.~\ref{fig:DPD_concept}, the novelty and contributions of this article can be summarized as follows:
\begin{itemize}
    \item New linear-in-parameters formulation for utilizing spline-interpolated I/Q LUTs in DPD systems, incorporating also the so-called injection-based DPD structure, is provided;
    \item New Hammerstein DPD solution utilizing the spline-interpolated I/Q LUT and decoupled gradient-based learning is proposed and derived;
    \item New memory polynomial DPD solution utilizing multiple parallel spline-interpolated I/Q LUTs and gradient-based learning is proposed and derived;
    \item Comprehensive computational complexity analysis of the methods is provided;
    \item Extensive performance-complexity assessments using versatile RF measurement examples at sub-6~GHz and 28~GHz bands are provided;
\end{itemize}
%\noident
\hspace{0mm}\rev{Compared to the existing literature, the new DPD formulation with spline-interpolated I/Q LUTs allows, in general, for (\textit{i}) using any typical linear estimator (gradient or least-squares) to learn or update the LUT entries and (\textit{ii}) reducing the main path processing complexity clearly when compared to ordinary canonical MP DPD. Specifically, the main path complexity and particularly the learning complexity are both reduced when compared to gradient-based canonical MP, owing to the use of the derived gradient-based learning in combination with the interpolated LUTs, since no basis function orthogonalization~\cite{abdelaziz2018decorrelation} nor self-orthogonalized learning procedure~\cite{7915706,molina2017digital} is needed with the proposed methods. Thus, even continuous DPD adapting/tracking is potentially viable. 
%Finally, the injection-based scheme allows for reduced LUT entry-size, gain ambiguities are better handled in cascaded Hammerstein scenario, less bits required in fixed-point implementation.
}

The rest of the paper is organized as follows. %In Section~\ref{sec:ham}, the adopted DPD architecture is described. 
\rev{Section~\ref{sec:DPDmodels} describes the I/Q spline interpolation scheme used throughout this paper, and presents the proposed SPH and SMP predistorter models. Section~\ref{sec:Learning} derives and presents then the gradient-descent parameter learning  algorithms for both DPD models.}  A complexity analysis and comparison of the proposed DPD solutions is provided in Section~\ref{sec:comp}. Section~\ref{sec:results} describes the RF measurement setups, and presents the corresponding measurement results and their analyses. Finally, conclusions are drawn in Section~\ref{sec:conc}. %, while the detailed derivations of the gradient-based parameter estimation algorithms are provided in Appendix A \rev{and Appendix B}.

Throughout the rest of this article, matrices are denoted by capital boldface letters, e.g., $\mathbf{A} \in \mathbb{C}^{(M \times N)}$, while vectors are denoted by lowercase boldface letters, e.g. , $\mathbf{v} \in \mathbb{C}^{M \times 1} = [v_1 \ v_2 \ \cdots \ v_M]^T$. Ordinary transpose and hermitian operators are represented as $(\cdot)^T$ and $(\cdot)^H$, respectively. %By default, vectors are complex-valued elements presented with lowercase boldface letters, i.e., $\mathbf{v} \in \mathbb{C}^{M \times 1} = [v_1 \ v_2 \ \cdots \ v_M]^T$. 
Additionally, the absolute value, floor, and ceil operators are represented as $\abs{\, \cdot \,}$, $\floor*{\cdot}$, and $\ceil*{\cdot}$, respectively. 

\section{Proposed DPD Models}
\label{sec:DPDmodels}
\rev{In this section, we introduce the proposed I/Q spline interpolation scheme, followed by the corresponding formulation of the SPH and SMP DPD models. For notational convenience, we formulate the mathematical presentation in the context of the indirect learning architecture (ILA) for postdistorter processing, with %$x_{\mathrm{SP}}[n]$ 
$z[n]$ and $r[n]$ denoting the postdistorter input and output, respectively. In the actual predistortion stage -- as illustrated also in Fig.~\ref{fig:DPD_concept} -- the input and output signals are $x[n]$ and $x_\mathrm{DPD}[n]$, respectively.}

\vspace{-3mm}
\subsection{Background and Basics}
\label{sec:background}
%
%To estimate the memory nonlinear PA model, different models can be taken, being the MP \cite{kim2001digital} or the GMP \cite{morgan2006generalized} two of the most popular approaches. Nonetheless, these designs can lead to high processing complexity, as high order polynomials are needed to obtain accurate estimations. In this paper, we propose the use of complex splines to greatly simplify the processing complexity, yet achieving a similar predistortion performance.

Building on piece-wise polynomials, spline based modeling and interpolation seeks to determine a smooth curve that approximates or conforms to a set of points, commonly known as control points \cite{de1978practical}. Consequently, the input signal range is divided into several pieces, and the polynomials model the nonlinear system behavior in the corresponding regions under continuity and smoothness constraints. With this approach, simple low-order functions can be adopted, per region, in contrast to methods where a single high-order function or polynomial seeks to model the whole input range. %For a comprehensive treatment of spline-based modeling for real-valued signals, please refer to \cite{de1978practical}.
%
%Traditionally, spline modeling has been applied to real-valued signals and systems, see, e.g., \cite{scarpiniti2013nonlinear, scarpiniti2014hammerstein,scarpiniti2015novel}. However, in the context of radio communications, complex I/Q signals are utilized, and hence

Traditionally, spline modeling has been applied to real-valued signals and systems \cite{de1978practical,scarpiniti2013nonlinear, scarpiniti2014hammerstein,scarpiniti2015novel}. However, in the context of radio communications, complex I/Q signals are utilized, and therefore the spline models need to be extended to the complex domain. %\rev{Specifically, in this paper, we consider complex baseband models of PA nonlinearities. To shortly illustrate how splines can be applied to PA modelling and digital predistortion, we start with the well-known memoryless polynomial model for the PA, written for input signal $x_{\mathrm{in}}[n]$ as  
\rev{Specifically, in this paper, we consider complex baseband models of RF nonlinearities, particularly those stemming from PA, for DPD purposes. To first shortly illustrate how splines can be applied to RF nonlinearity modelling at baseband, we start with the well-known memoryless polynomial, written for an arbitrary input signal $x_{\mathrm{in}}[n]$ as
\begin{align}
    \label{eq:poly_model}
    x_{\mathrm{out}}[n]=\sum_{\substack{p = 0, \hspace{1mm} 
                p\mathrm{\, odd}} }^{P} \alpha_p x_{\mathrm{in}}[n]\abs{x_{\mathrm{in}}[n]}^{p-1},    
\end{align}
where $\alpha_p \in \mathbb{C}$ are the corresponding polynomial coefficients~\cite{kim2001digital,morgan2006generalized}. Setting $\alpha_1=1$, without loss of generality, this can be re-written as
\begin{align}
    x_{\mathrm{out}}[n]&=x_{\mathrm{in}}[n](1+\alpha_3\abs{x_{\mathrm{in}}[n]}^2+\cdots+\alpha_P\abs{x_{\mathrm{in}}[n]}^{P-1}) \nonumber \\
    &=x_{\mathrm{in}}[n](1+F(\abs{x_{\mathrm{in}}[n]})),
    \label{eq:poly_model2}
\end{align}
where the function $F(\cdot)=F_{\mathrm{I}}(\cdot)+jF_{\mathrm{Q}}(\cdot)$ is a real-to-complex mapping. Thus, the baseband equivalent nonlinearity model consists of two real-valued functions $F_{\mathrm{I}}(\cdot)$ and $F_{\mathrm{Q}}(\cdot)$, both dependent only on the absolute value of the input signal.
}

\vspace{-2mm}
\subsection{Proposed I/Q Spline Interpolation Scheme}
\label{sec:splines}
%As is well known, the same model structure as shown in (\ref{eq:poly_model2}) for the PA direct model, can be used also for inverse modeling or DPD. 
In general, the above model structure shown in (\ref{eq:poly_model2}) can be used for both PA direct modeling as well as PA inverse modeling, i.e., DPD. In the context of DPD -- which is the focus of this article -- the nonlinear functions can be implemented efficiently with, for example, LUTs. 

\begin{figure}[t!]
  \begin{center}
    \includegraphics[width=0.9\columnwidth, trim = 34 20 39 20, clip]{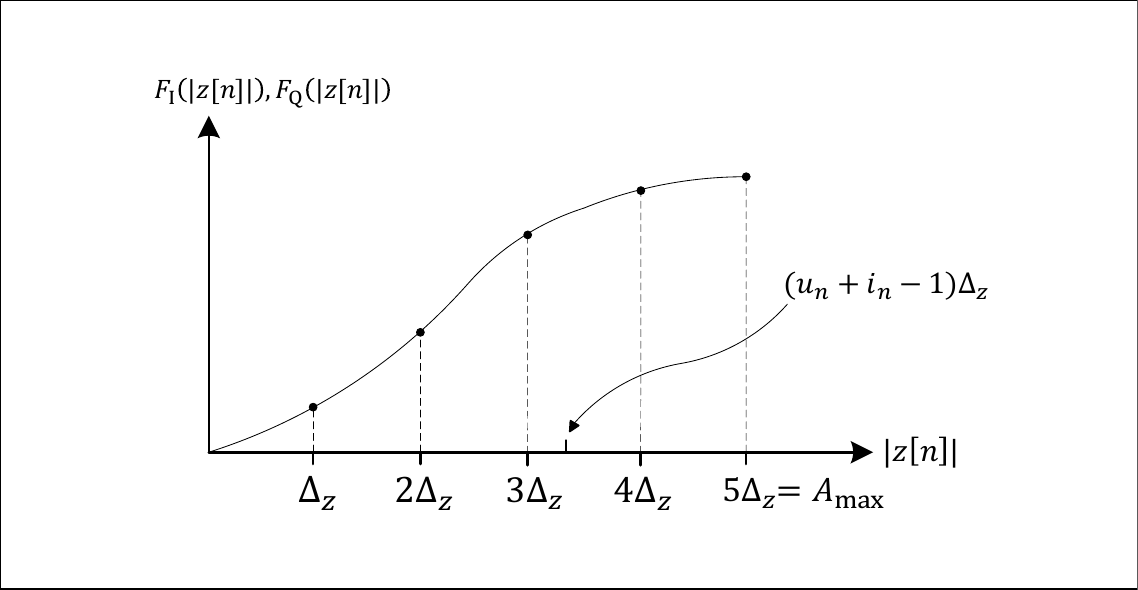}
  \end{center}
  \caption{\quad Conceptual illustration of the nonlinear model regions of $F_{\mathrm{I}}(\abs{z[n]})$ and $F_\mathrm{Q}(\abs{z[n]})$ with respect the input magnitude $\abs{z[n]}$ assuming $K=5$ regions. Also an example of the input envelope value is shown, in this particular case within region $i_n = 4$, where $u_n\in[0,\Delta_z)$ denotes the normalized input envelope within a region. %For visualization simplicity, y-axis values are assumed complex-valued. %Note that $u_n$ is here shown in its corresponding region for illustrational purposes; in reality it is only defined in the interval $[0,\Delta_x)$.
  }
  \label{fig:regions}
  \vspace{-2mm}
\end{figure}

To this end, in the linearization context of Fig.~\ref{fig:DPD_concept}, we formulate in this article spline-interpolated LUTs, i.e., small LUTs with spline interpolation to obtain the intermediate values.
By adopting the notations in Fig. \ref{fig:DPD_concept}, such spline-based modeling of the nonlinear functions $F_{\mathrm{I}}(\cdot)$ and $F_{\mathrm{Q}}(\cdot)$
is illustrated at conceptual level in Fig. \ref{fig:regions}, where the input is a unipolar signal $\abs{z[n]}$ with a maximum amplitude of $A_{\max}$. We adopt uniform equi-spaced splines with knot spacing (region width) of $\Delta_{z}>0$, thus resulting in a total of $K=A_{\max}/\Delta_{z}$ regions. These regions are built, and accessed at time instant $n$, through the span index $i_n$ and abscissa value $u_n$, defined as 
\begin{align}
    \label{eq:i_n}
    i_{n} & = \floor*{\frac{\abs{z[n]}}{\Delta_{z}}} + 1,\\[5pt]
    \label{eq:u_n}
    u_{n} & = \frac{\abs{z[n]}}{\Delta_{z}} - (i_{n}-1). 
\end{align}
%where $\Delta_{z}>0$ is the width or separation between the regions, and where we have assumed uniform equi-spaced splines for simplicity. 
Here, $i_{n}$ denotes the index of the selected region at time instant $n$, and $u_{n}$, $0 \! \leq \! u_n \! < \! \Delta_z$, represents the normalized value of the corresponding input envelope within the current region $i_n$. 
%\red{A conceptual illustration in the context of mapping between the complex input $z[n]$ and the output $s[n]$, defined in (\ref{eq:s}), as a function of the input envelope $\abs{z[n]}$ is shown in Fig. \ref{fig:regions}.} 
%It is noted that the range of the index variable $i_{n}$, and thus the number of regions, will depend on the range of the input envelope $\abs{z[n]}$ and $\Delta_{z}$, while $u_n$ will be always enclosed between $0$ and $\Delta_z$. %For notational convenience and later use, 
%Additionally, from the abscissa value $u_n$, 
%we further define the abscissa vector as $\mathbf{u}_n  = \begin{bmatrix}u_n^{P_{\mathrm{SP}}} & u_n^{P_{\mathrm{SP}}-1} & \cdots & 1\end{bmatrix}^T$ $\in \mathbb{R}^{(P_{\mathrm{SP}}+1)\times 1}$.
%, a parameter that will be used in posterior calculations. Fig. \ref{fig:regions} presents an intuitive representation of how the regions are modeled by the algorithm.

%  trim={<left> <lower> <right> <upper>}

In general, adopting uniform splines allows the spline-interpolated output signal to take a very simple form, discussed also in [27] in the context of real-valued systems. 
%Thus, %the spline interpolation scheme can be formulated by 
%following~\cite{scarpiniti2013nonlinear}, but extending the model to complex I/Q signals, 
The outputs of the I and Q splines can now be written as
\begin{align}
    F_{\mathrm{I}} (\abs{z[n]}) &= \mathbf{g}_n^T  \mathbf{c}^{\mathrm{re}}, \label{eq:F_I} \\
    F_{\mathrm{Q}} (\abs{z[n]}) &= \mathbf{g}_n^T  \mathbf{c}^{\mathrm{im}}, \label{eq:F_Q} 
\end{align}
where $\mathbf{c}^{\mathrm{re}}$ and $\mathbf{c}^{\mathrm{im}}$ contain the $Q$ control points of each spline. The vector $\mathbf{g}_n$ $\in \mathbb{R}^{Q \times 1}$, in turn, is defined as
\begin{align}
    \mathbf{g}_n = \begin{bmatrix}0 & \cdots & 0 & \mathbf{u}_{n}^T \, \mathbf{B}_{P_{\mathrm{SP}}} & 0 & \cdots& 0\end{bmatrix}^T, 
    \label{eq:g_n}
\end{align}
where
\begin{align}
    \mathbf{u}_n  = \begin{bmatrix}u_n^{P_{\mathrm{SP}}} & u_n^{P_{\mathrm{SP}}-1} & \cdots & 1\end{bmatrix}^T \in \mathbb{R}^{(P_{\mathrm{SP}}+1)\times 1},
\end{align}
and $\mathbf{B}_{P_{\mathrm{SP}}} \in \mathbb{R}^{(P_{\mathrm{SP}}+1) \times (P_{\mathrm{SP}}+1)}$ is the \emph{spline basis matrix} of order $P_{\mathrm{SP}}$. In (\ref{eq:g_n}), the term $\mathbf{u}_{n}^T \mathbf{B}_{P_{\mathrm{SP}}}$ of size $1 \times (P_{\mathrm{SP}}+1)$ is located such that the starting index is $i_n$. \red{Thus, at a given time instant $n$, only 
the control points %corresponding to the current span, i.e., 
$c_{i_{n}},  c_{i_{n}+1}, \hdots, c_{i_{n}+P_{\mathrm{SP}}}$ contribute to the output. \rev{It is noted that for simplicity, we assume in this work that the spline order $P_{\mathrm{SP}}$ does not depend on the region.} %Hence, intuitively, the nonlinear mapping between the input $z[n]$ and the output $s[n]$ is approximated by first linearly combining different monomial transformations $u_n^{P_{\mathrm{SP}}}, \ u_n^{P_{\mathrm{SP}}-1}, \ \cdots, \ 1$, through spline basis matrix $\mathbf{B}_{P_{\mathrm{SP}}}$ which are then further combined together via weighting by the control points. 
}

\rev{
Using (\ref{eq:F_I}) and (\ref{eq:F_Q}), while following the model structure in (\ref{eq:poly_model2}), the complex-valued output of the instantaneous nonlinear system, $s[n]$,
%calculated after the spline nonlinearity, that retains the phase information of the input signal via the multiplication by $x_{\mathrm{SP}}[n]$, 
can be constructed as
\begin{align}
    s[n] %&= z[n] +z[n]\mathbf{g}_n^T (\mathbf{c}^{\mathrm{re}} + j \mathbf{c}^{\mathrm{im}}) \\
    &= z[n] + z[n]\mathbf{g}_n^T (\mathbf{c}^{\mathrm{re}} + j \mathbf{c}^{\mathrm{im}})  \nonumber \\
    &= z[n] + z[n]\mathbf{g}_n^T \mathbf{c},
    \label{eq:s}
\end{align}
where $\mathbf{c} \in \mathbb{C}^{Q\times 1} = \begin{bmatrix}c_0 & c_1 & \cdots & c_{Q-1}\end{bmatrix}^T$ is the overall complex-valued LUT containing the control points for the I and Q components. The interpolation scheme is further detailed in Fig.~1(b). %\ref{fig:interp}. %, and $\mathbf{1}$ denotes a $Q \times 1$ column vector of all ones. The last equality follows from the property $\mathbf{g}_n^T \mathbf{1}=1$, which is called the partition of unity property of B-splines~\cite{prautzsch2013bezier}. %, representing essentially an LUT of size $Q$, with the control points being the table entries.
We also note that the total number of control points with $K$ regions and spline interpolation order $P_{\mathrm{SP}}$ is $Q = K + P_{\mathrm{SP}}$. %, where $i^{\mathrm{max}}$ is the number of regions considered in the model.} 
}

\rev{
Importantly, the spline output $\mathbf{g}_n^T \mathbf{c}$ in (\ref{eq:s}) is defined as a deviation from unit gain. We refer to such structure as an injection-based scheme. Specifically, with this formulation, if $\mathbf{c}$ is initialized as an all-zero vector, the nonlinear system output will be the original input signal, i.e. $s[n]=z[n]$.
%, as the multiplying term $\mathbf{g}_n^T \mathbf{1}$ results in 1, due to the partition of unity property of B-splines~\cite{prautzsch2013bezier}. 
By following this formulation, e.g., the gain ambiguities between the nonlinear spline and a cascaded FIR filter can be effectively removed -- an issue that is relevant in the following Hammerstein DPD system -- as the linear filter alone will handle the gain in the system. Additionally, the number of required bits in $\mathbf{c}$ in a fixed-point implementation is generally reduced, as this formulation reduces its dynamic range.
}

%\rev{
%Add here the different P for each region thing, and different filter or bank of filters ... ?
%}
\vspace{-2mm}
\subsection{Spline-Interpolated Hammerstein DPD}
\label{sec:SPH}

%  trim={<left> <lower> <right> <upper>}
\begin{figure*}[ht]
    \centering
    \begin{subfigure}[t]{0.49\textwidth }
        \includegraphics[width=\textwidth, trim = 24 2 20 21,clip]{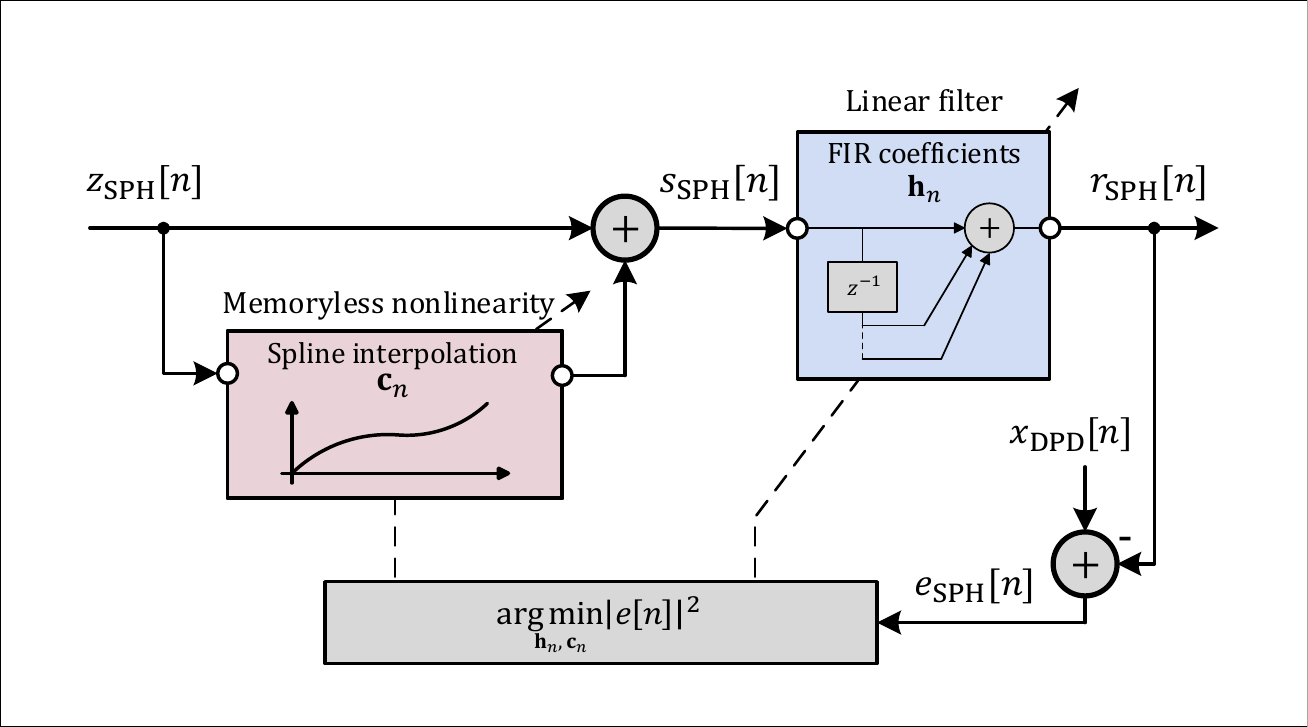}
        \caption{ }
        \label{fig:SPH}
    \end{subfigure}
    ~ %add desired spacing between images, e. g. ~, \quad, \qquad, \hfill etc.
      %(or a blank line to force the subfigure onto a new line)
    \begin{subfigure}[t]{0.45\textwidth}
        \includegraphics[width=\textwidth, trim = 10 20 20 23,clip]{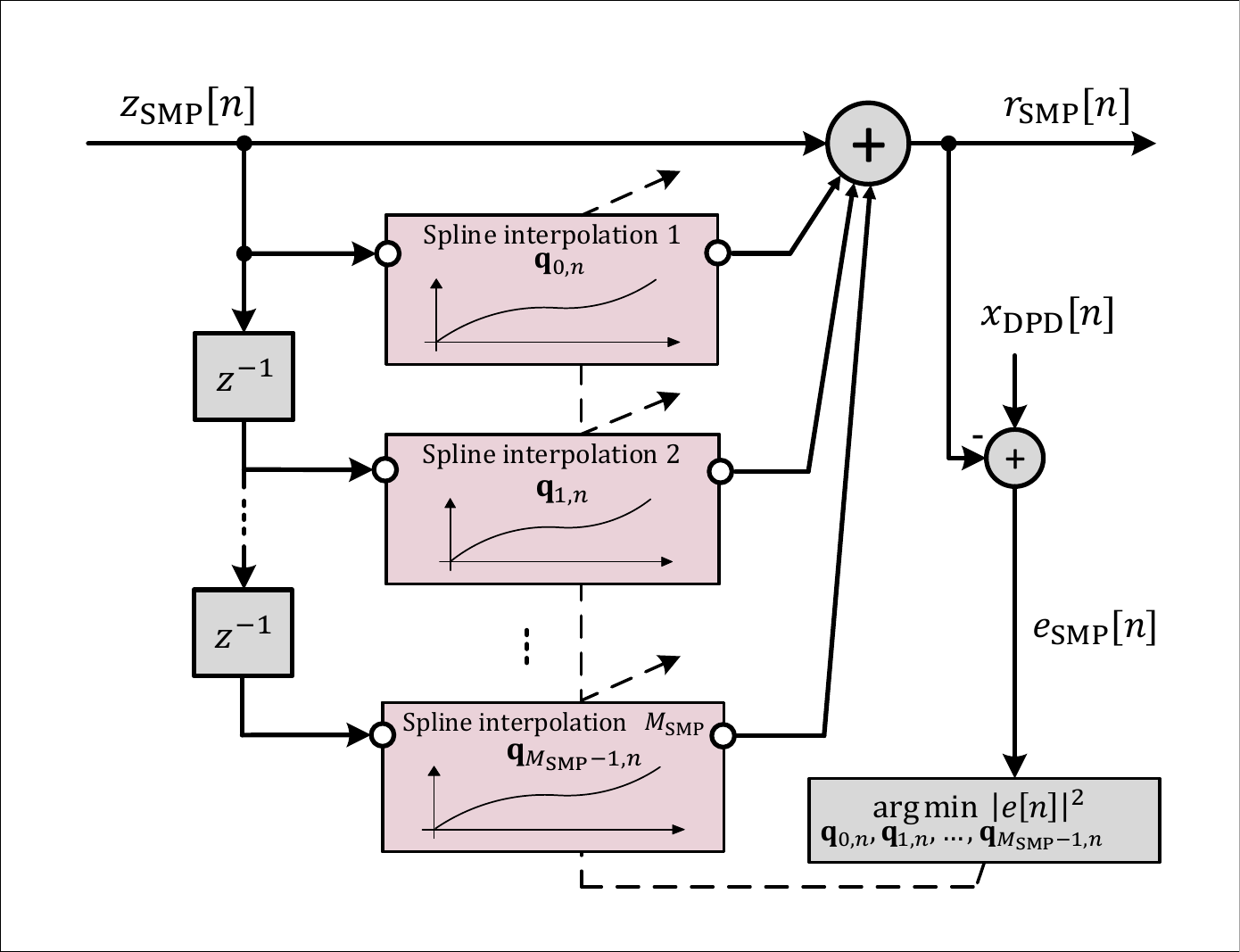}
        \caption{ }
        \label{fig:SMP}
    \end{subfigure}
    \caption{\quad \rev{Illustration of the postdistorter structures of a) the proposed SPH DPD system, and b) the proposed SMP DPD model. The \textit{Spline interpolation} block(s) in both models comprises the scheme shown in Fig.~1(b).}}
    \label{fig:SPmodels}
    \vspace{-2mm}
\end{figure*}

\rev{

This subsection introduces the proposed SPH scheme which builds on a Hammerstein structure where the involved nonlinearity is modelled with a complex spline-interpolated LUT. 
%
%In~\cite{5497423}, a resemblant structure that includes two independent Hammerstein systems in parallel compensating for the PA AM-AM and AM-PM responses, with Catmull-Rom spline interpolation, is presented. The model identification is based on a separable least squares technique, specifically using a Levenberg-Marquardt algorithm to identify the DPD coefficients. The main path and training complexity of this method is, however, high when compared to a simple Hammerstein approach, due to the processing required in both Hammerstein branches, the rectangular to polar and polar to rectangular conversion of the I/Q signals, and the matrix inversion in the DPD coefficient estimation.
%
%
Following the proposed interpolation scheme presented above, in (9), we thus express the output of the instantaneous nonlinear block in the Hammerstein structure as}
\begin{align}
    \rev{s_{\mathrm{SPH}}[n] =  z_{\mathrm{SPH}}[n] + z_{\mathrm{SPH}}[n] \mathbf{g}_n^T \mathbf{c}.}
    \label{eq:s_SPH}
\end{align}
It is noted that the term $\mathbf{g}_n$ depends on the B-spline basis matrix $\mathbf{B}_{P_{\mathrm{SP}}}$. This matrix can be precomputed for the given type of splines and polynomial order, and can be therefore considered as static. As a concrete example, in this article we focus on 3$^{\mathrm{rd}}$ order ($P_{\mathrm{SP}}=3$, cubic interpolation) B-splines, although other spline orders are tested and demonstrated as well. In this case, the basis matrix can be expressed as \cite{scarpiniti2013nonlinear}
\begin{align}
    \mathbf{B}_3 = \frac{1}{6} \begin{bmatrix}
    \frac{-1}{\Delta_{z}^3} & \frac{3}{\Delta_{z}^3} & \frac{-3}{\Delta_{z}^3} & \frac{1}{\Delta_{z}^3} \\[5pt]
    \frac{3}{\Delta_{z}^2} & \frac{-6}{\Delta_{z}^2} & \frac{3}{\Delta_{z}^2} & 0 \\[5pt]
    \frac{-3}{\Delta_{z}} & 0 & \frac{3}{\Delta_{z}} & 0 \\[5pt]
    1 & 4 & 1 & 0
    \end{bmatrix}.
    \label{eq:B3}
\end{align}

%\red{Note that if the set of control points $\mathbf{c}_n$ in (\ref{eq:s}) is a zero vector, the signal $s_{\mathrm{SP}}[n]$ is just the sum of the individual elements of the vector $\mathbf{u}_{n}^T \mathbf{B}_{P_{\mathrm{SP}}}$ multiplied by $x_{\mathrm{SP}}[n]$. In fact, this sum is always the unit for all values of $\Delta_x$ and $\mathbf{u}_n$. As a consequence, $s_{\mathrm{SP}}[n] = x_{\mathrm{SP}}[n]$ if no nonlinearity is considered.}

Next, after having derived the expression for the memoryless nonlinear signal model, the memory effects are incorporated through the FIR filter stage \rev{that is common to all regions}. Hence, the overall output signal $r_{\mathrm{SPH}}[n]$ can be directly expressed as
\begin{align}
    r_{\mathrm{SPH}}[n] = \mathbf{h}^T \mathbf{s}_n,
    \label{eq:r_SPH}
\end{align}
where $\mathbf{h} \in \mathbb{C}^{M_{\mathrm{SPH}} \times 1} = \begin{bmatrix}h_0 & h_1 & \cdots & h_{M_{\mathrm{SPH}}-1}\end{bmatrix}^T$ contains the filter coefficients, with $M_{\mathrm{SPH}}$ denoting the number of taps in the model, while $\mathbf{s}_n \in \mathbb{C}^{M_{\mathrm{SPH}} \times 1} = \begin{bmatrix}s_{\mathrm{SPH}}[n] & s_{\mathrm{SPH}}[n-1] & \cdots & s_{\mathrm{SPH}}[n-M_{\mathrm{SPH}}+1]\end{bmatrix}^T$. \rev{The overall processing structure is illustrated in Fig.~3(a).}

\rev{
\subsection{Spline-Interpolated MP LUT DPD}
\label{sec:SMP}
This subsection formulates the proposed SMP DPD model. Inspired by~\cite{6612754}, a memory polynomial type parallel branched structure is adopted to model the memory effects, while the actual parallel nonlinearities are each implemented through the complex spline-interpolated LUTs presented above. %, and derive a new gradient-descent type adaptive learning rule for the control points in each LUT. %coefficient updates are derived for each independent LUT. 
%This approach yields simple learning rules which are executed sample by sample,  %thus allowing for adaptive estimates of the learning parameters, 
%in contrast to other approaches that use more complex fitting techniques such as LS~\cite{molina2017digital}.
%
%
%The main advantage of this technique is the processing complexity reduction in both main and learning stages, as the LUTs only require a few arithmetical operations to perform the addressing and applying of its estimated gain.
%
% %  trim={<left> <lower> <right> <upper>}
% \begin{figure}[t!]
%   \begin{center}
%     \includegraphics[width=0.9\columnwidth, trim = 1 80 150 1, clip]{figures/MPLUTModel.pdf}
%   \end{center}
%   \caption{\quad \rev{MP LUT based DPD model}}
%   \label{fig:MPLUT}
% \end{figure}
%
%In this paper, we also present a novel gradient-descent type learning rule derivation for a specific MP LUT approach, where the LUT entries are interpolated by using the same spline-based interpolation scheme as used in the SPH model, presented in the previous section. By applying this approach, the new LUT coefficient updates are derived for each independent LUT, placed at each branch, according to the generated error signal and the instantaneous sample value at each specific time instant. This approach provides simpler learning rules which are executed sample by sample, thus allowing for adaptive estimates of the learning parameters, in contrast to other approaches that use more difficult fitting techniques, such as LS (Reference Molina paper).
%
To this end, the proposed SMP processing can thus be expressed as %a combination of the different parallel branches that include the interpolated LUT schemes, as
\begin{align}
    \label{eq:MPLUT}
    r_{\mathrm{SMP}}[n] = z_{\mathrm{SMP}}[n] + \sum_{m=0}^{M_{\mathrm{SMP}}-1} z_{\mathrm{SMP}}[n-m] \mathbf{g}_{n-m}^T \mathbf{q}_m,
\end{align}
where $M_{\mathrm{SMP}}$ denotes the considered memory order while $\mathbf{q}_m$, $m=0,1,\hdots, M_{\mathrm{SMP}}-1$, are the $M_{\mathrm{SMP}}$ LUTs of the model, containing the control points for the spline interpolation in each parallel branch. The proposed SMP processing structure, adopting also the injection principle but in generalized form, is illustrated in Fig.~3(b). %\ref{fig:SMP}

In general, in terms of the modeling capabilities, the SMP is a richer model compared to SPH, while it also naturally entails higher complexity. These models will be assessed and compared to classical DPD solutions in terms of complexity and performance in Sections \ref{sec:comp} and \ref{sec:results}.
%In systems presenting a memory configuration more complex than global memory, the SMP will provide a better modeling accuracy, as a complete LUT is given for each individual memory branch. However, the computational complexity of this model is also increased, leading to somewhat higher complexities when compared to the SPH model. Nonetheless, this metric will still remain simpler when compared to other classical DPD solutions, as is demonstrated later in the paper.
}

\section{Parameter Learning Rules}
\label{sec:Learning}

In this section, we derive efficient gradient-descent type learning rules for both proposed DPD approaches, to adaptively estimate and track the unknown parameters in each of the models. \rev{Notation-wise, to allow for sample-adaptive estimation, we denote the vectors to be estimated with a subindex $n$, i.e., $\mathbf{c}_n$ and $\mathbf{h}_n$ for SPH and $\mathbf{q}_{m,n}$,~$m=0,1,\hdots, M_{\mathrm{SMP}}-1$, for SMP, to indicate their time-dependence. %$\mathbf{c}_n$ and $\mathbf{h}_n$ from now on, to indicate their time-dependence.
}

\subsection{SPH Learning Rules}

%We next derive efficient gradient-descent type learning rules, to adaptively estimate and track the unknown parameters, namely the vectors $\mathbf{c}$ and $\mathbf{h}$ containing the spline control points and the FIR memory filter coefficients. \red{Notation-wise, to allow for sample-adaptive estimation, we denote the vectors by $\mathbf{c}_n$ and $\mathbf{h}_n$ from now on, to indicate their time-dependence.}
%
%Once calculated the output signal of the Hammerstein architecture, the algorithm can estimate the adaptive parameters describing the behavior of the PA, that is the set of control points $\mathbf{c}_n$ defining the nonlinearity, and the linear filter coefficients $\mathbf{h}_n$ accounting for the memory effects. This process is done by forcing the training output signal $y_{\mathrm{SP}}[n]$ to be equal to the PA input signal $x_{\mathrm{DPD}}[n]$, with the process described in the previous section. Thus, the more similar the estimated signal $y_{\mathrm{SP}}[n]$ lies with respect $x_{\mathrm{DPD}}[n]$, the more accurate $\mathbf{c}_n$ and $\mathbf{h}_n$ describe the inverse model of the PA.
%Those updates are then copied to the predistorter model.
%
To calculate the learning rule in the SPH case, the instantaneous error signal between $x_{\mathrm{DPD}}[n]$ and $r_{\mathrm{SPH}}[n]$, in the context of the considered ILA-type architecture is first extracted as
\begin{align}
    e_{\mathrm{SPH}}[n] = x_{\mathrm{DPD}}[n] - r_{\mathrm{SPH}}[n] = x_{\mathrm{DPD}}[n] - \mathbf{h}_n^T \mathbf{s}_n.
\end{align}
Then, to facilitate the gradient-descent learning \cite{haykin2008adaptive}, the cost function is defined as the instantaneous squared error, expressed as
%objective becomes now to minimize this expression by adjusting the adaptive parameters $\mathbf{c}_n$ and $\mathbf{h}_n$. One possibility is to follow the instantaneous gradient descent approach, with the quantities being adapted to the negative direction of the gradient of the cost function in each iteration \cite{scarpiniti2013nonlinear}, \cite{hansler2005acoustic}. Hence, the complex instantaneous cost function can be defined as:
\begin{align}
\label{eq:cost}
    J(\mathbf{h}_n,\mathbf{c}_n) = \abs{e_{\mathrm{SPH}}[n]}^2. % e^*[n].
\end{align}
The corresponding iterative learning rules are then obtained through the partial derivatives of $J(\mathbf{h}_n,\mathbf{c}_n)$ with respect both parameter vectors to adapt, expressed formally as
%\begin{align}
%\label{eq:w_n+1}
 %   \mathbf{h}_{n+1} & = \mathbf{h}_n - \mu_{\mathrm{h}}[n] \frac{\partial J(\mathbf{h}_n,\mathbf{c}_n)} {\partial \mathbf{h}_n},
%\end{align}
%%% HERE IS THE NABLA VERSION
\begin{align}
\label{eq:w_n+1}
    %\mathbf{h}_{n+1} & = \mathbf{h}_n - \mu_{\mathrm{h}}[n] \frac{\partial J(\mathbf{h}_n,\mathbf{c}_n)} {\partial \mathbf{h}_n},
    \mathbf{h}_{n+1} & = \mathbf{h}_n - \mu_{\mathrm{h}}[n] \mathbf{\nabla}_{\mathbf{h}_n}J(\mathbf{h}_n,\mathbf{c}_n) ,
\end{align}
\begin{align}
\label{eq:q_n+1}
    \mathbf{c}_{n+1} & = \mathbf{c}_n - \mu_{\mathrm{c}}[n]  \mathbf{\nabla}_{\mathbf{c}_n} J(\mathbf{h}_n,\mathbf{c}_n) ,
\end{align}
where \rev{$\mathbf{\nabla}_{\mathbf{x}}$ refers to the complex gradient operator \cite{haykin2008adaptive,4645581} of a real-valued function against complex-valued parameter vector $\mathbf{x}$.} Additionally, $\mu_\mathrm{h}[n]$ and $\mu_\mathrm{c}[n]$ are the learning rates for $\mathbf{h}_n$ and $\mathbf{c}_n$, respectively, at time instant $n$. After relatively straight-forward derivations, the resulting concrete learning rules read
%
%The reader is referred to \ref{ap:Apendix1} to find the expound of these mathematical expressions. As stated there, the resulting learning expressions read
\begin{align}
    \mathbf{h}_{n+1} & = \mathbf{h}_n + \mu_{\mathrm{h}}[n] e_{\mathrm{SPH}}[n] \mathbf{s}_n^*,  \label{eq:learning_rule_h} \\[10pt]
    %\vspace{10mm}
    \mathbf{c}_{n+1} & = \mathbf{c}_{n} + \mu_{\mathrm{c}}[n] e_{\mathrm{SPH}}[n] \mathbf{\Sigma}_{n}^T \mathbf{Z}_n^* \mathbf{h}_n^*, \label{eq:learning_rule_c}
\end{align}
where the diagonal matrix $\mathbf{Z}_n \in \mathbb{C}^{M_{\mathrm{SPH}} \times M_{\mathrm{SPH}}} = \operatorname{diag} \left\{ z_{\mathrm{SPH}}[n],\; z_{\mathrm{SPH}}[n-1], \; \cdots,\; z_{\mathrm{SPH}}[n-M_{\mathrm{SPH}}+1] \right\}$, and $\mathbf{\Sigma}_{n}$ contains $M_{\mathrm{SPH}}$ previous instances of $\mathbf{g}_n$, defined as $\mathbf{\Sigma}_{n} \in \mathbb{R}^{M_{\mathrm{SPH}} \times Q} = \begin{bmatrix}\mathbf{g}_{n} & \mathbf{g}_{n-1} & \cdots & \mathbf{g}_{n-M_{\mathrm{SPH}}+1}\end{bmatrix}^T$. 
These learning rules in (\ref{eq:learning_rule_h}) and (\ref{eq:learning_rule_c}) are executed in parallel such that both parameter vectors are updated simultaneously. For readers' convenience, an example illustration of the structure of the matrix $\mathbf{\Sigma}_{n}$ is given in (\ref{eq:sigma}), for $M_{\mathrm{SPH}}=4$, $Q=9$, and $P_{\mathrm{SP}}=3$, assuming representative example values of the index variable $i_n$.
%\vspace{3pt}
\begin{align}
\label{eq:sigma}
    \mathbf{\Sigma}_{n}=\begin{bmatrix}
    0 \quad 0 \quad \underbrace{[* \quad * \quad * \quad *]}_{\mathbf{u}_{n}^{T}\mathbf{B}_{P_{\mathrm{SP}}}} \quad 0 \quad 0 \quad 0\\
    \underbrace{[* \quad * \quad * \quad *]}_{\mathbf{u}_{n-1}^{T}\mathbf{B}_{P_{\mathrm{SP}}}} \quad 0 \quad 0 \quad 0 \quad 0 \quad 0\\
    0 \quad 0 \quad 0 \quad \underbrace{[* \quad * \quad * \quad *]}_{\mathbf{u}_{n-2}^{T}\mathbf{B}_{P_{\mathrm{SP}}}} \quad 0 \quad 0\\
    0 \quad 0 \quad 0 \quad 0 \quad 0 \quad \underbrace{[* \quad * \quad * \quad *]}_{\mathbf{u}_{n-3}^{T}\mathbf{B}_{P_{\mathrm{SP}}}}\\
    \end{bmatrix},
    \begin{array}{l}
    i_n = 3\\[17pt]
    i_{n-1} = 1\\[17pt]
    i_{n-2} = 4\\[17pt]
    i_{n-3} = 6\\[3pt]
    \textcolor{white}{.}
    \end{array}
\end{align}
Note that the term $\mathbf{u}_{n}^T \mathbf{B}_{P_{\mathrm{SP}}}$ is located in $\mathbf{\Sigma}_{n}$ at each iteration $n$ according to the span index $i_{n}$, as shown in \eqref{eq:g_n}. \rev{It is noted that the derived learning rules in (\ref{eq:learning_rule_h}) and (\ref{eq:learning_rule_c}) are novel, as the overall Hammerstein system is known to be not linear in its parameters.}

\subsection{SMP Learning Rules}

\rev{

We next derive gradient-based iterative learning rules for the SMP model.  Different to SPH case, the SMP model does not contain cascaded filters while the learning entity considers the $M_{\mathrm{SMP}}$ parallel spline-interpolated LUTs, specifically their control points $\mathbf{q}_{m,n}$,~$m=0,1,\hdots, M_{\mathrm{SMP}}-1$.
%LUT  the resulting learning is intrinsically more simple, as the spline-based nonlinearity no longer depends on the latter FIR filter present in the Hammerstein structure. However, these learnings need to be calculated for each individual branch, as the LUTs are independent from each other.

Following now a similar approach as earlier, the instantaneous error signal is first defined in the context of ILA-based learning as
\begin{align}
    e_{\mathrm{SMP}}[n] & = \! x_{\mathrm{DPD}}[n] - r_{\mathrm{SMP}}[n] \nonumber \\
    & = \! x_{\mathrm{DPD}}[n]  -  z_{\mathrm{SMP}}[n] \! - \! \! \! \sum_{m=0}^{M_{\mathrm{SMP}}-1} \! \! z_{\mathrm{SMP}}[n-m] \mathbf{g}_{n-m}^T \mathbf{q}_{m,n}.
\end{align}
For the gradient-descent learning, the cost function is defined as a function of the instantaneous error signal as
\begin{align}
    J(\mathbf{q}_{0,n},\mathbf{q}_{1,n},\cdots,\mathbf{q}_{M_{\mathrm{SMP}}-1,n}) = \abs{e_{\mathrm{SMP}}[n]}^2.
\end{align}
Then, by adopting again the complex gradient operator \cite{4645581}, the learning rule for the $m$th LUT can be written as
\begin{align}
\label{eq:q}
    \mathbf{q}_{m,n+1} = \mathbf{q}_{m,n} - \mu_{\mathrm{q}_m} \mathbf{\nabla}_{\mathbf{q}_{m,n}} J(\mathbf{q}_{0,n},\mathbf{q}_{1,n},\cdots,\mathbf{q}_{M_{\mathrm{SMP}}-1,n}),
\end{align}
while by following the complex differentiation steps, the final learning rule for the $m$th LUT reads
\begin{align}
    \label{eq:learning_rule_q}
    \mathbf{q}_{m,n+1} = \mathbf{q}_{m,n} + \mu_{\mathrm{q}_m}[n] \; e_{\mathrm{SMP}}[n] \; z_{\mathrm{SMP}}^*[n-m] \; \mathbf{g}_{n-m}.
\end{align}
These learning rules are adopted for all the involved $M_{\mathrm{SMP}}$ LUTs in parallel.
}

\section{Computational Complexity Analysis}
\label{sec:comp}

In this section, a computational complexity analysis and comparison between the proposed SPH, SMP and a widely-applied canonical MP DPD with self-orthogonalizing least-mean square (LMS) \cite{haykin2008adaptive} parameter adaptation is presented. LMS type adaptation is deliberately assumed also for MP DPD, for the fairness of the comparison. The complexity analysis is carried out in terms of real multiplications per linearized data sample, as multiplications are commonly more resource-intensive operations than additions in digital signal processing (DSP) implementations \cite{tehrani2010comparative}.

%The DPD main path is the most critical stage of a predistorter, as it needs to be executed continuously and in real time to predistort any transmitting signal. Low-complexity DPD algorithms allow to use fewer, cheaper, and smaller components in the TX chain, leading to important economical savings for semiconductor manufacturers and telecommunication operators.
%
%%% BELOW COMMENTED OUT TO SAVE SPACE
%
%In general, the DPD main processing/transmit path is the most critical stage, complexity wise, as it needs to be executed continuously and in real time to predistort the transmit signal. However, especially if frequent or even  continuous DPD coefficient learning/tracking is also required, the parameter learning complexity is also of large importance. Hence, in our complexity analysis, both the DPD main path and the DPD learning stage are addressed.

\rev{The quantitative complexity assessment of the proposed gradient-adaptive SPH DPD and SMP DPD follows the exact processing steps described in Sections II and III.} It is noted that the complexity expressions reported below basically represent an upper bound for the required arithmetical operations, as in real implementations some elementary or trivial operations such as multiplying by any integer power of $2$ or $1/2$ does not really reflect any actual complexity, while are included as normal operations in the expressions for simplicity. \rev{Additionally, it is noted that the modulus operator, needed in (\ref{eq:i_n}) and (\ref{eq:u_n}), is assumed to be calculated with the alpha max beta min algorithm~\cite{Frerking94a}. Finally, in the complexity analysis, we consider uniform splines with $\Delta_z = \Delta_x = 1$.
}
%\rev{For clarity, following the notations in Fig.~\ref{fig:DPD_concept}, it is noted that in the main DPD path, the predistorter input is the data signal $x[n]$ while in the learning, the postdistorter input is the feedback signal $z[n]$.}
%The next lines present a complexity analysis for the DPD models, following the flow of the particular algorithms implemented in each.

\subsection{Complexity of Proposed SPH Method}

 % of those (e.g. multiplications by 1, 0, or bit shifts) can be avoided when a real time implementation is to be made. 

With reference to Fig.~3(a) and the underlying processing elements, the generic complexity expressions can be stated in a straight-forward manner as follows:

\begin{itemize}
    \item \textit{DPD main path}, starting with the input signal $x[n]$. %, in terms of real multiplications per linearized sample. 
    The complexity of the predistorter intermediate signal, $s_{\mathrm{DPD}}[n]$, includes the processing in (\ref{eq:i_n}), (\ref{eq:u_n}), and (\ref{eq:s_SPH}) but with $x[n]$ as the input. These together with the FIR filtering in (\ref{eq:r_SPH}) to calculate $x_{\mathrm{DPD}}[n]$ yield the following complexity expressions \vspace{5pt}
    \begin{enumerate}
        \item $s_{\mathrm{DPD}}[n]$ $\rightarrow P_{\mathrm{SP}}^2 + 4P_{\mathrm{SP}} + 10$.\vspace{5pt}
        \item $x_{\mathrm{DPD}}[n]$ $\rightarrow 4M_{\mathrm{SPH}}$.\vspace{5pt}
    \end{enumerate}
    \item \textit{DPD learning}, for observed signal $z_{\mathrm{SPH}}[n]$. The generation of the error signal $e_{\mathrm{SPH}}[n]$ contains the same multiplication operations as in $s_{\mathrm{DPD}}[n]$ and $x_{\mathrm{DPD}}[n]$, due to the ILA architecture. The complexity of updating $\mathbf{h}_{n}$ and $\mathbf{c}_{n}$ corresponds to calculating (\ref{eq:learning_rule_h}) and (\ref{eq:learning_rule_c}), respectively. Overall, we thus get\vspace{5pt}
        \begin{enumerate}
        \item $e_{\mathrm{SPH}}[n]$ $\rightarrow P_{\mathrm{SP}}^2 + 4P_{\mathrm{SP}} + 4M_{\mathrm{SPH}}+ 10$.\vspace{5pt}
        \item $\mathbf{h}_{n+1}$ $\rightarrow 4M_{\mathrm{SPH}}+2$.\vspace{5pt}
        \item $\mathbf{c}_{n+1}$ $\rightarrow 2P_{\mathrm{SP}}M_{\mathrm{SPH}} + 4P_{\mathrm{SP}} + 6M_{\mathrm{SPH}} + 6$. \vspace{5pt}
    \end{enumerate}
\end{itemize}
\red{Interestingly, it is noted that the amount of multiplications in the DPD main path does not depend on the chosen number of control points $Q$, or equivalently the number of regions, as the spline-interpolation algorithm basically utilizes $P_{\mathrm{SP}}+1$ control points for any given region.}

\rev{
\subsection{Complexity of Proposed SMP Method}

With the SMP approach, as shown in~(\ref{eq:MPLUT}) for post-distortion, there is no separate linear filtering stage but the overall DPD output is composed as a sum of  $M_{\mathrm{SMP}}$ parallel spline-interpolated LUTs with input samples $x[n-m]$, $m=0,1,\hdots,M_{\mathrm{SMP}}-1$. % and the resulting contributions are then added up to obtain the predistorted signal $x_{\mathrm{DPD}}[n]$. The learning stage follows a similar process, calculating at the end the corresponding coefficient updates for each individual branched LUT.
Therefore, with reference to Fig.~3(b) and the underlying processing ingredients described in Sections II and III, the main path and parameter learning complexities can be stated as follows:
\begin{itemize}
    \item \textit{DPD main path}, starting with the input signal $x[n]$. The complexity involves calculating $x_{\mathrm{DPD}}[n]$, as in (\ref{eq:MPLUT}), with $x[n]$ as the input. By taking into account that at time instant $n$, only $\mathbf{g}_n$ needs to be calculated while $\mathbf{g}_{n-1}, \hdots, \mathbf{g}_{n-M_\mathrm{SMP}+1}$ are available from previous sample instant, we obtain the following overall complexity expression
    \begin{enumerate}
        \item $x_{\mathrm{DPD}}[n]$ $\rightarrow P_{\mathrm{SP}}^2 + 3 P_{\mathrm{SP}} + 2 P_{\mathrm{SP}} M_{\mathrm{SMP}} + 6M_{\mathrm{SMP}} + 4 $.\vspace{5pt}
    \end{enumerate}
    \item \textit{DPD learning path}, for observed signal $z_{\mathrm{SMP}}[n]$. Due to the ILA architecture, the involved complexity of calculating the error signal $e_{\mathrm{SMP}}[n]$ is, arithmetically, the same as calculating $x_{\mathrm{DPD}}[n]$. Additionally, the complexity of updating one of the LUTs or spline control point vectors, $\mathbf{q}_{m,n}$, corresponds to calculating (\ref{eq:learning_rule_q}). Thus, we get
    \begin{enumerate}
        \item $e_{\mathrm{SMP}}[n]$ $\rightarrow P_{\mathrm{SP}}^2 + 3 P_{\mathrm{SP}} + 2 P_{\mathrm{SP}} M_{\mathrm{SMP}} + 6M_{\mathrm{SMP}} + 4$.\vspace{5pt}
        \item $\mathbf{q}_{m,n+1} $ $\rightarrow 2 P_{\mathrm{SP}} + 8$.
    \end{enumerate}
\end{itemize}

}

\subsection{Complexity of Reference MP DPD}

% \textit{1) Linear interpolated LUT:} The linear interpolated LUT is basically a special case of the SPH or SMP model, with $P_{\mathrm{SP}} = 1$ and $M_{\mathrm{SPH}} = 0$ or $M_{\mathrm{SMP}} = 0$, and thus its computational complexity can be directly obtained through the expressions shown in the previous subsection.

When considering the LMS-adaptive MP DPD with monomial basis functions (BFs), in the context of ILA architecture in Fig.~1(a), we first write the postdistorter output sample as %model, a classical LMS-Newton algorithm is used to generate/update the DPD coefficients from the BFs of the PA output signal $y[n]/G$ and the predistorted signal $x_{\mathrm{DPD}}[n]$ of Fig. \ref{fig:ILA}
%\cite{paulo2008adaptive}
%. The first step to learn the LMS solution is to compute the output signal $y_{\mathrm{MP}}[n]$ as
\begin{align}
    r_{\mathrm{MP}}[n] = \mathbf{w}_{n}^T \mathbf{l}_{n} ,
\end{align}
where $\mathbf{w}_{n} \in \mathbb{C}^{m \times 1}$ is the MP DPD coefficient vector, with $m=\ceil{\frac{P_{\mathrm{MP}}}{2}} M_{\mathrm{MP}}$ denoting the number of coefficients, while $P_{\mathrm{MP}}$ and $M_{\mathrm{MP}}$ are the assumed polynomial order and memory length (per nonlinearity order), respectively. Additionally, the vector of the basis function samples $\mathbf{l}_n$ used to calculate the current output is as defined in~(\ref{eq:ln}), next page, where $z_{\mathrm{MP}}[n]$ denotes the observed feedback signal at postdistorter input.

\begin{figure*} %[t!]
\centering
\label{eq:ln1}
\footnotesize{
\begin{align}
\label{eq:ln}
    \mathbf{l}_n = & \Big[ z_{\mathrm{MP}}[n] \quad z_{\mathrm{MP}}[n] \, \abs{z_{\mathrm{MP}}[n]}^2 \; \cdots \; z_{\mathrm{MP}}[n] \, \abs{z_{\mathrm{MP}}[n]}^{P_{\mathrm{MP}}-1} \quad z_{\mathrm{MP}}[n-1] \quad z_{\mathrm{MP}}[n-1] \, \abs{z_{\mathrm{MP}}[n-1]}^2 \; \cdots \; z_{\mathrm{MP}}[n-1] \, \abs{z_{\mathrm{MP}}[n-1]}^{P_{\mathrm{MP}}-1} \nonumber \\ & \; \; z_{\mathrm{MP}}[n-M_{\mathrm{MP}}+1] \quad z_{\mathrm{MP}}[n-M_{\mathrm{MP}}+1] \, \abs{z_{\mathrm{MP}}[n-M_{\mathrm{MP}}+1]}^2 \; \cdots \; z_{\mathrm{MP}}[n-M_{\mathrm{MP}}+1] \, \abs{z_{\mathrm{MP}}[n-M_{\mathrm{MP}}+1]}^{P_{\mathrm{MP}}-1} \Big]^T.
\end{align}
}
\hrulefill
\end{figure*}

\begin{table*}[t!]
\centering
\setlength{\tabcolsep}{2pt}
\renewcommand{\arraystretch}{1.7}
\caption{\textsc{Complexity expressions in terms of real multiplications per sample for the proposed SPH, the proposed SMP and the reference canonical MP methods, covering both the DPD main path processing and the DPD parameter learning, with $m=\ceil{\frac{P_{\mathrm{MP}}}{2}} M_{\mathrm{MP}}$}}
\label{tab:complexity}
\begin{tabular}{|c|c|c|c|c|}
\hline
\multicolumn{2}{|c|}{\textbf{Operation}} & \textbf{SPH model} & \rev{ \textbf{SMP model} } & \textbf{MP model} \\ \hline
\multirow{3}{*}{\textbf{\ Predistortion \ }} & Nonlinearity & $P_{\mathrm{SP}}^2 + 4P_{\mathrm{SP}} + 10$ & \rev{ $P_{\mathrm{SP}}^2 + 3 P_{\mathrm{SP}} + 2 P_{\mathrm{SP}} M_{\mathrm{SMP}} + 6M_{\mathrm{SMP}} + 4$ } & $3 \ceil*{\frac{P_{\mathrm{MP}}}{2}}-2$ \\ \cline{2-5}
 & Filtering & $4M_{\mathrm{SPH}}$ & \rev{ 0 } & $4m$ \\ \cline{2-5}
 & \textbf{Total} & $P_{\mathrm{SP}}^2 + 4P_{\mathrm{SP}} + 4M_{\mathrm{SPH}}+10$ & \rev{ $P_{\mathrm{SP}}^2 + 3 P_{\mathrm{SP}} + 2 P_{\mathrm{SP}} M_{\mathrm{SMP}} + 6M_{\mathrm{SMP}} + 4$ } & $3 \ceil*{\frac{P_{\mathrm{MP}}}{2}} + 4m - 2$ \\ \hline
\multirow{3}{*}{\textbf{Learning}} & Error signal & $P_{\mathrm{SP}}^2 + 4P_{\mathrm{SP}} + 4M_{\mathrm{SPH}}+10$ & \rev{ $P_{\mathrm{SP}}^2 + 3 P_{\mathrm{SP}} + 2 P_{\mathrm{SP}} M_{\mathrm{SMP}} + 6M_{\mathrm{SMP}} + 4 $ } & $3 \ceil*{\frac{P_{\mathrm{MP}}}{2}} + 4m - 2$ \\ \cline{2-5} 
 & Update & $2P_{\mathrm{SP}}M_{\mathrm{SPH}} + 4P_{\mathrm{SP}} + 10M_{\mathrm{SPH}} \! + \! 8$ & \rev{ $M_{\mathrm{SMP}}(2P_{\mathrm{SP}}+8)$ } & $4m^2 + 4m + 2$ \\ \cline{2-5} 
& \textbf{Total} & $\ P_{\mathrm{SP}}(P_{\mathrm{SP}} + 2M_{\mathrm{SPH}} + 8) + 14M_{\mathrm{SPH}} \! + \! 18 \ $ & \rev{ $ \ P_{\mathrm{SP}}^2 + 3 P_{\mathrm{SP}} + 4 P_{\mathrm{SP}} M_{\mathrm{SMP}} + 14 M_{\mathrm{SMP}} + 4 \ $ } & $ \ 3 \ceil*{\frac{P_{\mathrm{MP}}}{2}} \! + \! 4m^2 \! + \! 8m \ $ \\ \hline
\end{tabular}
\end{table*}

Once $r_{\mathrm{MP}}[n]$ is calculated, the error signal can be directly obtained as
\begin{align}
    e_{\mathrm{MP}}[n] = x_{\mathrm{DPD}}[n] - r_{\mathrm{MP}}[n] = x_{\mathrm{DPD}}[n] - \mathbf{w}_{n}^T \mathbf{l}_{n},
\end{align}
and the coefficient update can be written as
\begin{align}
    \label{eq:MP_LMS}
    \mathbf{w}_{n+1} = \mathbf{w}_{n} + \mu_{\mathrm{w}}[n] e_{\mathrm{MP}}[n] \mathbf{R}^{-1} \mathbf{l}^*_{n},
\end{align}
where $\mu_{\mathrm{w}}[n]$ is the learning rate, and $\mathbf{R}^{-1}$ is the inverse of the autocorrelation matrix of the PA output basis function samples \cite{haykin2008adaptive}. {We assume that a block of $N_\mathrm{B}$ samples is used to calculate the sample estimate of $\mathbf{R}$, and include below the corresponding complexity for completeness of the study. Importantly, it is also noted that the self-orthogonalizing type transformation $\mathbf{R}^{-1}$ in (\ref{eq:MP_LMS}) is an important ingredient for stable operation, as the MP basis function samples in (\ref{eq:ln}) are known to be largely correlated \cite{1337325}. \rev{Alternatively, orthogonal polynomial type set of basis functions could be used \cite{1337325,5757879}, though with increased main path complexity.
The SPH and SMP DPD related learning rules in (\ref{eq:learning_rule_h})-(\ref{eq:learning_rule_c}) and  (\ref{eq:learning_rule_q}), on the other hand, do not suffer from such correlation challenge, and are shown in Section V to provide reliable linearization without any additional (self-)orthogonalization. This is one clear benefit, complexity-wise, compared to the existing gradient-adaptive reference DPD solutions.}}

Building on above, the \rev{self-orthogonalizing} LMS-adaptive MP DPD complexity can be detailed as follows
\begin{itemize}
    \item DPD main path, starting with the input signal $x[n]$, in terms of real multiplications per linearized sample:\vspace{5pt}
    \begin{enumerate}
        \item MP BF samples $\rightarrow 3 \ceil*{\frac{P_{\mathrm{MP}}}{2}} - 2$. \vspace{5pt}
        %\item Orthogonalization: $\rightarrow B\left( 4\sum_{i=1}^{m}i -2m \right)$ FLOPs.\vspace{5pt}
        \item $x_{\mathrm{DPD}}[n]$ $\rightarrow 4m$. \vspace{5pt}
    \end{enumerate}
    \item DPD training, for observed signal $z_\mathrm{MP}[n]$: \vspace{5pt}
        \begin{enumerate}
        \item MP BF samples $\rightarrow 3 \ceil*{\frac{P_{\mathrm{MP}}}{2}} - 2$.\vspace{5pt}
        %\item Orthogonalization $\rightarrow B\left( 4\sum_{i=1}^{m}i -2m \right)$ FLOPs.\vspace{5pt}
        \item $\mathbf{R}^{-1}$ $\rightarrow m^3$. \vspace{5pt}
        \item $r_{\mathrm{MP}}[n]$ $\rightarrow 4m$. \vspace{5pt}
        \item $e_{\mathrm{MP}}[n]$ $\rightarrow 3 \ceil*{\frac{P_{\mathrm{MP}}}{2}} + 4m - 2$. \vspace{5pt}
        \item $\textbf{w}_{n+1}$ $\rightarrow 4m^2 + 4m + 2$. \vspace{5pt}
    \end{enumerate}
\end{itemize}

%Here, $B$ represents the size of the block of samples used to compute the autocorrelation matrix $\mathbf{R}$. The calculation cost of $\mathbf{R}^{-1}$ is included here for the sake of detailing the whole flow of the algorithm, but it is excluded in the all numerical complexity comparison presented in this work, as it can be considered as a precomputed and fixed matrix, its calculation only needed in the first iterations.

% \begin{figure*}[ht]
% \begin{equation}\begin{split}\label{equation_mem_PIM}
%     s_{\mathrm{PIM}}[n] = &\sum_{k_{11}=0}^{2}\sum_{k_{12}=0}^{2-k_{11}}\cdots\sum_{k_{1M}=0}^{2-\sum_{i=1}^{M-1}k_{1i}}
% \sum_{k_{21}=0}^{1}\sum_{k_{22}=0}^{1-k_{21}}\cdots\sum_{k_{2M}=0}^{1-\sum_{i=1}^{M-1}k_{2i}} \sum_{l=-L_1}^{L_2}\gamma_{l,k_{11}, \cdots, k_{1M}, k_{21}, \cdots, k_{2M}} \times \\
% &s_1[n-l+M_1]^{k_{11}}s_1[n-l+M_1-1]^{k_{12}}\cdots s_1[n-l-M_2]^{2-\sum_{i=1}^{M}k_{1i}} \times \\ 
% &s_2^*[n-l+M_1]^{k_{21}}s_2^*[n-l+M_1-1]^{k_{22}}\cdots s_2^*[n-l-M_2]^{1-\sum_{i=1}^{M}k_{2i}}.
% %\vspace{10mm}
% \end{split}
% %\vspace{10mm}
% \end{equation}
% \vspace{2mm}
% %\setcounter{equation}{\value{mytempeqncnt}}
% \hrulefill
% \end{figure*}
% \setlength{\arrayrulewidth}{0.2mm}
% \setlength{\tabcolsep}{3pt}
% \renewcommand{\arraystretch}{1.5}
% %

\subsection{Summary and Comparison}

Table~\ref{tab:complexity} collects and summarizes the deduced expressions for the numbers of real multiplications per sample needed for the fundamental main path processing and parameter learning stages in the proposed SPH, SMP and the reference MP DPD methods. In this table, when it comes to MP DPD, we have excluded the complexity related to the calculation of the elements of $\mathbf{R}$ and its inverse, as those are something that can be considered carried out offline, or within the very first phases of the overall learning procedure. 
%first ILA iteration, and thus we exclude the computation of the elements f $R$ in our complexity analysis.
%
%Once studied the individual complexity of the SPH and MP models, a comparison between them can be carried out.  Note that, for the SPH model, all of these expressions give an upper bound for the final complexity, as in a hardware implementation many trivial operations can be avoided.

%
%\red{
%However, in all presented numerical results, these computationally trivial operations are excluded, showing therefore the exact number of operations to be executed in each particular case.
%}

\begin{table}[t!]
\centering
\setlength{\tabcolsep}{5pt}
\renewcommand{\arraystretch}{1.5}
\caption{\textsc{Numerical complexity values, in terms of real multiplications per sample, for $P_{\mathrm{SP}}=3$, $M_{\mathrm{SPH}} = M_{\mathrm{SMP}} = 4$, $Q_{\mathrm{SPH}} = Q_{\mathrm{SMP}} = 7$, $P_{\mathrm{MP}} = 11$, and $M_{\mathrm{MP}} = 4$.}}
\label{tab:flops}
\begin{tabular}{c|c|c|c|}
\cline{2-4}
 & \textbf{SPH model} & \multicolumn{1}{c|}{\rev{\textbf{SMP model}}} & \textbf{MP model} \\ \hline
\multicolumn{1}{|c|}{\textbf{No. of coefficients}} & \multicolumn{1}{c|}{14} & \multicolumn{1}{c|}{\rev{31}} & \multicolumn{1}{c|}{24} \\ \hline \hline
\multicolumn{1}{|c|}{\textbf{Nonlinearity}} & 24 & \rev{63} & 16 \\ \hline
\multicolumn{1}{|c|}{\textbf{Filtering}} & 16 & \rev{0} & 96 \\ \hline
\multicolumn{1}{|c|}{\textbf{Total main path}} & 40 & \rev{63} & 112 \\ \hline
\multicolumn{1}{|c|}{\textbf{Reduction (against MP)}} &  64.3\% & \rev{ 43.7\% } & -- \\ \hline \hline
\multicolumn{1}{|c|}{\textbf{Error signal}} & 40 & \rev{63} & 112 \\ \hline
\multicolumn{1}{|c|}{\textbf{Coeff. update}} & 84 & \rev{56} & 2402 \\ \hline
\multicolumn{1}{|c|}{\textbf{Total learning}} & 124 & \rev{119} & 2514 \\ \hline
\multicolumn{1}{|c|}{\textbf{Reduction (against MP)}} &  95.0\% & \rev{ 95.2\% } & -- \\ \hline
\end{tabular}
\end{table}

%%%% COMMENTING BELOW OUT TO SAVE SPACE, THIS WAS TABLE III EARLIER
% \begin{table}[t!]
% \centering
% \setlength{\tabcolsep}{5pt}
% \renewcommand{\arraystretch}{1.5}
% \caption{\textsc{Numerical complexity values, in terms of real multiplications per sample, for $P_{\mathrm{SP}}=3$, $M_{\mathrm{SPH}} = 17$, $Q_{\mathrm{SPH}} = 10$, $M_{\mathrm{SMP}} = 4$, $Q_{\mathrm{SPH}} = 7$, $P_{\mathrm{MP}} = 11$, and $M_{\mathrm{MP}} = 5$.}}
% \label{tab:params}
% \begin{tabular}{c|c|c|c|}
% \cline{2-4}
%  & \textbf{SPH model} & \multicolumn{1}{c|}{\rev{\textbf{SMP model}}} & \textbf{MP model} \\ \hline
% \multicolumn{1}{|c|}{\textbf{No. of coefficients}} & \multicolumn{1}{c|}{30} & \multicolumn{1}{c|}{\rev{31}} & \multicolumn{1}{c|}{30} \\ \hline \hline
% \multicolumn{1}{|c|}{\textbf{Nonlinearity}} & 24 & \rev{63} & 16 \\ \hline
% \multicolumn{1}{|c|}{\textbf{Filtering}} & 68 & \rev{0} & 120 \\ \hline
% \multicolumn{1}{|c|}{\textbf{Total main path}} & 92 & \rev{63} & 136 \\ \hline
% \multicolumn{1}{|c|}{\textbf{Reduction (against MP)}} &  32.3\% & \rev{ 53.7\% } & -- \\ \hline \hline
% \multicolumn{1}{|c|}{\textbf{Error signal}} & 92 & \rev{63} & 136 \\ \hline
% \multicolumn{1}{|c|}{\textbf{Coeff. update}} & 292 & \rev{56} & 3722 \\ \hline
% \multicolumn{1}{|c|}{\textbf{Total learning}} & 384 & \rev{119} & 3858 \\ \hline
% \multicolumn{1}{|c|}{\textbf{Reduction (against MP)}} &  90\% & \rev{ 97\% } & -- \\ \hline
% \end{tabular}
% \end{table}

Next, to obtain concrete numerical complexity numbers and to carry out a comparison, we study an example case where the SPH and SMP DPD spline polynomial order is $P_{\mathrm{SP}} = 3$. %, which is sufficient to reach a predistortion performance similar to that of the MP model in most scenarios (see Section \ref{sec:results}). 
Additionally, the number of control points per LUT is chosen to be $Q = 7$ for both SPH and \rev{SMP} models, %the width of the regions is assumed to be $\Delta_z = 1$ (uniform splines), 
and the considered memory length is $M_{\mathrm{SPH}} = M_{\mathrm{SMP}} = 4$. These constitute a total number of $14$ free parameters to be estimated in the SPH model \rev{and $31$ free parameters in the SMP case}. Then, the MP DPD polynomial order is chosen as $P_{\mathrm{MP}} = 11$, and the considered memory length per filter is $M_{\mathrm{MP}} = 4$. This configuration leads to $24$ free parameters in the MP DPD. %, constituting thus a fair starting point of the complexity comparison with roughly the same number of free parameters. 
Similar type parametrizations are used also in the actual DPD measurements and experiments, in Section \ref{sec:results}.

\begin{figure*}
\centering
    %\label{fig:FR1_setup}
    \begin{subfigure}[t]{1\textwidth}
        \includegraphics[width=1\textwidth, trim = 10 34 35 35,clip]{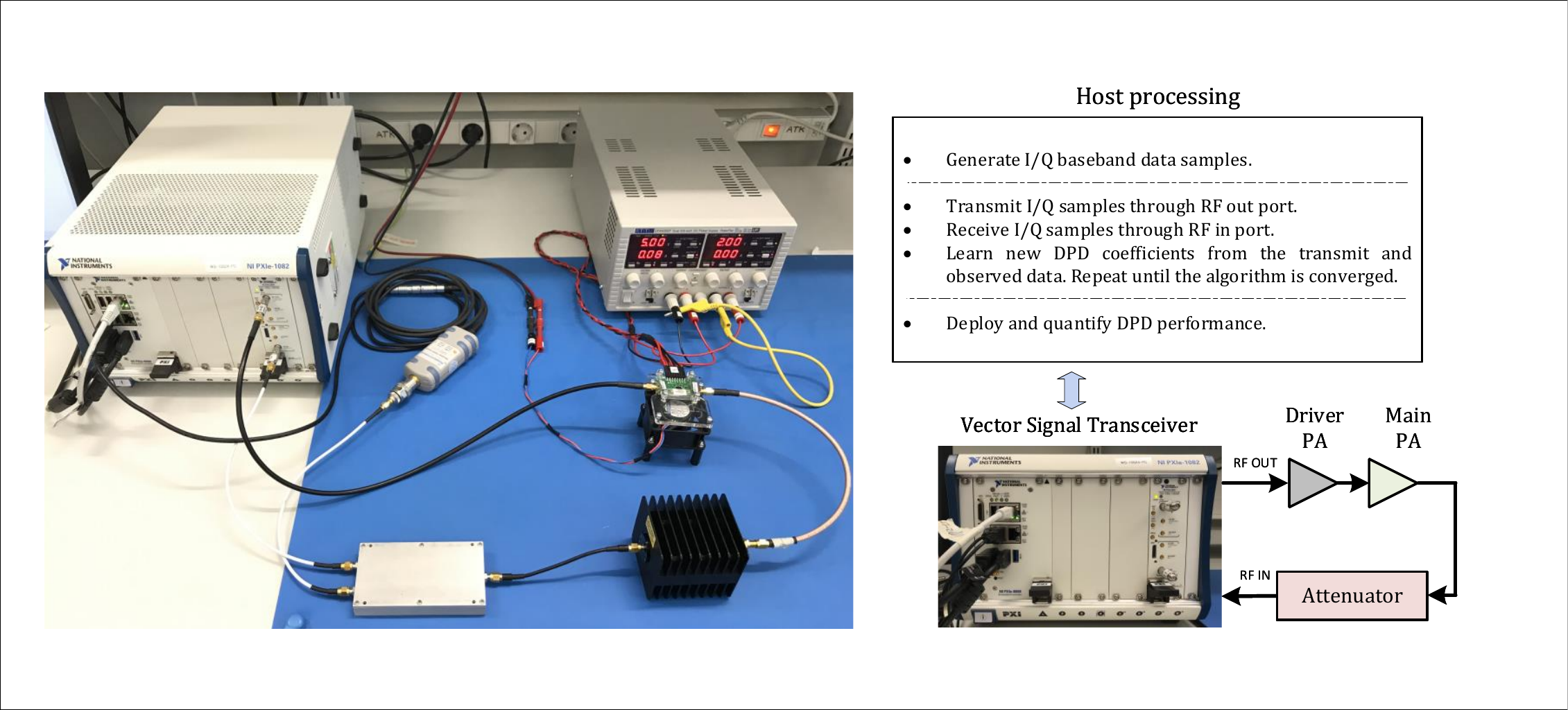}
        \caption{\footnotesize{RF measurement setup at FR-1.}}
        \label{fig:Setup1}
    \end{subfigure}
    
    \vspace{0.5cm}
    
    \begin{subfigure}{0.3\linewidth}
        \centering
        \includegraphics[width=0.7\textwidth, trim = 20 20 20 20,clip]{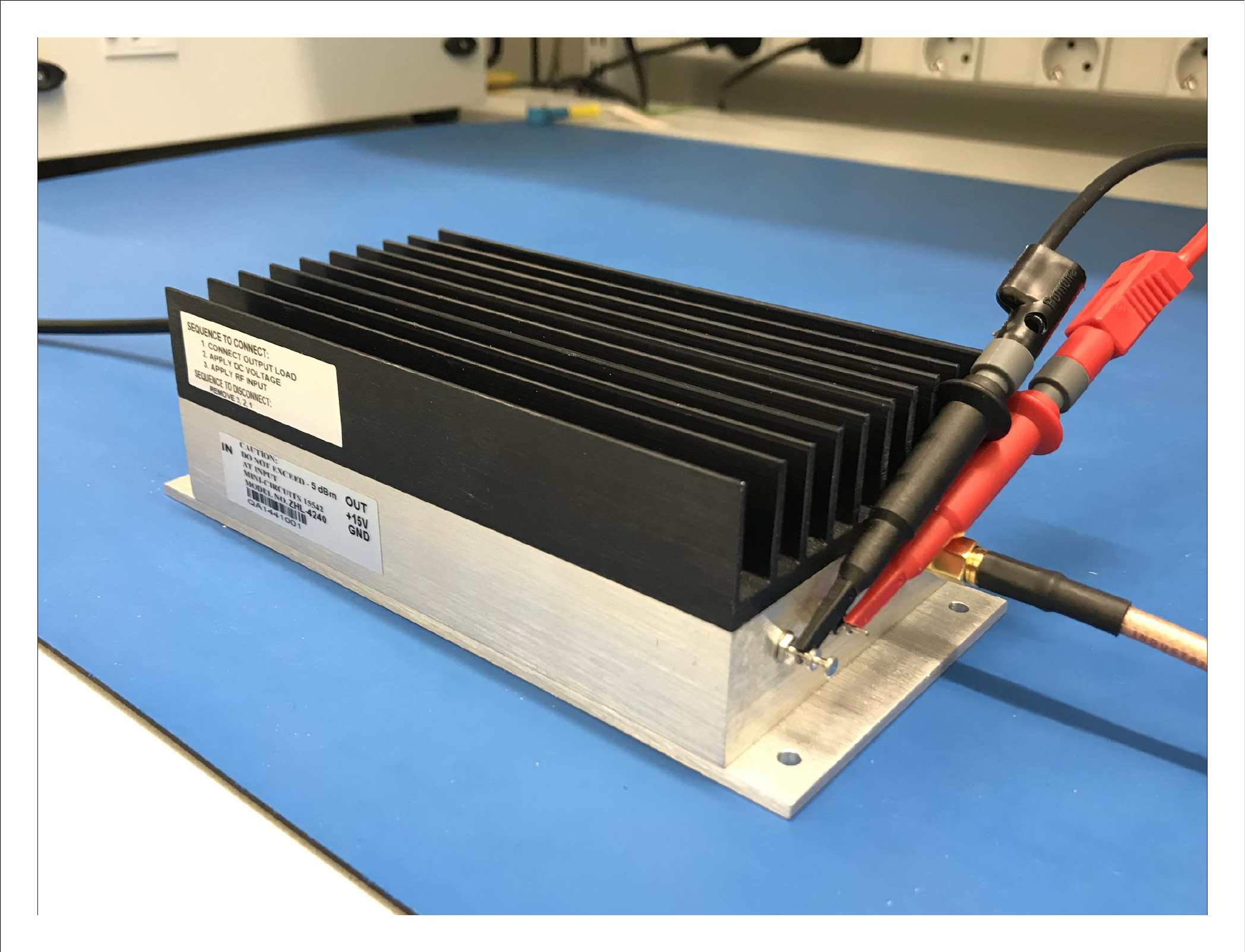}
        \caption{\footnotesize{General purpose wideband PA.}}
        \label{fig:Gen_PA}
    \end{subfigure}
    ~
    \begin{subfigure}{0.3\linewidth}
        \centering
        \includegraphics[width=0.7\textwidth, trim = 20 20 20 20,clip]{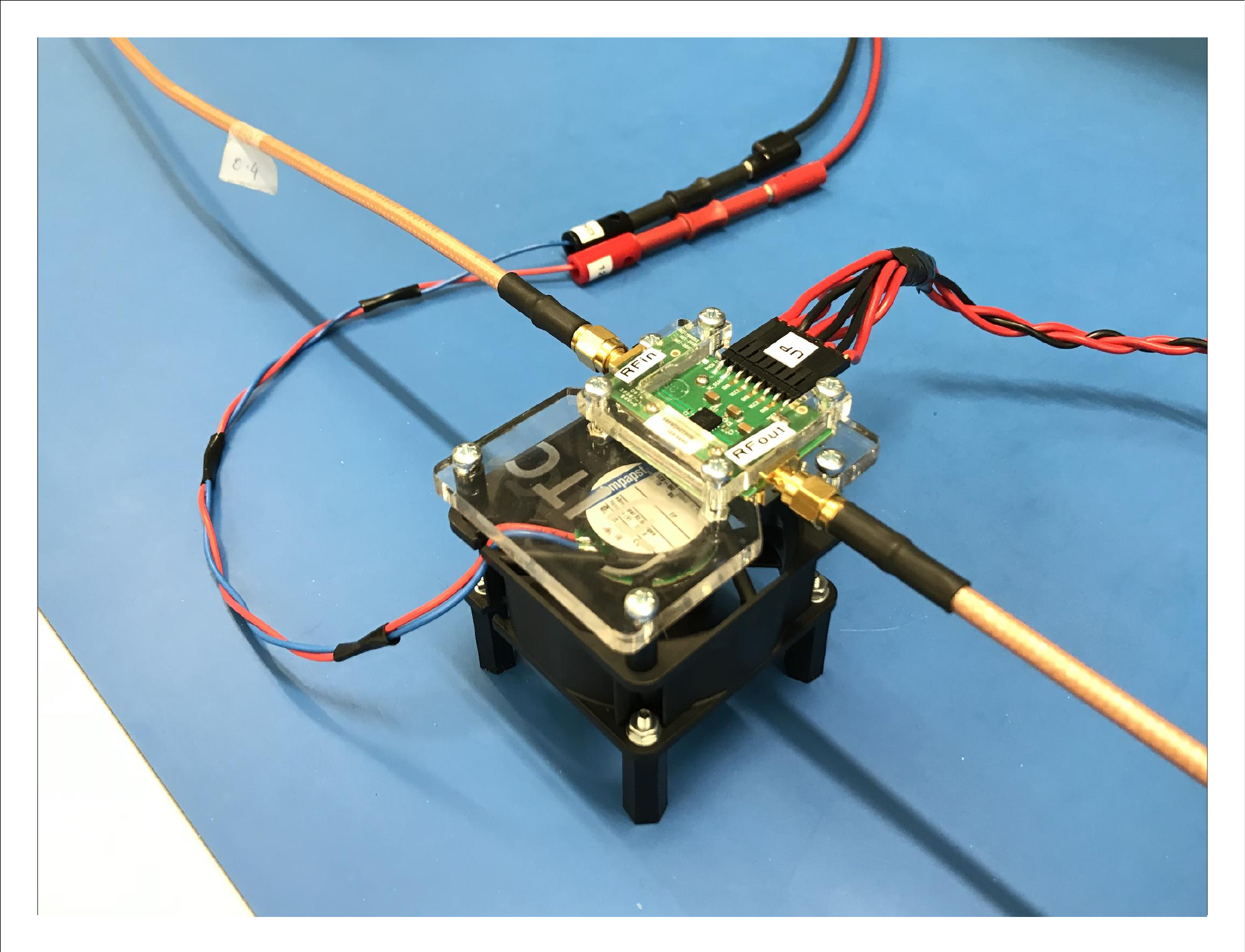}
        \caption{\footnotesize{Skyworks NR Band 78 PA.}}
        \label{fig:Skyworks_PA}
    \end{subfigure}
    %\hfill
    % \begin{subfigure}{0.3\linewidth}
    %     \centering
    %     \includegraphics[width=0.7\textwidth, trim = 20 20 20 20,clip]{figures/Nokia_PA.pdf}
    %     \caption{\footnotesize{Proprietary LTE-A Band 3 LDMOS PA.}}
    %     \label{fig:Nokia_PA}
    % \end{subfigure}
    \caption{\quad Overall RF measurement setup at FR-1 and the sub-6~GHz PA modules used in the Experiments 1-2.}
    \label{fig:ab}
    \vspace{-2mm}
\end{figure*}
% ------------------------

The resulting exact numerical processing complexities, expressed in terms of real multiplications per linearized sample, are presented in Table~\ref{tab:flops}. In these numerical values, when it comes to the SPH and SMP DPD, we have excluded the trivial operations, i.e., multiplications by zeros, ones and integer powers of two or half, stemming from the structure of $\mathbf{B}_3$ in~(\ref{eq:B3}). Overall, the results in Table~\ref{tab:flops} demonstrate the large complexity reduction provided by the proposed spline-based DPD approaches, as nearly 64\% (SPH) \rev{and 44\% (SMP)} less real multiplications per sample are needed in the DPD main path to predistort the input signal. Furthermore, the required parameter learning complexity is also very remarkably reduced, \rev{by approximately 95\% in both SPH and SMP cases} in terms of real multiplications per sample, indicating that solutions like these might already facilitate even continuous learning in selected applications.
%
% The reader can find the complexity comparison between linear interpolation and cubic interpolation in Tables \ref{tab:Perf_20} and \ref{tab:Perf_80}, excluded in this section for simplicity.
%
Additionally, owing to the largely reduced learning complexity, the feasibility of implementing both the DPD parameter learning as well as the main path processing in the same chip increases.

\section{Experimental Results}
\label{sec:results}
In order to evaluate and validate the proposed DPD concepts, three separate linearization experiments are carried out. Two of the measurement scenarios are related to FR-1 (sub-6~GHz) PAs and classical conducted measurements, including a general purpose wideband PA and a 5G NR Band 78 small-cell BS PA. The third experiment is then related to FR-2 and over-the-air (OTA) measurements where a state-of-the-art 28~GHz active antenna array with 64 integrated PAs and antenna units is linearized.
\rev{For complexity assessment, we use the derived results in Table~\ref{tab:complexity}, while again exclude the trivial operations, i.e., multiplications by zeros, ones and integer powers of two or half, stemming from the structure of the B-spline basis matrix $\mathbf{B}_{P_\mathrm{SP}}$.
%$\mathbf{B}_3$ in~(\ref{eq:B3}). 
Additionally, we also provide the corresponding amounts of floating point operations (FLOPs) per sample. One complex multiplication is assumed to cost 6 FLOPs, while one complex-real multiplication and one complex sum both cost 2 FLOPs\cite{Flops}.}

%using a different DPD setup and amplifier in each of them. %The first PA is a general purpose wideband PA (Fig. \ref{fig:Gen_PA}), the second is a 5G Band 78 BS PA (Fig. \ref{fig:Skyworks_PA}), the third is a LTE-A Band 3, 100 Watts peak power BS PA (Fig. \ref{fig:Nokia_PA}), and the fourth device is a 
%64-antenna active array transmitter with mutually different PAs. The RF measurements carried out in this study prove that the complexity-performance trade-off of the proposed model is very favorable when compared to other state-of-the-art solutions.

\subsection{FR-1 Measurement Environment and Figures of Merit}

The FR-1 measurement setup utilized for the first two experiments is illustrated in Fig. \ref{fig:ab}(a), and  consists of a \red{National Instruments PXIe-5840} vector signal transceiver (VST), facilitating arbitrary waveform generation and analysis between 0--6~GHz with instantaneous bandwidth of 1~GHz. This instrument is used as both the transmitter and the observation receiver, and includes also an additional host-processing based computing environment where all the digital waveform and DPD processing can be executed. In a typical conducted measurement, the baseband complex I/Q waveform is generated by \textsc{Matlab} in the VST host environment, and fed to the device under test (DUT) through the VST transmit chain. %, incorporating also an external feeding or driver amplifier in high power measurements. 
The DUT output is then observed via the VST receiver, through an external attenuator. All DPD parameter learning and actual DPD main math processing stages are executed in the host environment. Finally, the actual DPD performance measurements are carried out where different random modulating data is used, compared to the learning phase.

%In order to measure and quantify the performance of the DPD methods, selected metrics or figures of merit are needed. In this work, 
As the DPD system figures of merit, we adopt the well-established error vector magnitude (EVM) and ACLR metrics, as defined for 5G NR in \cite{3GPPTS38104}. The EVM focuses on the passband transmit signal quality, and is defined as
%\begin{align}
%    \operatorname{NMSE\;(dB)} = 10 \operatorname{log_{10}}\frac{\sum\limits_{n=1}^N \abs{y_{\mathrm{meas}}[n] - y_{\mathrm{ideal}}[n]}^2}{\sum\limits_{n=1}^N \abs{y_{\mathrm{meas}}[n]}^2},
%\end{align}
%\begin{align}
%    \operatorname{EVM\;(\%)} = \sqrt{\frac{\sum\limits_{n=1}^N\abs{y_{\mathrm{meas,eq}}[n]-y_{\mathrm{ideal}}[n]}^2}{\sum\limits_{n=1}^N\abs{y_{\mathrm{ideal}}[n]}^2}} \times 100,
%\end{align}
\begin{align}
    \operatorname{EVM\;(\%)} = \sqrt{\frac{P_{\mathrm{error,\;eq.}}}{P_{\mathrm{ref.}}}} \times 100,
\end{align}
where $P_{\mathrm{error,\;eq.}}$ denotes the power of the error signal calculated between the ideal subcarrier symbols and the corresponding observed subcarrier samples at the PA output after zero forcing equalization removing the effects of the possible linear distortion \cite{3GPPTS38104}. Furthermore, $P_{\mathrm{ref.}}$ denotes the corresponding power of the ideal (reference) symbols.
%$y_{\mathrm{meas,eq}}[n]$ denotes the observed PA output signal, filtered to the occupied channel bandwidth and equalized with a zero-forcing equalizer to remove the effects of linear distortion, while $y_{\mathrm{ideal}}[n]$ refers to the corresponding ideal reference waveform samples. 
The ACLR, in turn, is defined as the ratio of the transmitted power within the desired channel ($P_{\mathrm{desired\;ch}}$) and that in the left or right adjacent channel ($P_{\mathrm{adj.\;ch.}}$), expressed as
\begin{align}
    \operatorname{ACLR\;(dB)} = 10 \operatorname{log_{10}} \frac{P_{\mathrm{desired\;ch.}}}{P_{\mathrm{adj.\;ch.}}},
\end{align}
measuring thus the out-of-band performance. While ACLR is, by definition, a relative measure, an explicit out-of-band spectral density limit, in terms of dBm/MHz measured with a sliding 1~MHz window in the adjacent channel region, is also defined for certain base-station types \cite{3GPPTS38104}, referred to as the absolute basic limit in 3GPP terminology. Thus, the PA output spectral density in dBm/MHz is also quantified in the measurements, particularly in the context of local area and medium-range BS PAs \cite{3GPPTS38104}.
%Additionally, the ACLR can be presented in terms of the independent left and right channel ratios, or alternatively as the highest value of those.

All the forth-coming experiments utilize 5G NR Release-15 standard compliant OFDM downlink waveform and channel bandwidths \cite{3GPPTS38104}, while the adopted carrier frequencies in each experiment are selected according to the available 5G NR bands and the available PA samples. In all experiments, the \rev{initial PAPR of the digital waveform is 9.5 dB, when measured at the 0.01\% point of the instantaneous PAPR complementary cumulative distribution function (CCDF), and is then limited to 7 dB through well-known iterative clipping and filtering based processing}, while also additional time-domain windowing is applied to suppress the inherent OFDM signal sidelobes. \rev{These impose an EVM floor of some 4\% to the transmit signal.} More specific waveform parameters such as the subcarrier spacing (SCS) and the occupied physical resource block (PRB) count are stated along the experiments. 
%Additionally, to decrease the noise variance in the observed signals, a statistical averaging has been applied to the received I/Q samples, where the final data is the result of $10$ averaged measurements.

%  trim={<left> <lower> <right> <upper>}
\begin{figure}[t!]
  \begin{center}
    \includegraphics[width=1\columnwidth, trim = 0 0 0 0, clip]{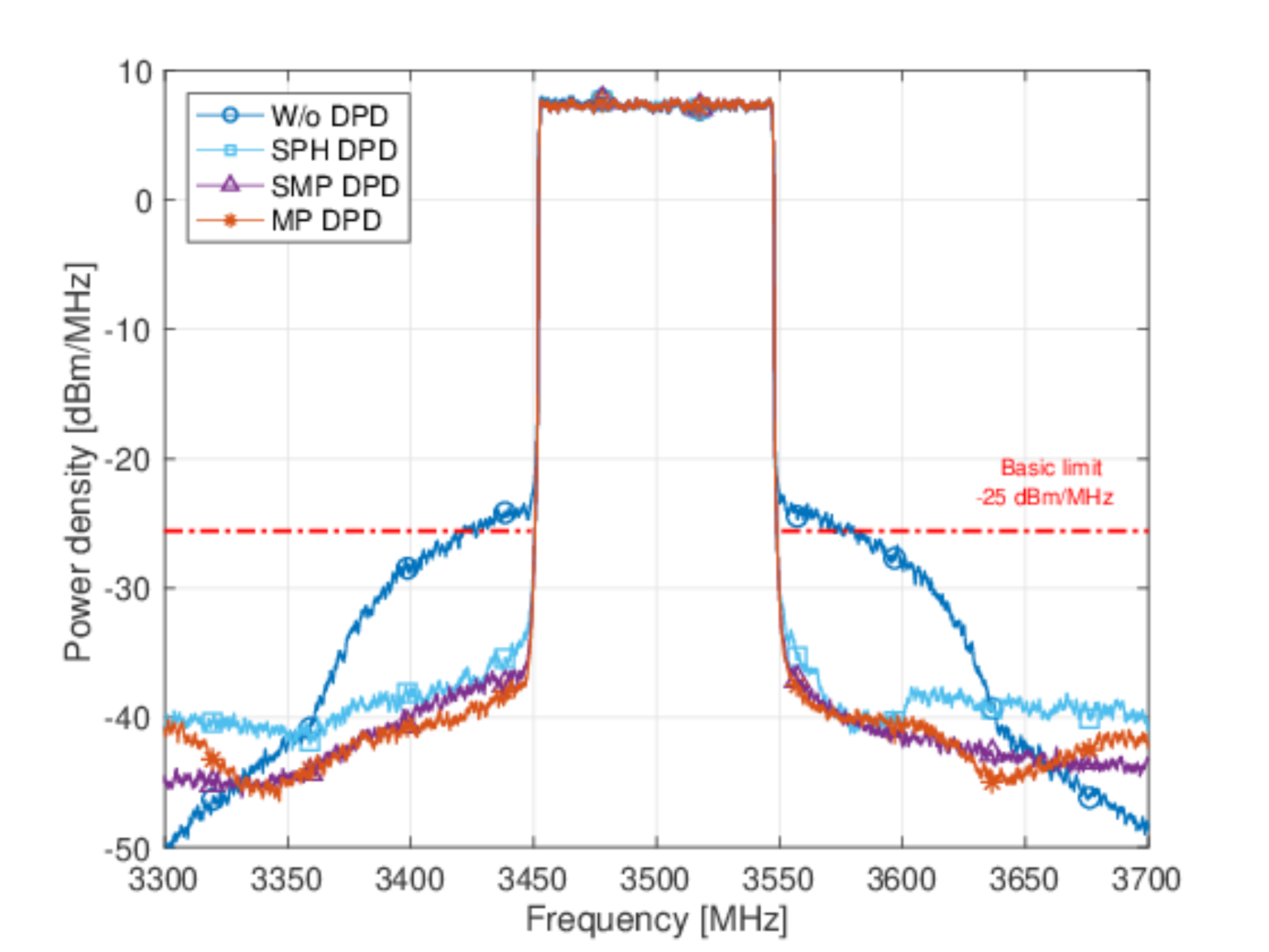}
  \end{center}
  \caption{\quad Example illustration of linearization results in Experiment~1 (general purpose PA measured at 3.5~GHz), with 100~MHz channel bandwidth and PA output power of +27~dBm, \rev{while adopting $P_{\mathrm{SP}}=3$, $Q_{\mathrm{SPH}} = Q_{\mathrm{SMP}} = 7$, $M_{\mathrm{SPH}} = 3$, $M_{\mathrm{SMP}} = 4$, $P_{\mathrm{MP}}=11$, and $M_{\mathrm{MP}}=4$.}}
  \label{fig:PSD_gen}
  \vspace{-2mm}
\end{figure}

\subsection{Experiment 1: General Purpose PA}

The first experiment focuses on a general purpose wideband PA (\red{Mini-Circuits} ZHL-4240)\red{, illustrated in Fig.~\ref{fig:ab}(b)}, as the actual amplification stage. The amplifier has a gain of 41~dB, and a 1-dB compression point of +31~dBm, being basically applicable in small-cell and medium-range base-stations. The transmit signal is a 5G NR downlink OFDM waveform, with 30~kHz subcarrier spacing and 264 active PRBs \cite{3GPPTS38104}, yielding an aggressive passband width of 95.04~MHz. The RF center frequency is 3.5~GHz and the assumed channel bandwidth is 100~MHz. The I/Q samples are transmitted through the VST RF output port directly to the PA, facilitating a maximum output power of +27~dBm. The proposed and the reference DPD schemes are then adopted, and the performance quantification measurements are carried out. In all results, five ILA learning iterations are adopted while the signal length within each ILA iteration is 100,000 samples. In this experiment, the VST observation receiver runs at 491.52~MHz (4x oversampling).

\begin{figure}[t!]
    \centering
    \begin{subfigure}[t]{1\columnwidth}
        \includegraphics[width=1\columnwidth, trim = 25 0 25 5,clip]{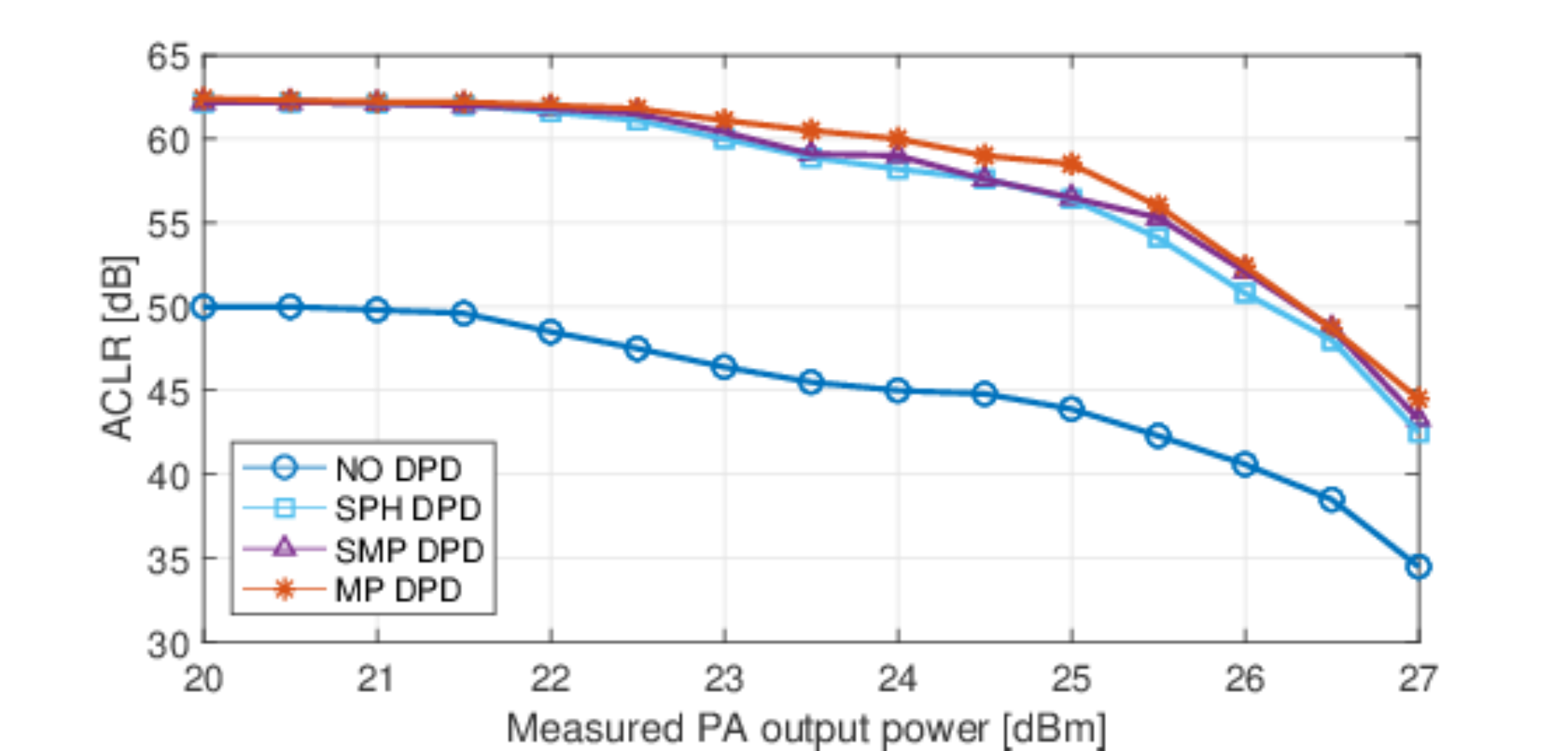}
        \caption{}
        \label{fig:aclr_gen}
    \end{subfigure}
    ~ %add desired spacing between images, e. g. ~, \quad, \qquad, \hfill etc.
      %(or a blank line to force the subfigure onto a new line)
    \begin{subfigure}[t]{1\columnwidth}
        \includegraphics[width=1\columnwidth, trim = 25 0 25 5,clip]{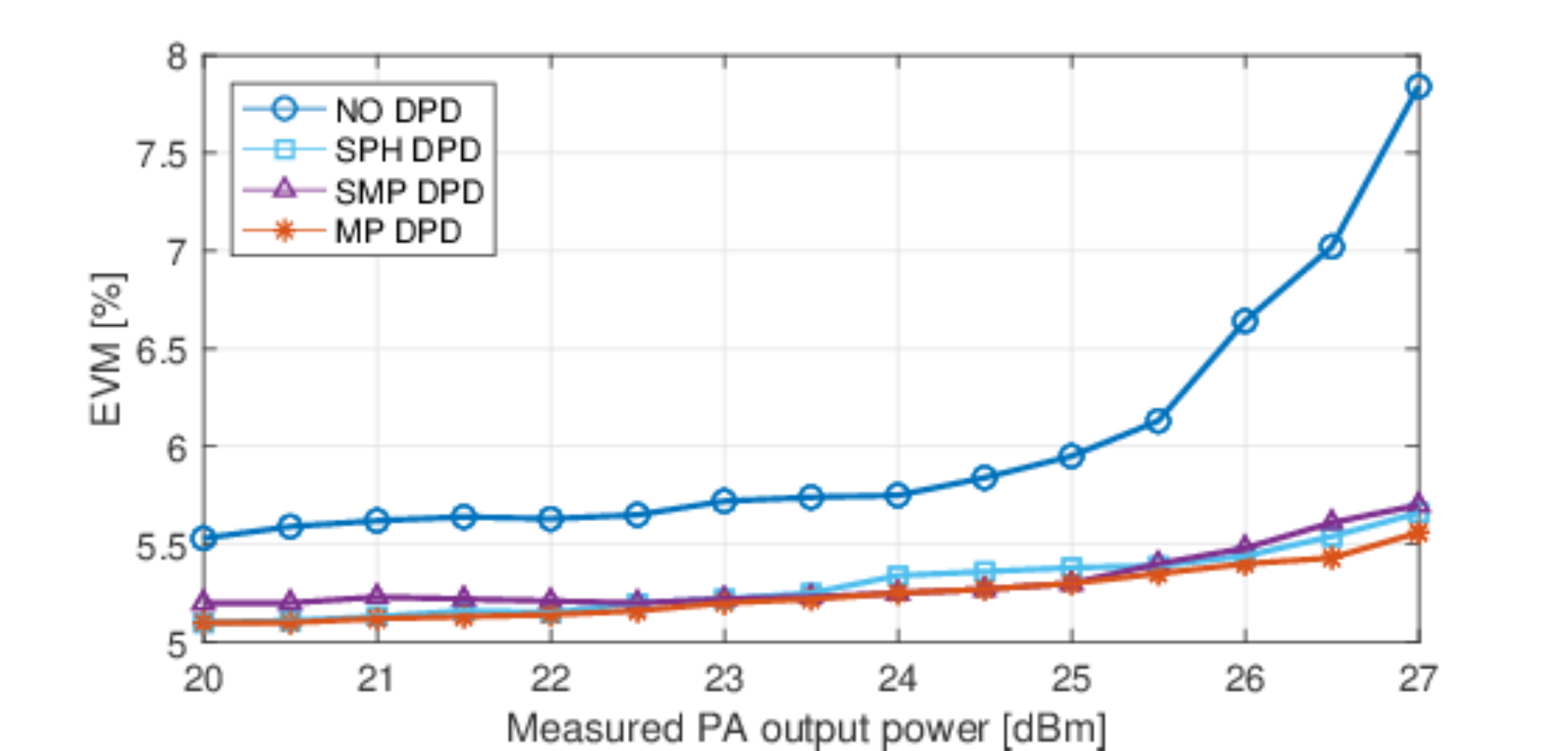}
        \caption{}
        \label{fig:evm_gen}
    \end{subfigure}
    \caption{\quad Measured ACLR and EVM performance in Experiment~1 as a function of the PA output power, while adopting $P_{\mathrm{SP}}=3$, $Q_{\mathrm{SPH}} = Q_{\mathrm{SMP}} = 7$, $M_{\mathrm{SPH}} = 3$, $M_{\mathrm{SMP}} = 4$, $P_{\mathrm{MP}}=11$, and $M_{\mathrm{MP}}=4$.}
    \label{fig:gen}
    \vspace{-2mm}
\end{figure}

\begin{figure}[h!]
    \centering
    \begin{subfigure}[t]{1\columnwidth}
        \includegraphics[width=1\columnwidth, trim = 25 0 25 5,clip]{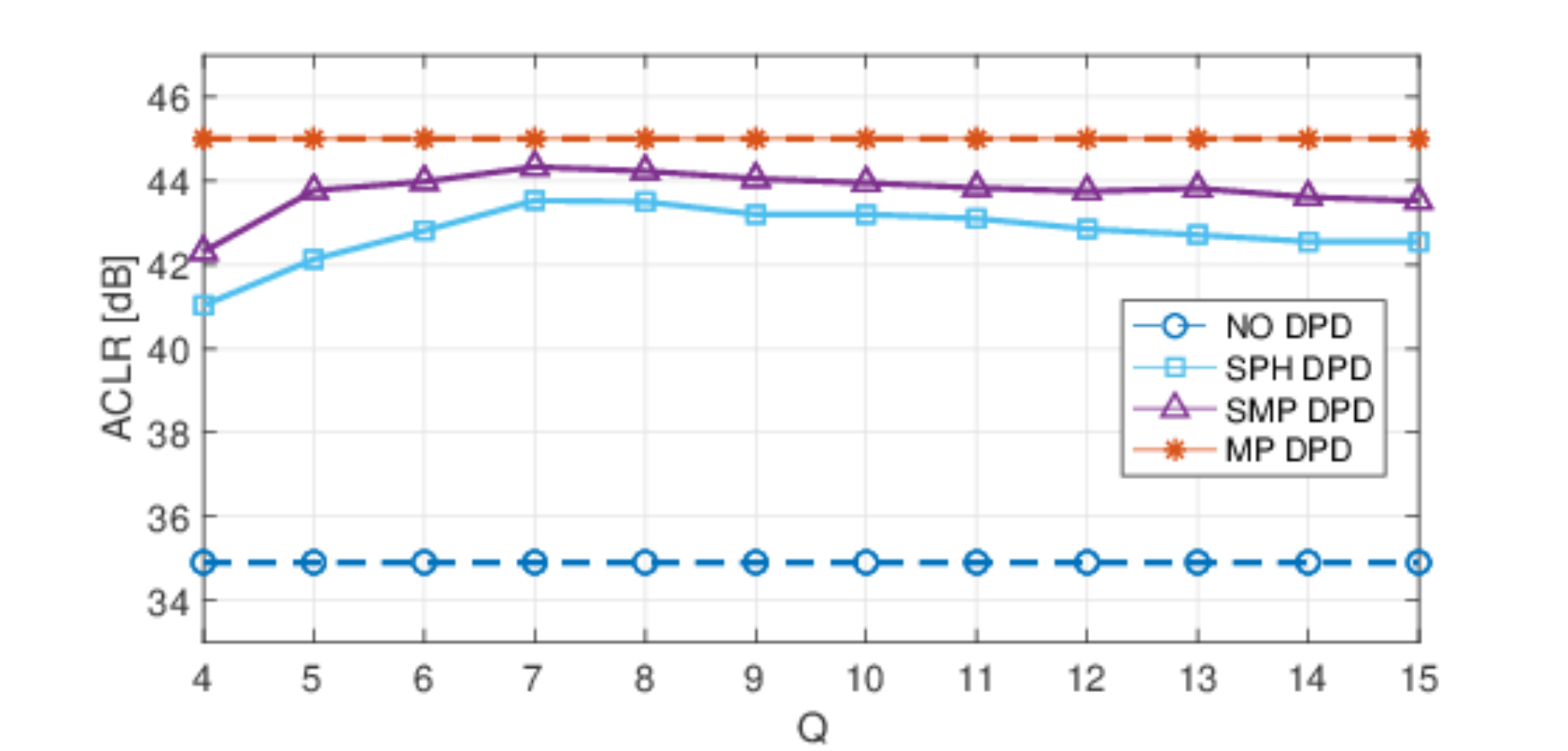}
        \caption{}
        \label{fig:QvsACLR}
    \end{subfigure}
    ~ %add desired spacing between images, e. g. ~, \quad, \qquad, \hfill etc.
      %(or a blank line to force the subfigure onto a new line)
    \begin{subfigure}[t]{1\columnwidth}
        \includegraphics[width=1\columnwidth, trim = 25 0 25 5,clip]{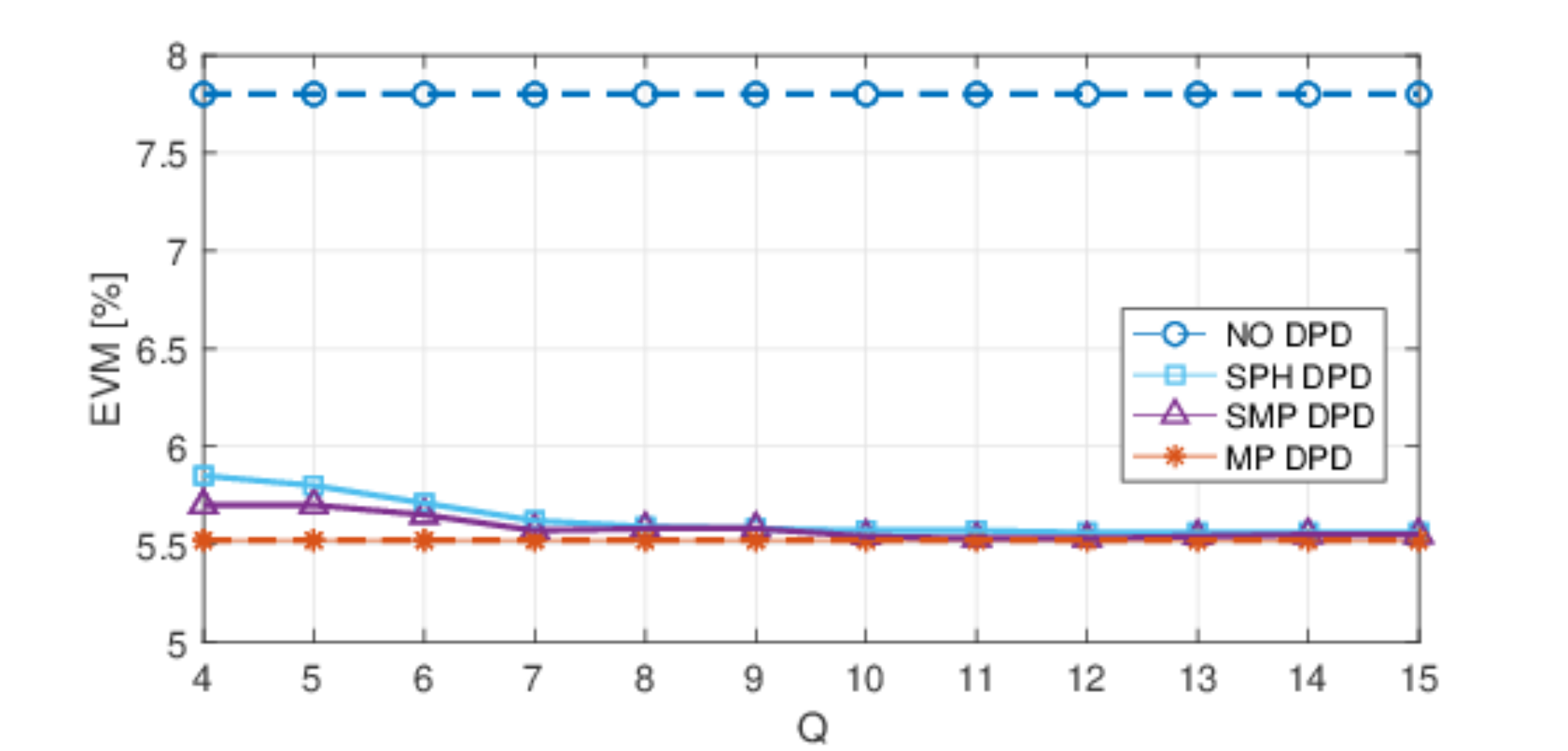}
        \caption{}
        \label{fig:QvsEVM}
    \end{subfigure}
    \caption{\quad \rev{Measured ACLR and EVM performance in Experiment~1 at PA output power of +27 dBm, as a function of the number of LUT control points in the SPH and SMP models, $Q$, while adopting $P_{\mathrm{SP}}=3$, $M_{\mathrm{SPH}} = 3$, $M_{\mathrm{SMP}} = 4$, $P_{\mathrm{MP}}=11$, and $M_{\mathrm{MP}}=4$.}}
    \label{fig:QvsMetrics}
    \vspace{-2mm}
\end{figure}

Fig. \ref{fig:PSD_gen} shows a snap-shot linearization example, at PA output power of +27~dBm, when $P_\mathrm{SP}=3$ is chosen as the spline order in both the SPH and SMP models, while the number of control points is fixed to $Q_{\mathrm{SPH}} = Q_{\mathrm{SMP}} = 7$ and the memory filter orders are $M_\mathrm{SPH}=3$ and $M_{\mathrm{SMP}} = 4$. Additionally, an LMS-based MP DPD of order $P_\mathrm{MP}=11$ with memory filters of order $M_\mathrm{MP}=4$ is also adopted and presented for reference. \rev{We can observe that the performances of the proposed SPH and SMP DPDs are very close to each other}, and to that of the MP DPD, despite the substantially reduced complexity. The figure also illustrates that all DPD methods basically satisfy the absolute basic limit requirement of -25~dBm/MHz, which if less stringent than the classical 45~dB ACLR limit, and applies in medium-range BS cases with TX powers of higher than +24~dBm up to +38~dBm \cite{3GPPTS38104}.

\begin{table*}[!t]
\centering
\setlength{\tabcolsep}{2pt}
\renewcommand{\arraystretch}{1.5}
\caption{\textsc{Summary of DPD main path processing complexity and linearization performance in Experiment 1, PA output power is} +27~dBm}
\label{tab:Perf_gen}
\begin{tabular}{c|c|c|c|c|c|c|c|c|c|}
\cline{2-10}
 & \multicolumn{7}{c|}{\textbf{Running complexity}} & \multicolumn{2}{c|}{\textbf{Model performance}} \\ \cline{2-10} 
\textbf{} & \ \ $\textbf{P}$\ \ & \ \ $\textbf{M}$\ \ & \ \ $\textbf{Q}$\ \ & \ \ $ \Delta_{z,x} $ \ \ & \ \ \textbf{\# of coefficients}\ \ & \textbf{ \ \ FLOPs/sample\ \ } & \textbf{ \ \ Mul./sample\ \ } & \textbf{\ \ EVM (\%)\ \ } & \textbf{\ \ Max. dBm/MHz\ \ } \\ \hline
\multicolumn{1}{|c|}{\textbf{No DPD}} & - & - & - & - & - & - & - & 7.82 & -23.80 \\ \hline
\multicolumn{1}{|c|}{\multirow{3}{*}{\textbf{SPH DPD}}} & 2 & 3 & 7 & 1 & 12 & 55 & 28 & 5.61 & -32.30 \\ \cline{2-10} 
\multicolumn{1}{|c|}{} & 3 & 3 & 7 & 1 & 13 & 69 & 36 & 5.54 & -36.30 \\ \cline{2-10} 
\multicolumn{1}{|c|}{} & 4 & 3 & 7 & 1 & 14 & 89 & 45 & 5.55 & -36.80 \\ \hline
\multicolumn{1}{|c|}{\multirow{3}{*}{\rev{ \textbf{\ \ SMP DPD \ \ }}}} & \rev{2} & \rev{4} & \rev{7} & \rev{1} & \rev{30} & \rev{65} & \rev{50} & \rev{5.55} & \rev{-37.20} \\ \cline{2-10} 
\multicolumn{1}{|c|}{} & \rev{3} & \rev{4} & \rev{7} & \rev{1} & \rev{31} & \rev{99} & \rev{63} & \rev{5.57} & \rev{-37.80} \\ \cline{2-10} 
\multicolumn{1}{|c|}{} & \rev{4} & \rev{4} & \rev{7} & \rev{1} & \rev{32} & \rev{143} & \rev{77} & \rev{5.57} & \rev{-37.80} \\ \hline
%\multicolumn{1}{|c|}{\textbf{\ \ Lin. interp. \ \ }} & 1 & 0 & 15 & 1 & 16 & 23 & 11 & 6.00 & -34.20 \\ \hline
\multicolumn{1}{|c|}{\textbf{MP DPD}} & 11 & 4 & - & - & 24 & 255 & 112 & 5.47 & -38.20 \\ \hline
\end{tabular}
\end{table*}

% \begin{table*}[t!]
% \centering
% \setlength{\tabcolsep}{2pt}
% \renewcommand{\arraystretch}{1.5}
% \caption{\textsc{Summary of DPD main path processing complexity and linearization performance results obtained in Experiment 1, PA output power is} +27~dBm}
% \label{tab:Perf_gen}
% \begin{tabular}{c|c|c|c|c|c|c|c|c|c|}
% \cline{2-10}
%  & \multicolumn{7}{c|}{\textbf{DPD running complexity}} & \multicolumn{2}{c|}{\textbf{DPD performance}} \\ \cline{2-10} 
% \textbf{} & \ \ $\textbf{P}$ \ \ & \ \ $\textbf{M}$ \ \ & \ \ $\textbf{Q}$ \ \ & \ \! $ \Delta_x $ \! \ & \ \  \textbf{\# of coefficients} \ \ & \ \  \textbf{FLOPs/sample} \ \ & \ \ \textbf{Mul./sample}\ \ & \ \ \textbf{EVM (\%)} \ \  & \ \  \textbf{Max. dBm/MHz} \ \ \\ \hline
% \multicolumn{1}{|c|}{\textbf{No DPD}} & - & - & - & - & - & 0 & 0 & 7.82 & -23.80 \\ \hline
% \multicolumn{1}{|c|}{\multirow{4}{*}{\textbf{SPH DPD}}} & 1 & - & 15 & 1 & 16 & 30 & 15 & 6.54 & -29.20 \\ \cline{2-10} 
% \multicolumn{1}{|c|}{} & 2 & 3 & 15 & 1 & 20 & 63 & 43 & 5.61 & -32.30 \\ \cline{2-10} 
% \multicolumn{1}{|c|}{} & 3 & 3 & 15 & 1 & 21 & 68 & 46 & 5.54 & -36.30 \\ \cline{2-10} 
% \multicolumn{1}{|c|}{} & 4 & 3 & 15 & 1 & 22 & 73 & 49 & 5.55 & -36.80 \\ \hline
% \multicolumn{1}{|c|}{\textbf{\ \ MP DPD \ \ }} & 9 & 3 &  - & - & 20 & 172 & 93 & 5.47 & -38.20 \\ \hline
% \end{tabular}
% \end{table*}

Fig. \ref{fig:gen} then presents the behavior of the measured EVM and ACLR performance metrics, as functions of the PA output power, following the same DPD parameterization. Again, we can observe that the proposed SPH, SMP, and the MP DPD behave very similarly. \rev{Similar type of observation follows also from Fig.~\ref{fig:QvsMetrics}, showing again the EVM and ACLR metrics but this time at fixed PA output power of +27~dBm while then varying the number of LUT control points in the proposed SPH and SMP models. From this figure we can also observe that the LUT based DPD performance is optimized with some $Q=7$ or $Q=8$ control points, in this example, while in general it is likely that the optimization of the value of $Q$ is to be done separately for different PA types. } %  starting from a linear region and heading towards a more saturated point, where the metrics are degraded due to high output power and therefore nonlinear distortion. 

Finally, Table~\ref{tab:Perf_gen} then collects and summarizes the obtained DPD results in Experiment 1 while also showing the DPD main path processing complexities. Here also other spline interpolation orders $P_\mathrm{SP}$ are considered and shown. We can conclude that the proposed spline-based DPD models offer a favorable performance-complexity trade-off compared to the reference MP DPD approach.

\subsection{Experiment 2: 5G NR Band 78 Small-Cell PA}

% \begin{table*}[t!]
% \centering
% \setlength{\tabcolsep}{2pt}
% \renewcommand{\arraystretch}{1.5}
% \caption{\textsc{Summary of DPD main path processing complexity and linearization performance results obtained in Experiment 2, PA output power is} +24~dBm}
% \label{tab:Perf_sky}
% \begin{tabular}{c|c|c|c|c|c|c|c|c|c|}
% \cline{2-10}
%  & \multicolumn{7}{c|}{\textbf{DPD running complexity}} & \multicolumn{2}{c|}{\textbf{DPD performance}} \\ \cline{2-10} 
% \textbf{} & \ \ $\textbf{P}$ \ \ & \ \ $\textbf{M}$ \ \ & \ \ $\textbf{Q}$ \ \ & \ \! $ \Delta_x $ \! \ & \ \  \textbf{\# of coefficients} \ \ & \ \  \textbf{FLOPs/sample} \ \ & \ \ \textbf{Mul./sample}\ \ & \ \ \textbf{EVM (\%)} \ \ & \ \  \textbf{Max. dBm/MHz} \ \ \\ \hline
% \multicolumn{1}{|c|}{\textbf{No DPD}} & - & - & - & - & - & 0 & 0 & 8.64 & -18.20 \\ \hline
% \multicolumn{1}{|c|}{\multirow{4}{*}{\textbf{SPH DPD}}} & 1 & - & 15 & 1 & 16 & 30 & 15 & 6.02 & -26.90 \\ \cline{2-10}
% \multicolumn{1}{|c|}{} & 2 & 4 & 15 & 1 & 21 & 71 & 35 & 5.80 & -29.40 \\ \cline{2-10}
% \multicolumn{1}{|c|}{} & 3 & 4 & 15 & 1 & 22 & 77 & 38 & 5.57 & -32.60 \\ \cline{2-10}
% \multicolumn{1}{|c|}{} & 4 & 4 & 15 & 1 & 23 & 81 & 41 & 5.58 & -32.60 \\ \hline
% \multicolumn{1}{|c|}{\textbf{\ \ MP DPD \ \ }} & 11 & 4 & - & - & 30 & 255 & 136 & 5.54 & -33.20 \\ \hline
% \end{tabular}
% \end{table*}

\begin{table*}[ht]
\centering
\setlength{\tabcolsep}{2pt}
\renewcommand{\arraystretch}{1.5}
\caption{\textsc{Summary of DPD main path processing complexity and linearization performance in Experiment 2, PA output power is} +24~dBm}
\label{tab:Perf_sky}
\begin{tabular}{c|c|c|c|c|c|c|c|c|c|}
\cline{2-10}
 & \multicolumn{7}{c|}{\textbf{Running complexity}} & \multicolumn{2}{c|}{\textbf{Model performance}} \\ \cline{2-10} 
\textbf{} & \ \ $\textbf{P}$\ \ & \ \ $\textbf{M}$\ \ & \ \ $\textbf{Q}$\ \ & \ \ $ \Delta_{z,x} $ \ \ & \ \ \textbf{\# of coefficients}\ \ & \textbf{\ \ FLOPs/sample\ \ } & \textbf{\ \ Mul./sample\ \ } & \textbf{\ \ EVM (\%)\ \ } & \textbf{\ \ Max. dBm/MHz\ \ } \\ \hline
\multicolumn{1}{|c|}{\textbf{No DPD}} & - & - & - & - & - & - & - & 8.64 & -18.20 \\ \hline
\multicolumn{1}{|c|}{\multirow{2}{*}{\textbf{SPH DPD}}} & 2 & 4 & 7 & 1 & 13 & 63 & 32 & 5.70 & -31.40 \\ \cline{2-10} 
\multicolumn{1}{|c|}{} & 3 & 4 & 7 & 1 & 14 & 77 & 40 & 5.57 & -33.20 \\ \hline
\multicolumn{1}{|c|}{\multirow{2}{*}{\rev{\textbf{\ \ SMP DPD \ \ }}}} & \rev{2} & \rev{5} & \rev{7} & \rev{1} & \rev{37} & \rev{73} & \rev{56} & \rev{5.60} & \rev{-32.90} \\ \cline{2-10} 
\multicolumn{1}{|c|}{} & \rev{3} & \rev{5} & \rev{7} & \rev{1} & \rev{38} & \rev{111} & \rev{75} & \rev{5.55} & \rev{-33.10} \\ \hline
%\multicolumn{1}{|c|}{\textbf{\ \ Lin. interp. \ \ }} & 1 & 0 & 15 & 1 & 16 & 23 & 11 & 6.02 & -31.90 \\ \hline
\multicolumn{1}{|c|}{\textbf{MP DPD}} & 11 & 5 & - & - & 30 & 255 & 136 & 5.54 & -33.20 \\ \hline
\end{tabular}
\end{table*}

% \begin{table*} %[t!]
% \centering
% \setlength{\tabcolsep}{2pt}
% \renewcommand{\arraystretch}{1.5}
% \caption{\textsc{Summary of DPD main path processing complexity and linearization performance results obtained in Experiment 3, PA output power is} +48~dBm}
% \label{tab:Perf_Nokia}
% \begin{tabular}{c|c|c|c|c|c|c|c|c|c|}
% \cline{2-10}
%  & \multicolumn{7}{c|}{\textbf{DPD running complexity}} & \multicolumn{2}{c|}{\textbf{DPD performance}} \\ \cline{2-10} 
% \textbf{} & \ \ $\textbf{P}$ \ \ & \ \ $\textbf{M}$ \ \ & \ \ $\textbf{Q}$ \ \ & \ \! $ \Delta_x $ \! \ & \ \  \textbf{\# of coefficients} \ \ & \ \  \textbf{FLOPs/sample} \ \ & \ \ \textbf{Mul./sample}\ \ & \ \ \textbf{EVM (\%)} \ \ & \ \  \textbf{ACLR (L/R) (dB)} \ \ \\ \hline
% \multicolumn{1}{|c|}{\textbf{No DPD}} & - & - & - & - & - & 0 & 0 & 8.62  & 35.40 / 35.80 \\ \hline
% \multicolumn{1}{|c|}{\multirow{4}{*}{\textbf{SPH DPD}}} & 1 & - & 15 & 1 & 16 & 30 & 15 & 7.54 & 40.20 / 41.80 \\ \cline{2-10}
% \multicolumn{1}{|c|}{} & 2 & 6 & 15 & 1 & 23 & 87 & 43 & 6.02 & 45.10 / 45.20 \\ \cline{2-10} 
% \multicolumn{1}{|c|}{} & 3 & 6 & 15 & 1 & 24 & 92 & 46 & 5.60 & 45.90 / 46.10 \\ \cline{2-10} 
% \multicolumn{1}{|c|}{} & 4 & 6 & 15 & 1 & 25 & 97 & 49 & 5.57 & 45.20 / 47.30 \\ \hline
% \multicolumn{1}{|c|}{\textbf{\ \ MP DPD \ \ }} & 11 & 4 & - & - & 30 & 255 & 136 & 5.46 & 51.40 / 49.90 \\ \hline
% \end{tabular}
% \end{table*}

The second experiment includes the Skyworks SKY66293-21 PA module, illustrated in Fig.~\ref{fig:ab}(c), which is a low-to-medium power PA oriented to be used either in small-cell base-stations or in large antenna array transmitters. The PA module is specifically designed to operate in the NR Band n78 (3300-3800 MHz), having a gain of 34~dB, and a 1-dB compression point of +31.5~dBm.
Similar 5G NR downlink signal corresponding to the 100~MHz channel bandwidth scenario, as in the Experiment~1, is adopted, while the considered RF center-frequency is 3.65~GHz. The test signal is again transmitted via the RF TX port of the VST directly to the PA module, while the considered PA output power is +24~dBm, corresponding to the maximum transmit power of a Local Area BS according to the NR regulations \cite{3GPPTS38104}. Again, five ILA learning iterations are adopted while the signal length within each ILA iteration is 100,000 samples. The VST observation receiver runs at 491.52~MHz (4x oversampling).

\begin{figure}[t!]
  \begin{center}
    \includegraphics[width=1\columnwidth, trim = 0 0 0 0, clip]{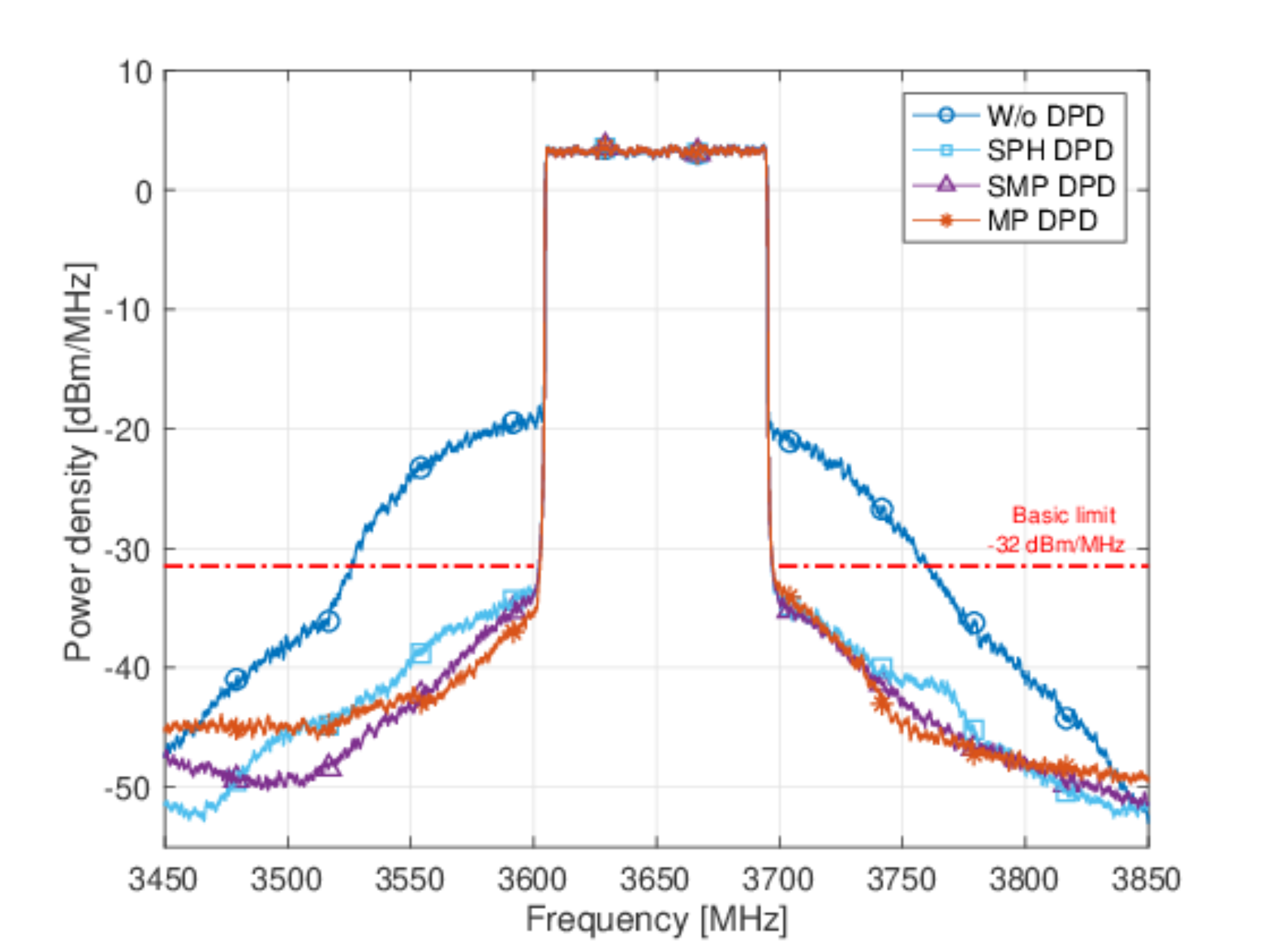}
  \end{center}
  \caption{{\quad Example illustration of linearization results in Experiment~2 (NR small-cell PA measured at 3.65~GHz), with 100~MHz channel bandwidth and PA output power of +24~dBm, \rev{while adopting $P_{\mathrm{SP}}=3$, $Q_{\mathrm{SPH}} = Q_{\mathrm{SMP}} = 7$, $M_{\mathrm{SPH}} = 4$, $M_{\mathrm{SMP}} = 5$, $P_{\mathrm{MP}}=11$, and $M_{\mathrm{MP}}=5$.}}} %Note that +24 dBm output power corresponds to +4 dBm/MHz with this bandwidth configuration.}
  \label{fig:PSD_sky}
  \vspace{-2mm}
\end{figure}

Fig. \ref{fig:PSD_sky} and Table \ref{tab:Perf_sky} illustrate and summarize the obtained linearization results for the proposed and the reference DPD methods. Again, also comparative complexity numbers are stated in Table \ref{tab:Perf_sky}. As stated in \cite{3GPPTS38104}, a 5G NR Local Area BS can operate within an absolute basic limit of -32~dBm/MHz in the adjacent channel region, assuming the considered PA output power of +24~dBm. As shown in Fig.~\ref{fig:PSD_sky} and Table \ref{tab:Perf_sky}, the SPH, \rev{SMP}, and the MP DPD satisfy this limit, indicating successful linearization. Again, as can be observed in Table \ref{tab:Perf_sky}, a remarkable complexity reduction is obtained through the proposed spline-based DPD approaches, compared to the reference MP DPD, while all provide a very similar linearization performance.
%for this PA with the $3^{\mathrm{rd}}$ order SPH and both reference models. From this figure, it is shown that both the amplitude and phase distortion are successfully corrected with 

%  trim={<left> <lower> <right> <upper>}

\subsection{Experiment 3: FR-2 Environment and 28 GHz Active Array}

%  trim={<left> <lower> <right> <upper>}
\begin{figure}[h]
  \begin{center}
    \includegraphics[width=1\columnwidth, trim = 20 20 20 20, clip]{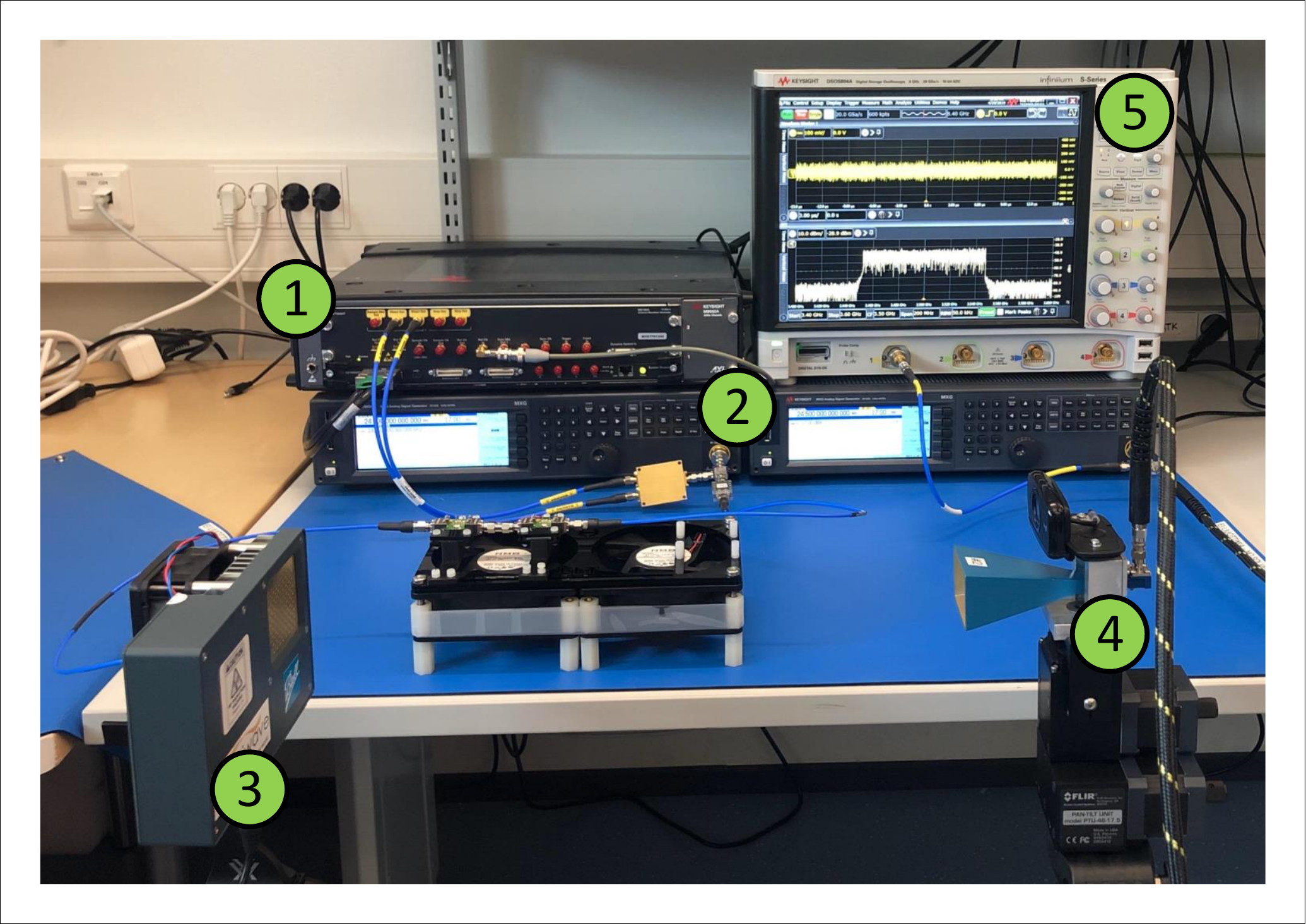}
  \end{center}
  \caption{\quad RF measurement setup in Experiment 3 including the Keysight M8190 arbitrary waveform generator (1), Keysight N5183B-MXG LO signal generators (2), Anokiwave AWMF-0129 active antenna array (3) working at 28~GHz center frequency (NR Band n257), horn antenna as receiver (4), and Keysight DSOS804A digitizer (5).}
  \label{fig:mm_setup}
  \vspace{-2mm}
\end{figure}
% Old caption:
% RF measurement setup in Experiment 4 including the 64-element Anokiwave AWMF-0129 active antenna array, working at 28~GHz center frequency (NR Band n257).

In order to further demonstrate the applicability of the proposed spline-based DPD concepts, the third and final experiment focuses on timely 5G NR mmWave/FR-2 deployments \cite{3GPPTS38104} with active antenna arrays. Unwanted emission modeling and DPD-based linearization of active arrays with large numbers of PA units is, generally, an active research field, with good examples of recent papers being, e.g., \cite{OOB_Mollen,JSTSP,DPD_DigitalMIMO,Swedes_review,DPD_MM_6,OTA_combining_DPD}. Below we first describe shortly the FR-2 measurement setup, and then present the actual linearization results.

%a 64-element active antenna array operating at 28 GHz carrier frequency and transmitting 100 and 200 MHz wide 5G NR signals is shown. This last experiment is thus representative of a 5G deployment in the frequency range 2 (FR2) , which is a very timely example.
%In the following, we shortly provide some insights regarding the fundamental ideas underlying the linearization of such transmitters, while the interested reader is kindly referred to  for further details.

% Subfigure with both PSDs:
%  trim={<left> <lower> <right> <upper>}
\begin{figure*}[t]
    \centering
    \begin{subfigure}[t]{0.49\textwidth }
        \includegraphics[width=\textwidth, trim = 0 0 0 0,clip]{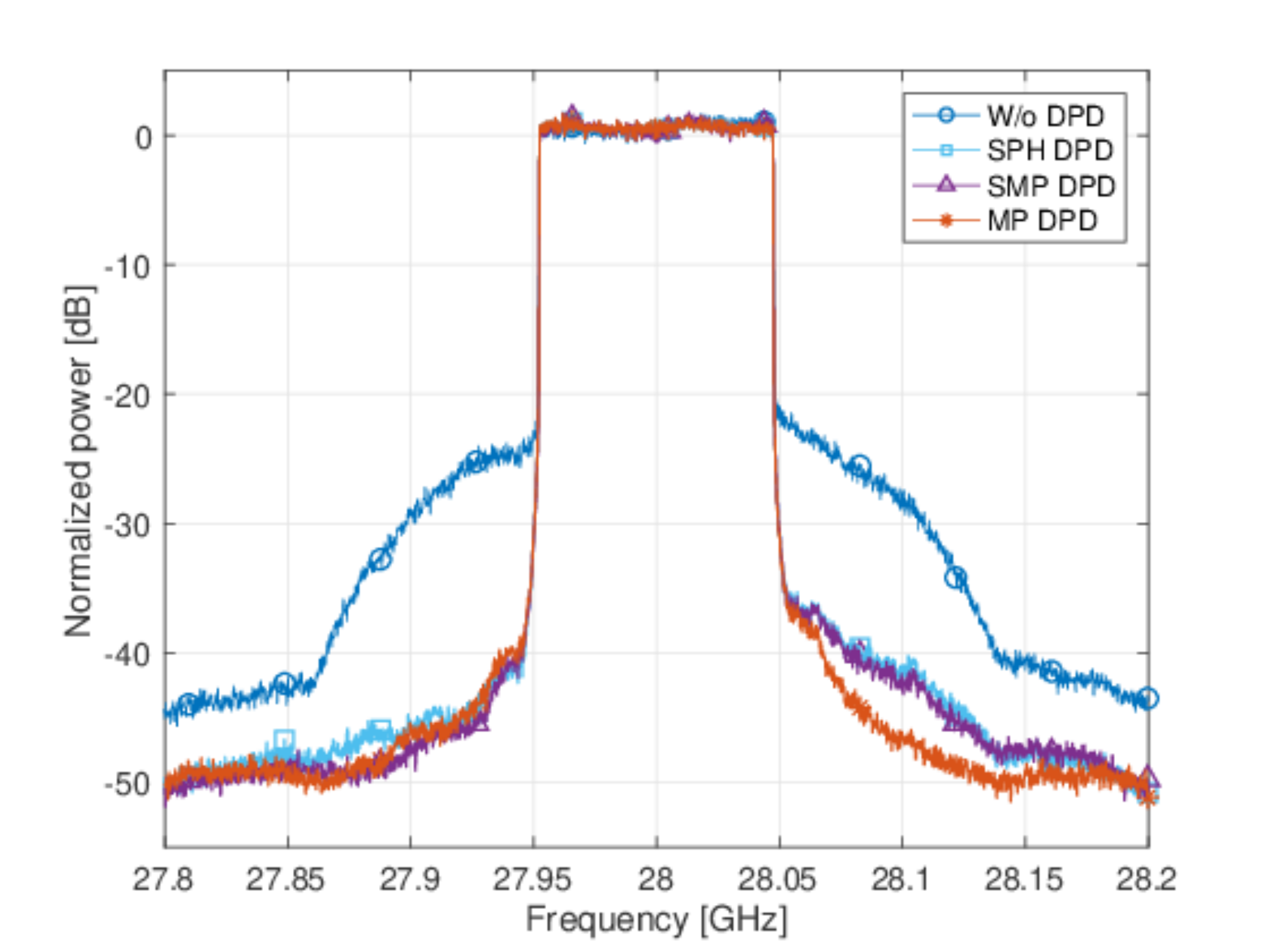}
        \caption{ }
        \label{fig:PSD_mm1}
    \end{subfigure}
    ~ %add desired spacing between images, e. g. ~, \quad, \qquad, \hfill etc.
      %(or a blank line to force the subfigure onto a new line)
    \begin{subfigure}[t]{0.49\textwidth}
        \includegraphics[width=\textwidth, trim = 0 0 0 0,clip]{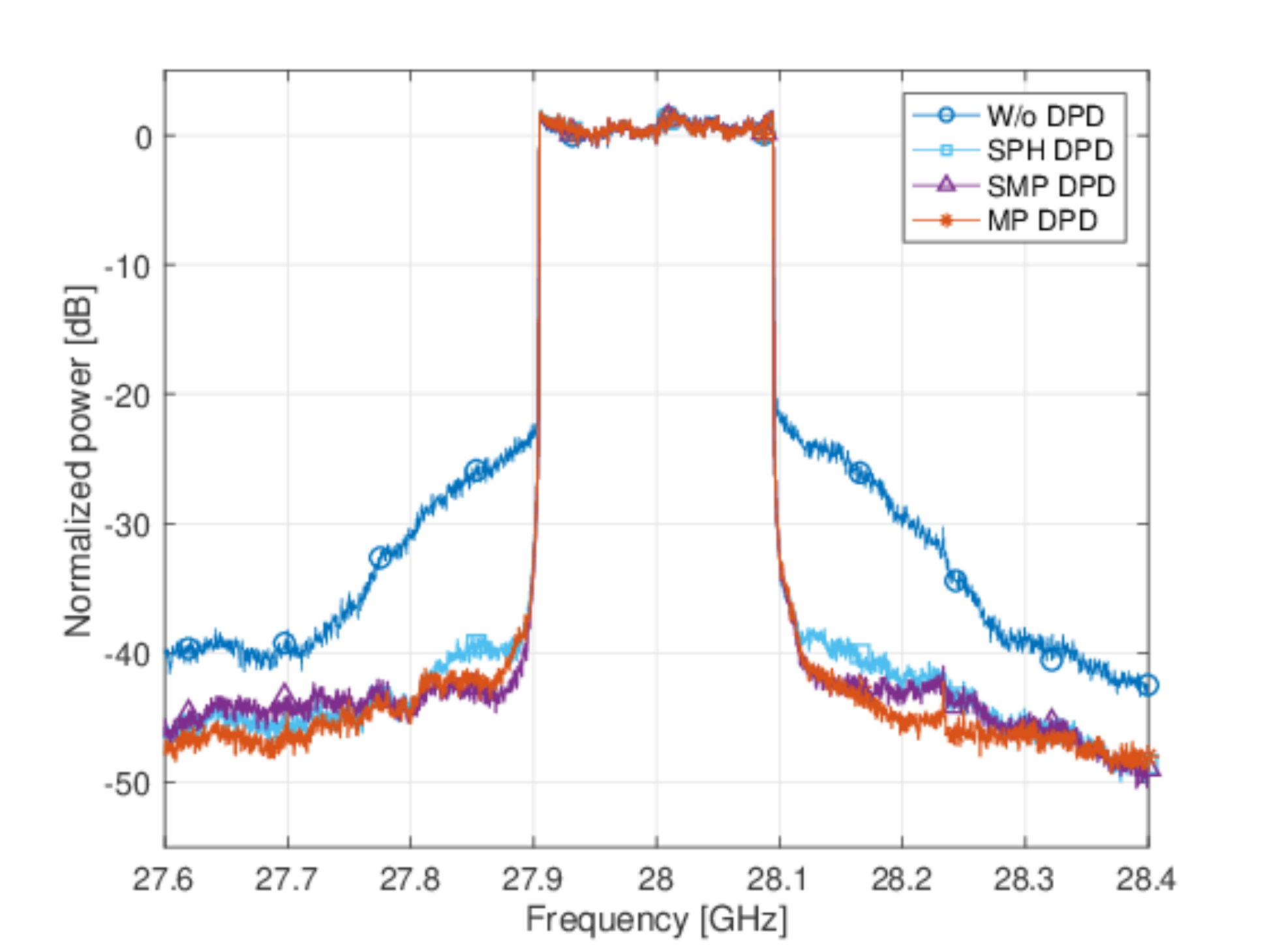}
        \caption{ }
        \label{fig:PSD_mm2}
    \end{subfigure}
    \caption{\quad Illustration of OTA linearization of the Anokiwave AWMF-0129 active antenna array, when (a) NR 100 MHz and (b) NR 200 MHz transmit signals are applied, measured at EIRP of +42.5~dBm \rev{The SPH and SMP DPD spline order is $P_{\mathrm{SP}}=3$, while $Q_{\mathrm{SPH}} = Q_{\mathrm{SMP}} = 7$, $M_\mathrm{SPH}=3$, and $M_{\mathrm{SMP}}=4$. The MP DPD order is $P_\mathrm{MP}=11$ while $M_\mathrm{MP}=4$.}}
    \label{fig:PSD_mm}
    \vspace{-2mm}
\end{figure*}

\begin{figure*}[t]
    \centering
    \begin{subfigure}[t]{0.49\textwidth }
        \includegraphics[width=\textwidth, trim = 0 0 0 0,clip]{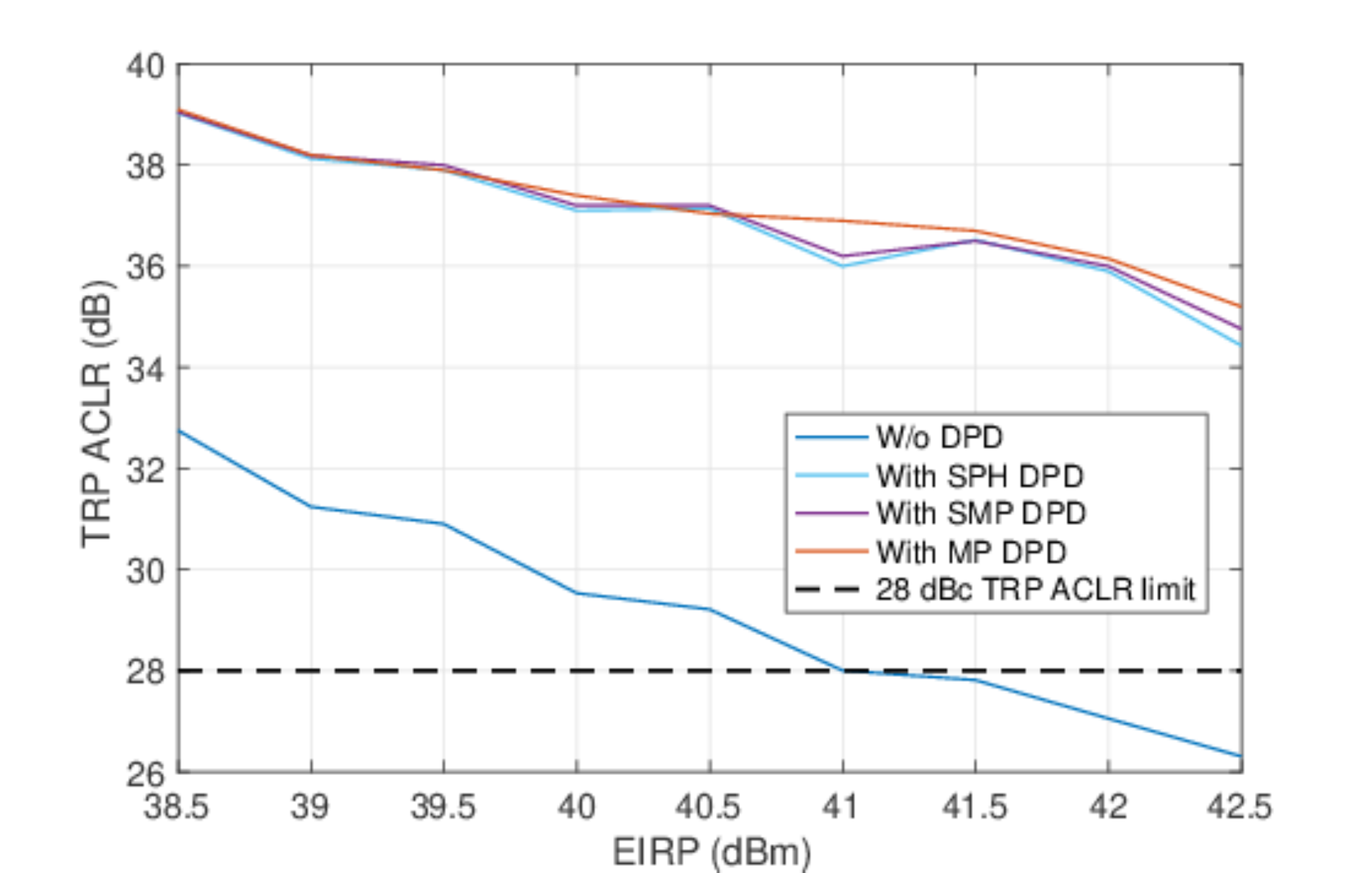}
        \caption{ }
        \label{fig:PS_ACLR}
    \end{subfigure}
    ~ %add desired spacing between images, e. g. ~, \quad, \qquad, \hfill etc.
      %(or a blank line to force the subfigure onto a new line)
    \begin{subfigure}[t]{0.49\textwidth}
        \includegraphics[width=\textwidth, trim = 0 0 0 0,clip]{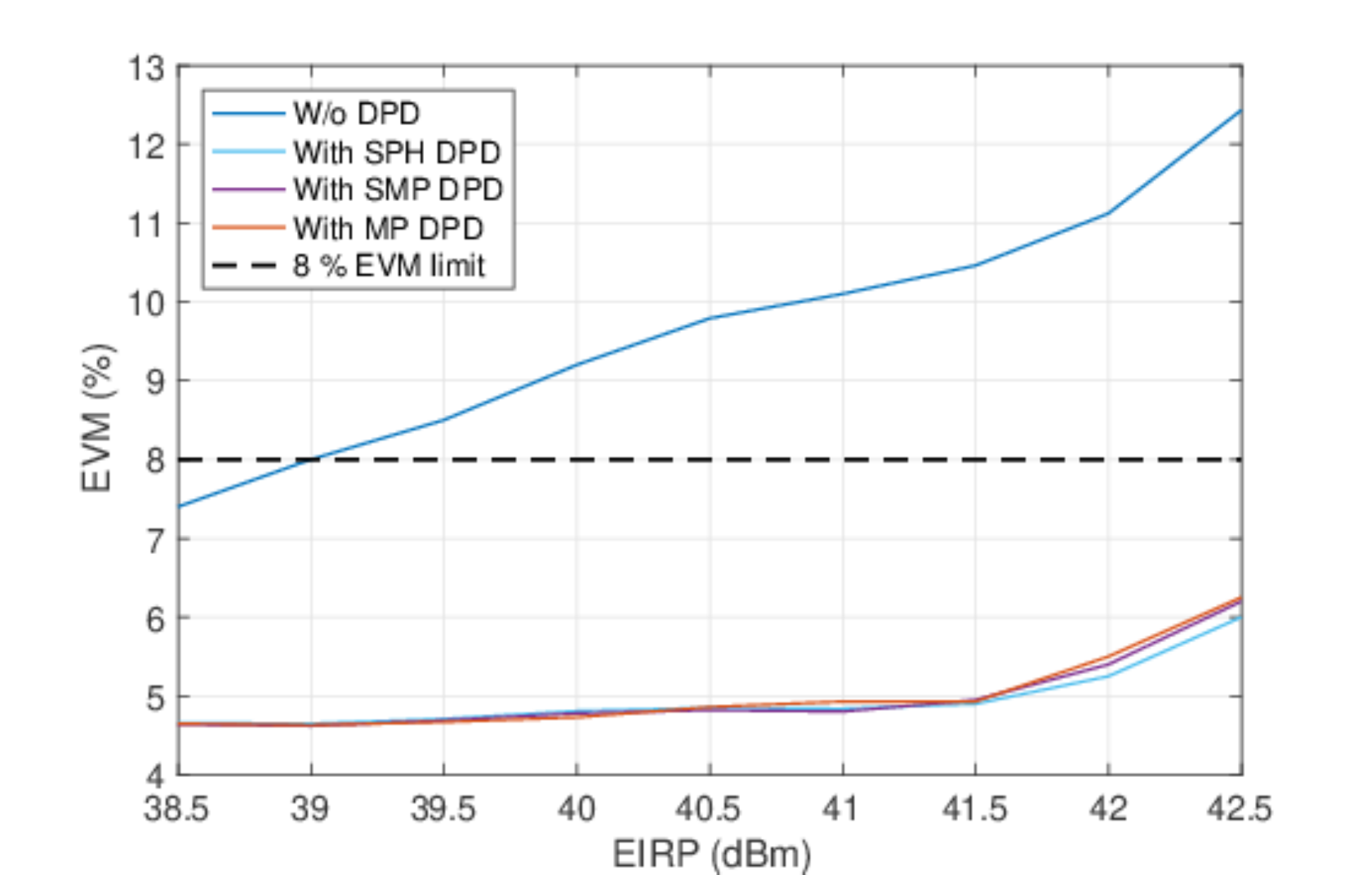}
        \caption{ }
        \label{fig:PS_EVM}
    \end{subfigure}
    \caption{\quad \rev{OTA linearization performance of the Anokiwave AWMF-0129 active antenna array as a function of the EIRP of the proposed DPD models in terms of a) TRP ACLR, and b) EVM.}}
    \label{fig:PS}
    \vspace{-2mm}
\end{figure*}

\subsubsection{FR-2 Measurement Setup} 

The overall mmWave/FR-2 measurement setup is depicted in Fig.~\ref{fig:mm_setup}, incorporating an Anokiwave AWMF-0129 active antenna array together with other relevant instruments and equipment for signal generation and analysis, facilitating measurements at 28~GHz center-frequency with up to 3~GHz of instantaneous bandwidth. On the transmit chain side, the setup consists of a Keysight M8190 arbitrary waveform generator that is used to generate directly the I/Q samples of a wideband modulated IF signal centered at 3.5 GHz. The signal is then upconverted to the 28~GHz carrier frequency by utilizing the Keysight N5183B-MXG that generates the corresponding local oscillator signal running at 24.5~GHz, together with external mixers and filters. The modulated RF waveform is then pre-amplified by means of two \red{Analog Devices'} driver PAs, HMC499LC4 and HMC943ALP5DE, with 17 dB and 23 dB gain, respectively, such that the 
integrated PAs of the Anokiwave AWMF-0129 active antenna array are driven towards saturation. 

The transmit signal propagates over-the-air (OTA) and is captured by a horn antenna at the observation receiver, such that the receiving antenna system is well aligned with the main transmit beam. At the receiver side, another Keysight N5183B-MXG and a mixing stage are used to downconvert the signal back to IF. Then, the Keysight DSOS804A oscilloscope is utilized as the actual digitizer, including also built-in filtering, and the signal is taken to baseband and processed in a host PC, where the DPD learning and predistortion are executed. The OTA measurement system is basically following the measurement procedures described in \cite{3GPPTS38104,3GPPTS38141}, specifically the measurement option utilizing the beam-based directions. In these measurements, the observation receiver provides I/Q samples at 7x oversampled rate.

\begin{table*}[t]
\centering
\setlength{\tabcolsep}{2pt}
\renewcommand{\arraystretch}{1.5}
\caption{\textsc{Summary of linearization performance of the Anokiwave AWMF-0129 active antenna array, with 100~MHz and 200~MHz 5G NR channel bandwidths, measured at} +42.5~dBm EIRP}
\label{tab:mm_perf}
\begin{tabular}{c|c|c|c|c|c|c|c|c|c|c|}
\cline{2-11}
 & \multicolumn{6}{c|}{\textbf{DPD running complexity}} & \multicolumn{2}{c|}{\textbf{DPD perf., 100 MHz}} & \multicolumn{2}{c|}{\textbf{DPD perf., 200 MHz}} \\ \cline{2-11} 
 & \ \  \textbf{P} \ \  &  \ \ \textbf{M} \ \  & \ \  \textbf{Q} \ \  & \ \  $\Delta_{z,x}$ \ \  & \ \  \textbf{FLOPs/sample} \ \  & \ \  \textbf{Mul./sample} \ \  & \ \ \textbf{EVM (\%)} \ \  & \ \  \textbf{TRP ACLR (dB)} \ \  & \ \  \textbf{EVM (\%)} \ \  & \ \  \textbf{TRP ACLR (dB)} \ \  \\ \hline
\multicolumn{1}{|c|}{\textbf{No DPD}} & - & - & - & - & 0 & 0 & 12.10 & 26.10 & 12.43 & 26.30 \\ \hline
\multicolumn{1}{|c|}{\textbf{SPH DPD}} & 3 & 3 & 7 & 1 & 69 & 36 & 6.20 & 34.40 & 6.25 & 34.10 \\ \hline
\multicolumn{1}{|c|}{\rev{\textbf{ \ \ SMP DPD \ \ }}} & \rev{3} & \rev{4} & \rev{7} & \rev{1} & \rev{99} & \rev{63} & \rev{6.15} & \rev{34.80} & \rev{6.20} & \rev{34.40} \\ \hline
\multicolumn{1}{|c|}{\textbf{MP DPD}} & 11 & 4 & - & - & 255 & 112 & 6.00 & 35.20 & 6.13 & 35.00 \\ \hline
\end{tabular}
\end{table*}

% Separate figures:
% %  trim={<left> <lower> <right> <upper>}
% \begin{figure}[t!]
%   \begin{center}
%     \includegraphics[width=1\columnwidth, trim = 0 0 0 0, clip]{figures/PSD_mm.pdf}
%   \end{center}
%   \caption{OTA linearization performance of the Anokiwave AWMF-0129, when a NR 100 MHz transmit signal is applied.}
%   \label{fig:PSD_mm}
% \end{figure}

% \begin{figure}[t!]
%   \begin{center}
%     \includegraphics[width=1\columnwidth, trim = 0 0 0 0, clip]{figures/PSD_mm2.pdf}
%   \end{center}
%   \caption{OTA linearization performance of the Anokiwave AWMF-0129, when a NR 200 MHz transmit signal is applied.}
%   \label{fig:PSD_mm2}
% \end{figure}

\subsubsection{Active Array Linearization}

Linearization of active phased-array transmitters is generally a challenging task, since a single DPD unit must linearize a bank of mutually different PAs. There are multiple ways of acquiring the observation signal for DPD parameter learning, as discussed e.g. in \cite{JSTSP,Swedes_review,DPD_DigitalMIMO, OTA_combining_DPD, DPD_MM_6}. In this work, we assume and adopt the so-called combined observation signal approach and utilize specifically the OTA-combined received signal for DPD parameter learning \cite{JSTSP,Swedes_review, OTA_combining_DPD}, while otherwise following exactly the same learning algorithms as in the Experiments 1 and 2. 

%The adopted learning principle relies on the fact that the nonlinear distortion stemming from phased-array transmitters is more significant in the direction of the main beam, while in other directions it gets partly diluted due to less coherent propagation \cite{OOB_Mollen}. Consequently, by exploiting the spatial characteristics of the distortion, a very efficient learning principle that consist on generating a replica of the signal in the main beam direction can be utilized to learn the DPD coefficients \cite{JSTSP,DPD_DigitalMIMO,Swedes_review}. Such learning signal models the nonlinear distortion stemming from the antenna array transmitter from the beamformed-channel perspective, and the resulting DPD results in minimizing the emissions in the beam direction, while in the rest of directions the beampattern and the DPD ensure that the emissions remain at a sufficiently low level \cite{JSTSP,DPD_DigitalMIMO}. Such approach results in a single-input single-output system, and can be thus learned utilizing traditional  learning algorithms intended for single-antenna transmitters. The learning signal can be obtained in multiple ways, for instance, by means of a combined feedback observation receiver that phase-aligns and combines the individual PA output signals in the RF domain \cite{JSTSP} or by sending OTA feedback from a test receiver in the far field \cite{Swedes_review}, as is the case in this experiment.

In the DPD measurements, we adopt 5G NR FR-2 OFDM signal with SCS of 60~kHz, and consider active PRB counts of 132 and 264, mapping to 100~MHz and 200~MHz channel bandwidths, respectively \cite{3GPPTS38104}. In this case, 5 ILA iterations are adopted, each containing 50,000 samples. Example OTA linearization results are illustrated in Fig.~\ref{fig:PSD_mm}, measured at an EIRP of +42.5~dBm, where the received spectra with the proposed SPH, \rev{SMP} and the reference MP DPD are shown, while the no-DPD case is also shown for comparison. The parametrization of the SPH and \rev{SMP} DPD is $P_{\mathrm{SP}}=3$ and $M_{\mathrm{SPH}}=3$, and $M_{\mathrm{SMP}}=4$, while MP DPD is configured with $P_\mathrm{MP}=11$ and $M_{\mathrm{MP}}=4$. As mentioned already in the introduction, the OTA ACLR requirements at FR-2 are quite clearly relaxed, compared to the classical 45 dB number at FR-1, with 28~dB defined as the \rev{TRP-based} ACLR limit in the current NR Release-15 specifications \cite{3GPPTS38104}. Additionally, 64-QAM is currently the highest supported modulation scheme at FR-2, heaving 8\% as the required EVM. 

In both channel bandwidth cases, considered in Fig.~\ref{fig:PSD_mm}, \rev{the initial EVM and TRP ACLR metrics are around 12.5\% and 26~dB}, respectively, when measuring at EIRP of +42.5~dBm and when no DPD is applied. Hence, linearization is indeed required if the same output power is to be maintained, while Fig.~\ref{fig:PSD_mm} demonstrates that all considered DPD methods can successfully linearize the active array. Table~\ref{tab:mm_perf} shows the exact measured numerical \rev{TRP} ACLR and EVM values, indicating good amounts of linearization gain and that the EVM and \rev{TRP} ACLR requirements can be successfully met. It is also noted that the initial \rev{TRP} ACLR of some 26~dB corresponds already to a very nonlinear starting point. 

Finally, Fig.~\ref{fig:PS} features a power sweep performed with the antenna array, illustrating the TRP ACLR and EVM as a function of the EIRP with and without DPD. It can be clearly observed that in this particular experiment, when no DPD is applied, it is the EVM metric that is limiting the maximum achievable EIRP such that both TRP ACLR and EVM requirements are still fulfilled. Specifically, without DPD processing, this limits the maximum EIRP to some +39~dBm, while when DPD processing is applied, both requirements are fulfilled at least up to the considered maximum EIRP of +42.5~dBm -- and clearly also somewhat beyond. In this particular linearization experiment, it can be noted that the SPH DPD is an intriguing approach, due to its very low computational complexity, while still being clearly able to linearize the array.

%%% BELOW COMMENTED OUT TO SAVE SPACE
%
%For completeness, it is finally noted that the specifications \cite{3GPPTS38104} also state the so-called maximum limit at FR-2, however, it was not measured in this study.}

%however, in our future work we pursue further measurements with even more saturated PA units.

\section{Conclusions}
\label{sec:conc}

In this paper, novel complex spline-interpolated LUT concepts and corresponding DPD methods with gradient-adpative learning rules were proposed for power amplifier linearization.
%, with particular emphasis on reduced main path and parameter learning complexities.  
%while still allowing for good linearization performance. The proposed predistortion concept divides the input envelope range in to several pieces or regions, and relies on new complex spline interpolation to estimate or approximate the instantaneous nonlinear behavior. This is then complemented with two different approaches for memory modeling purposes, namely 1) a cascaded FIR filter, composing thus as a whole a spline-interpolated Hammerstein like DPD solution -- called SPH DPD -- and 2) a memory polynomial like branched structure, where the memory is incorporated through an input delay line and multiple parallel spline-interpolated LUTs -- called SMP DPD. Gradient based iterative parameter learning algorithms were also derived, allowing to estimate the unknown spline control points and the unknown FIR filter parameters in a decoupled manner, in the SPH DPD case, while allowing parallel estimation of the control points of all LUTs in the SMP DPD case.
%
A vast amount of different measurement-based experiments were provided, covering successful linearization of different %local-area, medium range and wide-area/macro 
PA samples at sub-6~GHz bands. Additionally, a 28~GHz state-of-the-art active antenna array was successfully linearized. The measured linearization performance results, together with the provided explicit complexity analysis, show that the proposed spline-interpolated DPD concepts can provide very appealing complexity-performance trade-offs, compared to, e.g., ordinary canonical MP DPD. \rev{Specifically, the SMP DPD was shown to provide in all measurement examples linearization performance very close to that of ordinary MP DPD, while having substantially lower main path and DPD learning complexity. Additionally, the SPH DPD offers further reduction in the main path processing complexity, while was also shown to be performing fairly close to the other DPD systems, particularly in the timely 28~GHz active array linearization experiment.} 

\bibliographystyle{IEEEtran}
\bibliography{InPaper.bib}

% Generated by IEEEtran.bst, version: 1.12 (2007/01/11)
\begin{thebibliography}{10}
\providecommand{\url}[1]{#1}
\csname url@samestyle\endcsname
\providecommand{\newblock}{\relax}
\providecommand{\bibinfo}[2]{#2}
\providecommand{\BIBentrySTDinterwordspacing}{\spaceskip=0pt\relax}
\providecommand{\BIBentryALTinterwordstretchfactor}{4}
\providecommand{\BIBentryALTinterwordspacing}{\spaceskip=\fontdimen2\font plus
\BIBentryALTinterwordstretchfactor\fontdimen3\font minus
  \fontdimen4\font\relax}
\providecommand{\BIBforeignlanguage}[2]{{%
\expandafter\ifx\csname l@#1\endcsname\relax
\typeout{** WARNING: IEEEtran.bst: No hyphenation pattern has been}%
\typeout{** loaded for the language `#1'. Using the pattern for}%
\typeout{** the default language instead.}%
\else
\language=\csname l@#1\endcsname
\fi
#2}}
\providecommand{\BIBdecl}{\relax}
\BIBdecl

\bibitem{2018DahlmanL5G}
E.~Dahlman, S.~Parkvall, and J.~Skold, \emph{{5G NR: The Next Generation
  Wireless Access Technology}}, 1st~ed.\hskip 1em plus 0.5em minus 0.4em\relax
  Academic Press, 2018.

\bibitem{ghannouchi2009behavioral}
F.~M. Ghannouchi and O.~Hammi, ``Behavioral modeling and predistortion,''
  \emph{{IEEE} Microw. Mag.}, vol.~10, no.~7, pp. 52--64, Dec. 2009.

\bibitem{taxonomy}
Y.~Rahmatallah and S.~Mohan, ``Peak-to-average power ratio reduction in {OFDM}
  systems: A survey and taxonomy,'' \emph{IEEE Commun. Surveys Tuts.}, vol.~15,
  no.~4, pp. 1567--1592, Mar. 2013.

\bibitem{6153399}
S.~{Afsardoost}, T.~{Eriksson}, and C.~{Fager}, ``Digital predistortion using a
  vector-switched model,'' \emph{{IEEE} Trans. Microw. Theory Tech.}, vol.~60,
  no.~4, pp. 1166--1174, April 2012.

\bibitem{Swedes_review}
C.~{Fager} \emph{et~al.}, ``Linearity and efficiency in 5{G} transmitters: New
  techniques for analyzing efficiency, linearity, and linearization in a 5{G}
  active antenna transmitter context,'' \emph{IEEE Microw. Mag.}, vol.~20,
  no.~5, pp. 35--49, May 2019.

\bibitem{4497808}
O.~{Hammi}, F.~M. {Ghannouchi}, and B.~{Vassilakis}, ``A compact
  envelope-memory polynomial for {RF} transmitters modeling with application to
  baseband and {RF}-digital predistortion,'' \emph{{IEEE} Microw. Wireless
  Compon. Lett.}, vol.~18, no.~5, pp. 359--361, May 2008.

\bibitem{abdelaziz2018decorrelation}
M.~Abdelaziz, L.~Anttila, A.~Kiayani, and M.~Valkama, ``Decorrelation-based
  concurrent digital predistortion with a single feedback path,'' \emph{{IEEE}
  Trans. Microw. Theory Tech.}, vol.~66, no.~1, pp. 280--293, Jan. 2018.

\bibitem{6612754}
Y.~{Ma}, Y.~{Yamao}, Y.~{Akaiwa}, and C.~{Yu}, ``{FPGA} implementation of
  adaptive digital predistorter with fast convergence rate and low complexity
  for multi-channel transmitters,'' \emph{{IEEE} Trans. Microw. Theory Tech.},
  vol.~61, no.~11, pp. 3961--3973, Nov 2013.

\bibitem{kim2001digital}
J.~Kim and K.~Konstantinou, ``Digital predistortion of wideband signals based
  on power amplifier model with memory,'' \emph{Electron. Lett.}, vol.~37,
  no.~23, p.~1, Nov. 2001.

\bibitem{zhu_2011}
A.~Zhu, ``Digital predistortion and its combination with crest factor
  reduction,'' in \emph{Digital Front-End in Wireless Communications and
  Broadcasting: Circuits and Signal Processing}, F.-L. Luo, Ed.\hskip 1em plus
  0.5em minus 0.4em\relax Cambridge: Cambridge University Press, 2011, ch.~9,
  pp. 244--279.

\bibitem{tehrani2010comparative}
A.~S. Tehrani \emph{et~al.}, ``A comparative analysis of the
  complexity/accuracy tradeoff in power amplifier behavioral models,''
  \emph{{IEEE} Trans. Microw. Theory Tech.}, vol.~58, no.~6, pp. 1510--1520,
  June 2010.

\bibitem{morgan2006generalized}
D.~R. Morgan, Z.~Ma, J.~Kim, M.~G. Zierdt, and J.~Pastalan, ``A generalized
  memory polynomial model for digital predistortion of {RF} power amplifiers,''
  \emph{{IEEE} Trans. Signal Process.}, vol.~54, no.~10, pp. 3852--3860, Oct.
  2006.

\bibitem{mkadem2016multi}
F.~Mkadem, A.~Islam, and S.~Boumaiza, ``Multi-band complexity-reduced
  generalized-memory-polynomial power-amplifier digital predistortion,''
  \emph{{IEEE} Trans. Microw. Theory Tech.}, vol.~64, no.~6, pp. 1763--1774,
  June 2016.

\bibitem{4161079}
A.~{Zhu} and T.~J. {Brazil}, ``An overview of volterra series based behavioral
  modeling of {RF}/microwave power amplifiers,'' in \emph{2006 IEEE Annual
  Wireless and Microwave Technology Conference}, Dec. 2006, pp. 1--5.

\bibitem{1362669}
------, ``Behavioral modeling of {RF} power amplifiers based on pruned volterra
  series,'' \emph{{IEEE} Microw. Wireless Compon. Lett.}, vol.~14, no.~12, pp.
  563--565, Dec. 2004.

\bibitem{Schetzen2006}
M.~Schetzen, \emph{The Volterra and Wiener Theories of Nonlinear
  Systems}.\hskip 1em plus 0.5em minus 0.4em\relax Melbourne, FL, USA: Krieger
  Publishing Co., Inc., 2006.

\bibitem{6353238}
C.~{Yu}, L.~{Guan}, E.~{Zhu}, and A.~{Zhu}, ``Band-limited volterra
  series-based digital predistortion for wideband {RF} power amplifiers,''
  \emph{{IEEE} Trans. Microw. Theory Tech.}, vol.~60, no.~12, pp. 4198--4208,
  Dec. 2012.

\bibitem{4350085}
N.~{Safari}, N.~{Holte}, and T.~{Roste}, ``Digital predistortion of power
  amplifiers based on spline approximations of the amplifier characteristics,''
  in \emph{2007 IEEE 66th Veh. Technol. Conf.}, Sep. 2007, pp. 2075--2080.

\bibitem{4264109}
N.~{Safari}, P.~{Fedorenko}, J.~S. {Kenney}, and T.~{Roste}, ``Spline-based
  model for digital predistortion of wide-band signals for high power amplifier
  linearization,'' in \emph{2007 IEEE/MTT-S International Microwave Symposium},
  June 2007, pp. 1441--1444.

\bibitem{9018116}
J.~{Kral}, T.~{Gotthans}, R.~{Marsalek}, M.~{Harvanek}, and M.~{Rupp}, ``On
  feedback sample selection methods allowing lightweight digital predistorter
  adaptation,'' \emph{IEEE Trans. Circuits Syst. I}, vol.~67, no.~6, June 2020.

\bibitem{8255667}
C.~{Cheang}, P.~{Mak}, and R.~P. {Martins}, ``A hardware-efficient feedback
  polynomial topology for dpd linearization of power amplifiers: Theory and
  fpga validation,'' \emph{IEEE Trans. Circuits Syst. I}, vol.~65, no.~9, pp.
  2889--2902, 2018.

\bibitem{6963510}
F.~M. {Barradas}, T.~R. {Cunha}, P.~M. {Lavrador}, and J.~C. {Pedro},
  ``Polynomials and {LUTs} in {PA} behavioral modeling: {A} fair theoretical
  comparison,'' \emph{{IEEE} Trans. Microw. Theory Tech.}, vol.~62, no.~12, pp.
  3274--3285, Nov. 2014.

\bibitem{3GPPTS38104}
{3GPP Tech. Spec. 38.104}, ``{NR; Base Station (BS) radio transmission and
  reception},'' v15.4.0 (Release 15), Dec. 2018.

\bibitem{liang2014quadratic}
K.-F. Liang, J.-H. Chen, and Y.-J.~E. Chen, ``A quadratic-interpolated
  {LUT}-based digital predistortion technique for cellular power amplifiers,''
  \emph{{IEEE} Trans. Circuits Syst. {II}}, vol.~61, no.~3, pp. 133--137, Mar.
  2014.

\bibitem{molina2017digital}
A.~Molina, K.~Rajamani, and K.~Azadet, ``Digital predistortion using lookup
  tables with linear interpolation and extrapolation: Direct least squares
  coefficient adaptation,'' \emph{{IEEE} Trans. Microw. Theory Tech.}, vol.~65,
  no.~3, pp. 980--987, Nov. 2017.

\bibitem{jardin2007filter}
P.~Jardin and G.~Baudoin, ``Filter lookup table method for power amplifier
  linearization,'' \emph{{IEEE} Trans. Veh. Technol.}, vol.~56, no.~3, pp.
  1076--1087, May 2007.

\bibitem{5497423}
X.~{Wu}, N.~{Zheng}, X.~{Yang}, J.~{Shi}, and H.~{Chen}, ``A spline-based
  hammerstein predistortion for 3{G} power amplifiers with hard
  nonlinearities,'' in \emph{2010 2nd Int. Conf. Future Comp. Commun.}, vol.~3,
  May 2010.

\bibitem{4336139}
O.~{Hammi}, F.~M. {Ghannouchi}, S.~{Boumaiza}, and B.~{Vassilakis}, ``A
  data-based nested {LUT} model for {RF} power amplifiers exhibiting memory
  effects,'' \emph{{IEEE} Microw. Wireless Compon. Lett.}, vol.~17, no.~10, pp.
  712--714, Oct. 2007.

\bibitem{1599577}
{H. Zhi-yong} \emph{et~al.}, ``An improved look-up table predistortion
  technique for {HPA} with memory effects in {OFDM} systems,'' \emph{IEEE
  Trans. Broadcasting}, vol.~52, no.~1, pp. 87--91, Feb. 2006.

\bibitem{7915706}
Z.~{Wang} \emph{et~al.}, ``Low computational complexity digital predistortion
  based on direct learning with covariance matrix,'' \emph{{IEEE} Trans.
  Microw. Theory Tech.}, vol.~65, no.~11, pp. 4274--4284, May 2017.

\bibitem{de1978practical}
C.~De~Boor, \emph{A Practical Guide to Splines}.\hskip 1em plus 0.5em minus
  0.4em\relax Springer, New York, 1978.

\bibitem{scarpiniti2013nonlinear}
M.~Scarpiniti, D.~Comminiello, R.~Parisi, and A.~Uncini, ``Nonlinear spline
  adaptive filtering,'' \emph{Signal Process.}, vol.~93, no.~4, pp. 772--783,
  Apr. 2013.

\bibitem{scarpiniti2014hammerstein}
------, ``Hammerstein uniform cubic spline adaptive filters: Learning and
  convergence properties,'' \emph{Signal Process.}, vol. 100, pp. 112--123,
  July 2014.

\bibitem{scarpiniti2015novel}
------, ``Novel cascade spline architectures for the identification of
  nonlinear systems,'' \emph{{IEEE} Trans. Circuits Syst. {I}}, vol.~62, no.~7,
  pp. 1825--1835, July 2015.

\bibitem{haykin2008adaptive}
S.~Haykin, \emph{Adaptive Filter Theory}.\hskip 1em plus 0.5em minus
  0.4em\relax Prentice Hall, 2001.

\bibitem{4645581}
D.~H. {Brandwood}, ``A complex gradient operator and its application in
  adaptive array theory,'' \emph{IEE Proceedings F - Commun., Radar and Signal
  Process.}, vol. 130, no.~1, pp. 11--16, Feb. 1983.

\bibitem{Frerking94a}
M.~Frerking, \emph{Digital Signal Processing in Communications Systems}.\hskip
  1em plus 0.5em minus 0.4em\relax Springer, 1994.

\bibitem{1337325}
R.~{Raich}, {Hua Qian}, and G.~T. {Zhou}, ``Orthogonal polynomials for power
  amplifier modeling and predistorter design,'' \emph{IEEE Trans. Veh.
  Technol.}, vol.~53, no.~5, pp. 1468--1479, Sep. 2004.

\bibitem{5757879}
R.~{Dallinger}, H.~{Ruotsalainen}, R.~{Wichman}, and M.~{Rupp}, ``Adaptive
  pre-distortion techniques based on orthogonal polynomials,'' in \emph{44th
  Asilomar Conf. Signals, Syst., and Computers}, 2010, pp. 1945--1950.

\bibitem{Flops}
R.~J. Schilling and S.~L. Harris, \emph{Fundamentals of Digital Signal
  Processing Using MATLAB}.\hskip 1em plus 0.5em minus 0.4em\relax Cengage
  Learning, 2010.

\bibitem{OOB_Mollen}
C.~Mollen, E.~G. Larsson, U.~Gustavsson, T.~Eriksson, and R.~W. Heath,
  ``Out-of-band radiation from large antenna arrays,'' \emph{{IEEE} Commun.
  Mag.}, vol.~56, no.~4, pp. 196--203, April 2018.

\bibitem{JSTSP}
M.~Abdelaziz, L.~Anttila, A.~Brihuega, F.~Tufvesson, and M.~Valkama, ``{Digital
  Predistortion for Hybrid MIMO Transmitters},'' \emph{{IEEE} J. Sel. Topics
  Signal Process.}, vol.~12, no.~3, pp. 445--454, June 2018.

\bibitem{DPD_DigitalMIMO}
A.~{Brihuega}, L.~{Anttila}, M.~{Abdelaziz}, and M.~{Valkama}, ``{Digital
  Predistortion in Large-Array Digital Beamforming Transmitters},'' in
  \emph{52nd Asilomar Conf. Signals, Syst., Comp.}, Oct. 2018, pp. 611--618.

\bibitem{DPD_MM_6}
X.~Liu, Q.~Zhang, W.~Chen, H.~Feng, L.~Chen, F.~M. Ghannouchi, and Z.~Feng,
  ``{Beam-Oriented Digital Predistortion for 5G Massive MIMO Hybrid Beamforming
  Transmitters},'' \emph{{IEEE} Trans. Microw. Theory Tech.}, vol.~66, no.~7,
  pp. 3419--3432, July 2018.

\bibitem{OTA_combining_DPD}
N.~Tervo \emph{et~al.}, ``Digital predistortion of amplitude varying phased
  array utilising over-the-air combining,'' in \emph{2017 IEEE MTT-S
  International Microwave Symposium (IMS)}, June 2017, pp. 1165--1168.

\bibitem{3GPPTS38141}
{3GPP Tech. Spec. 38.141-2}, ``{NR; Base Station (BS) conformance testing, Part
  2},'' v15.1.0 (Release 15), March 2019.

\end{thebibliography}
\vspace{0mm}
\begin{IEEEbiography}[{\includegraphics[width=1in,height=1.25in,clip,keepaspectratio]{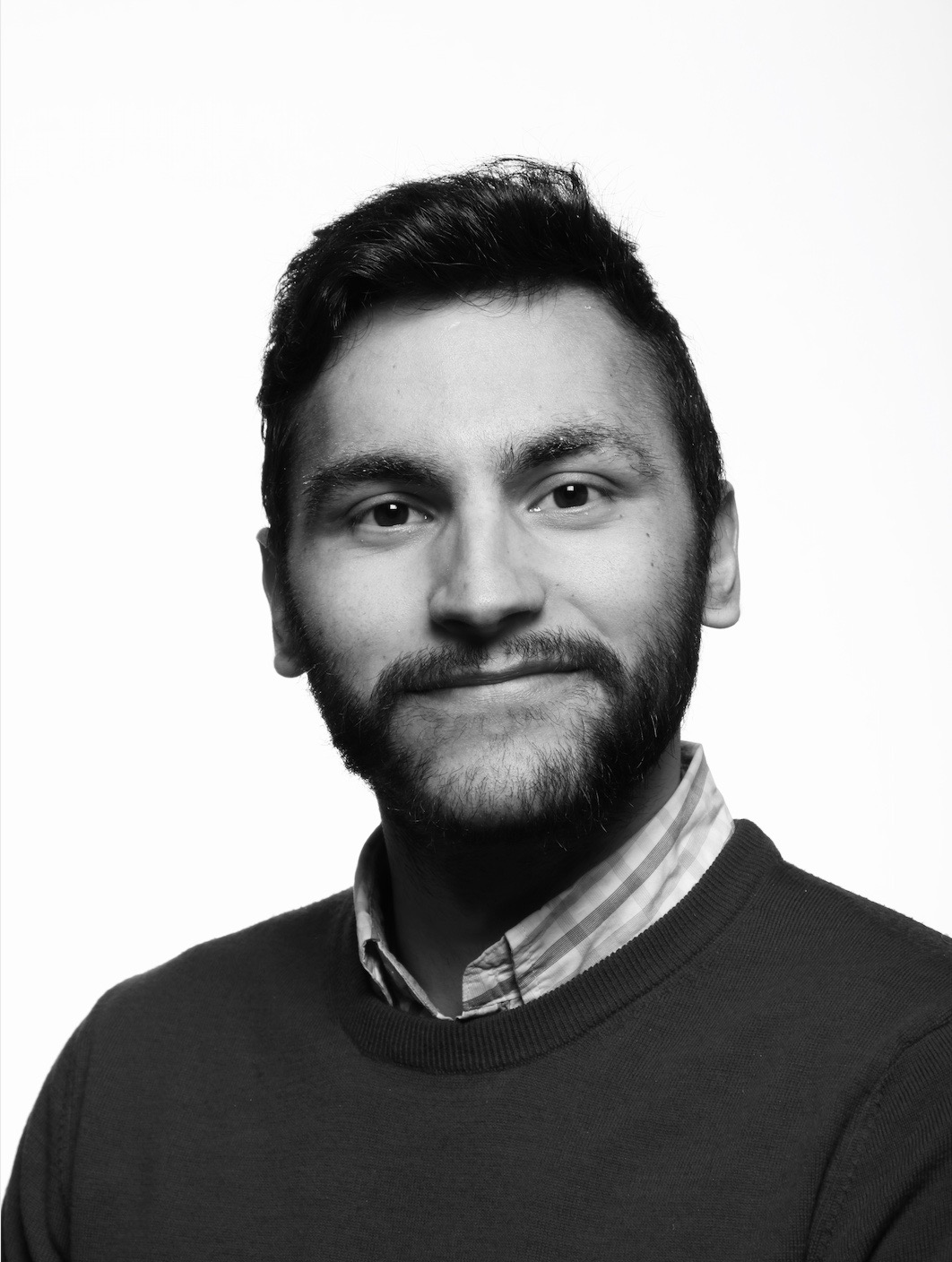}}]  {Pablo Pascual Campo} received his B.Sc. and M.Sc. degrees in Telecommunications and Electrical Engineering in 2012 and 2014, respectively, from Universidad Polit\'{e}cnica de Madrid, Madrid, Spain. He is currently pursuing his D.Sc. degree at Tampere University, Department of Electrical Engineering, Tampere, Finland. His research interests include digital predistortion, full-duplex systems and applications, and signal processing for wireless communications at the mmWave bands.
\end{IEEEbiography}
\begin{IEEEbiography}[{\includegraphics[width=1in,height=1.25in,clip,keepaspectratio]{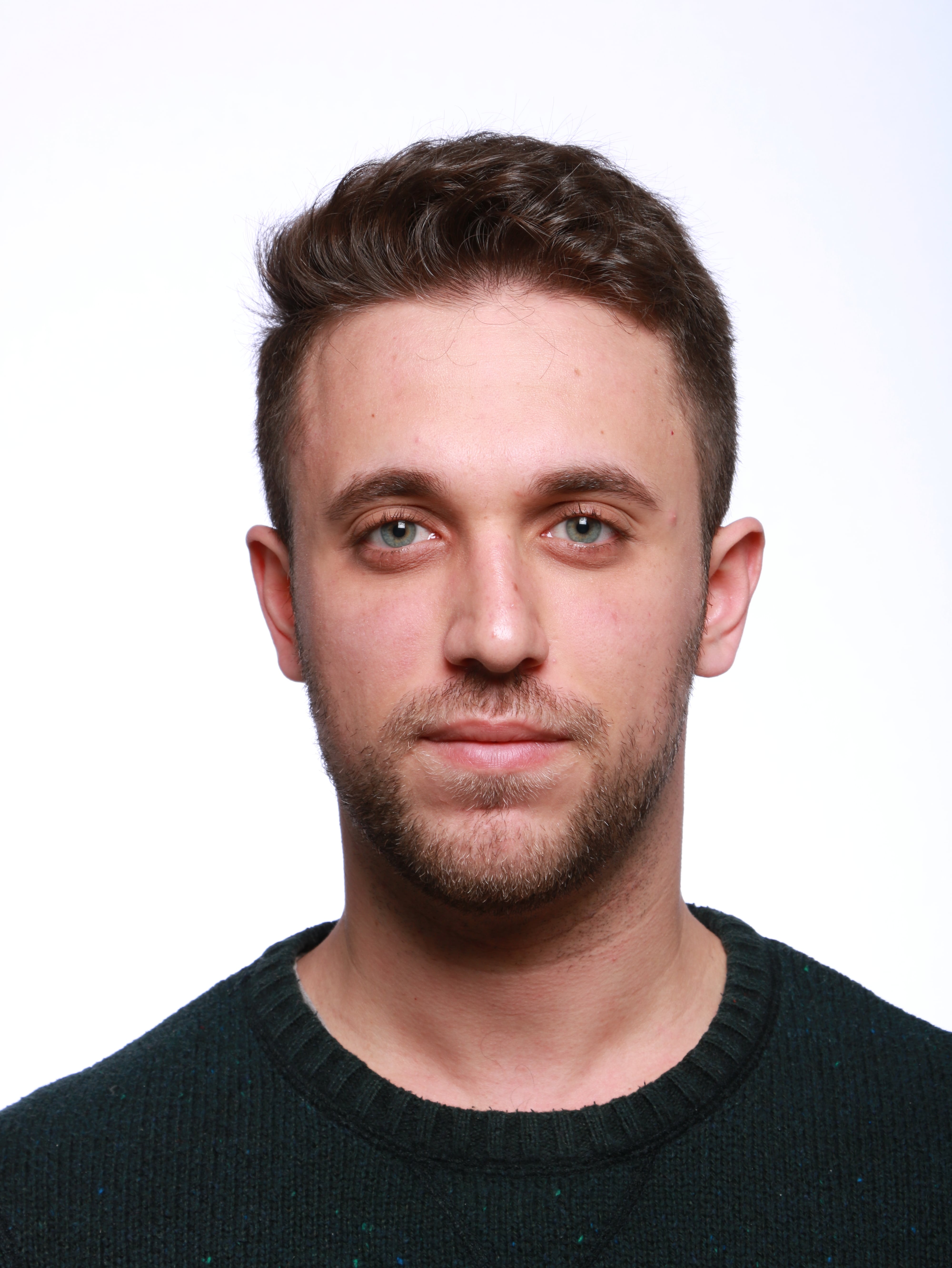}}]{Alberto Brihuega} (S'18) received the B.Sc. and M.Sc. degrees in Telecommunications Engineering from Universidad Polit\'ecnica de Madrid, Spain, in 2015 and 2017, respectively. He is currently working towards the Ph.D. degree with Tampere University, Finland, where he is a researcher with the Department of Electrical Engineering. His research interests include statistical and adaptive digital signal processing for compensation of hardware impairments in large-array antenna transceivers.
\end{IEEEbiography}
\begin{IEEEbiography}[{\includegraphics[width=1in,height=1.25in,clip,keepaspectratio]{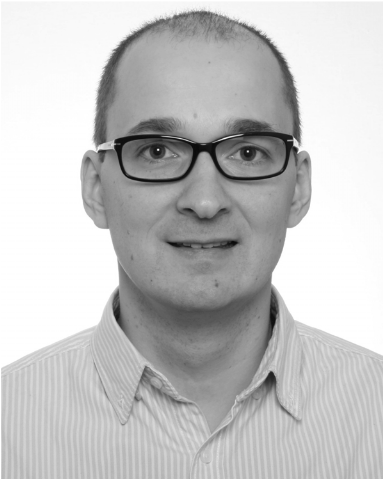}}]{Lauri Anttila} received the D.Sc. (Tech.) degree (with distinction) in 2011 from Tampere University of Technology (TUT), Finland. Since 2016, he has been a University Researcher at the Department of Electrical Engineering, Tampere University (formerly TUT). In 2016-2017, he was a Visiting Research Fellow at the Department of Electronics and Nanoengineering, Aalto University, Finland. His research interests are in radio communications and signal processing, with a focus on the radio implementation challenges in systems such as 5G, full-duplex radio, and large-scale antenna systems.
\end{IEEEbiography}
\begin{IEEEbiography}[{\includegraphics[width=1in,height=1.25in,clip,keepaspectratio]{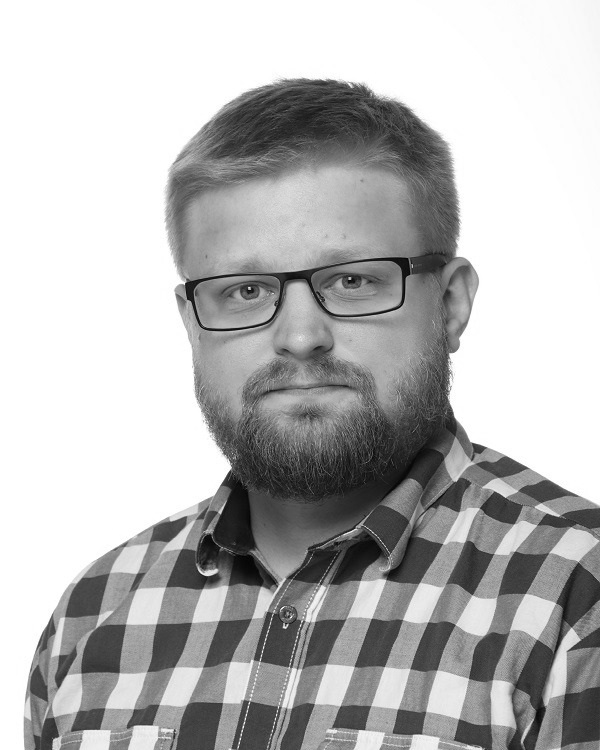}}]{Matias Turunen} is currently pursuing the M.Sc. degree in electrical engineering at Tampere University (TAU), Tampere, Finland, while also working as a Research Assistant with the Department of Electrical Engineering at TAU. His current research interests include in-band full-duplex radios with an emphasis on analog RF cancellation, OFDM radar, and 5G New Radio systems. 
\end{IEEEbiography}
\begin{IEEEbiography}[{\includegraphics[width=1in,height=1.25in,clip,keepaspectratio]{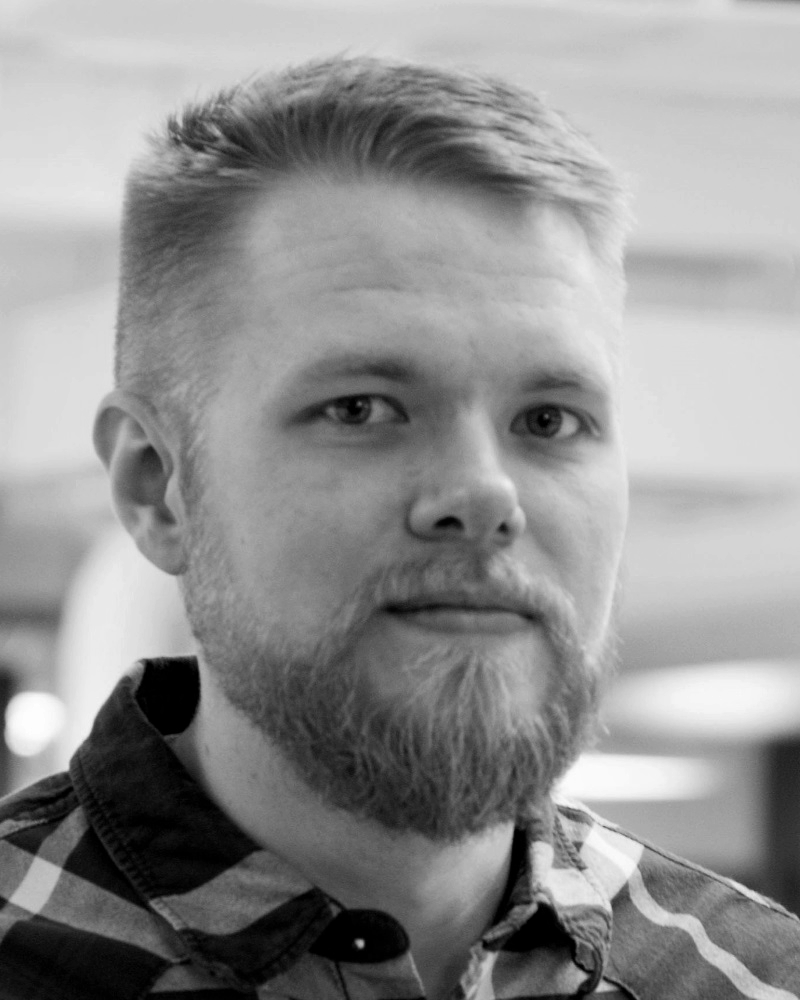}}]{Dani Korpi} received D.Sc. degree (Hons.) in electrical engineering from Tampere University of Technology, Finland, in 2017. Currently, he is a Senior Specialist with Nokia Bell Labs in Espoo, Finland. His doctoral thesis received an award for the best dissertation of the year in Tampere University of Technology, as well as the Finnish Technical Sector’s Award for the best doctoral dissertation of 2017. His research interests include inband full-duplex radios, machine learning for wireless communications, and beyond 5G radio systems.
\end{IEEEbiography}
\begin{IEEEbiography}[{\includegraphics[width=1in,height=1.25in,clip,keepaspectratio]{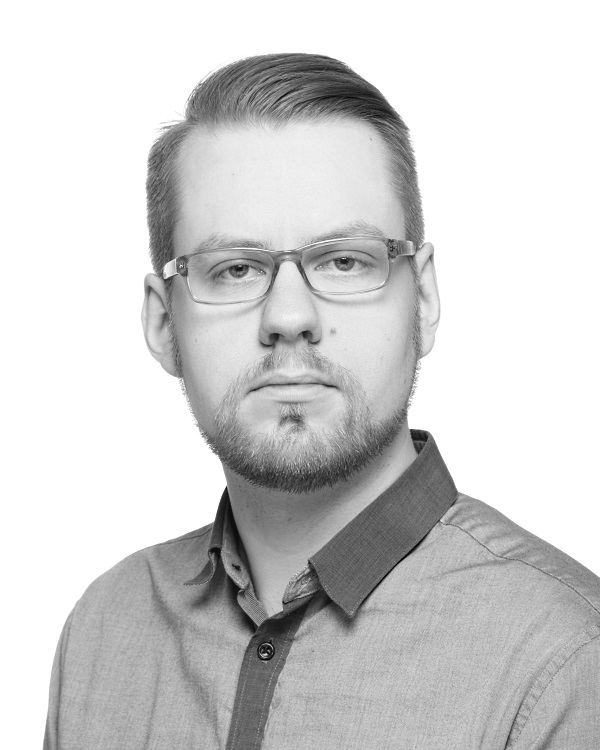}}]{Markus All\'{e}n} received the B.Sc., M.Sc. and D.Sc. degrees in communications engineering from Tampere University of Technology, Finland, in 2008, 2010 and 2015, respectively. He is currently with the Department of Electrical Engineering at Tampere University, Finland, as a University Instructor. His current research interests include software-defined radios, 5G-related RF measurements and digital signal processing for radio transceiver linearization.
\end{IEEEbiography}
\begin{IEEEbiography}[{\includegraphics[width=1in,height=1.25in,clip,keepaspectratio]{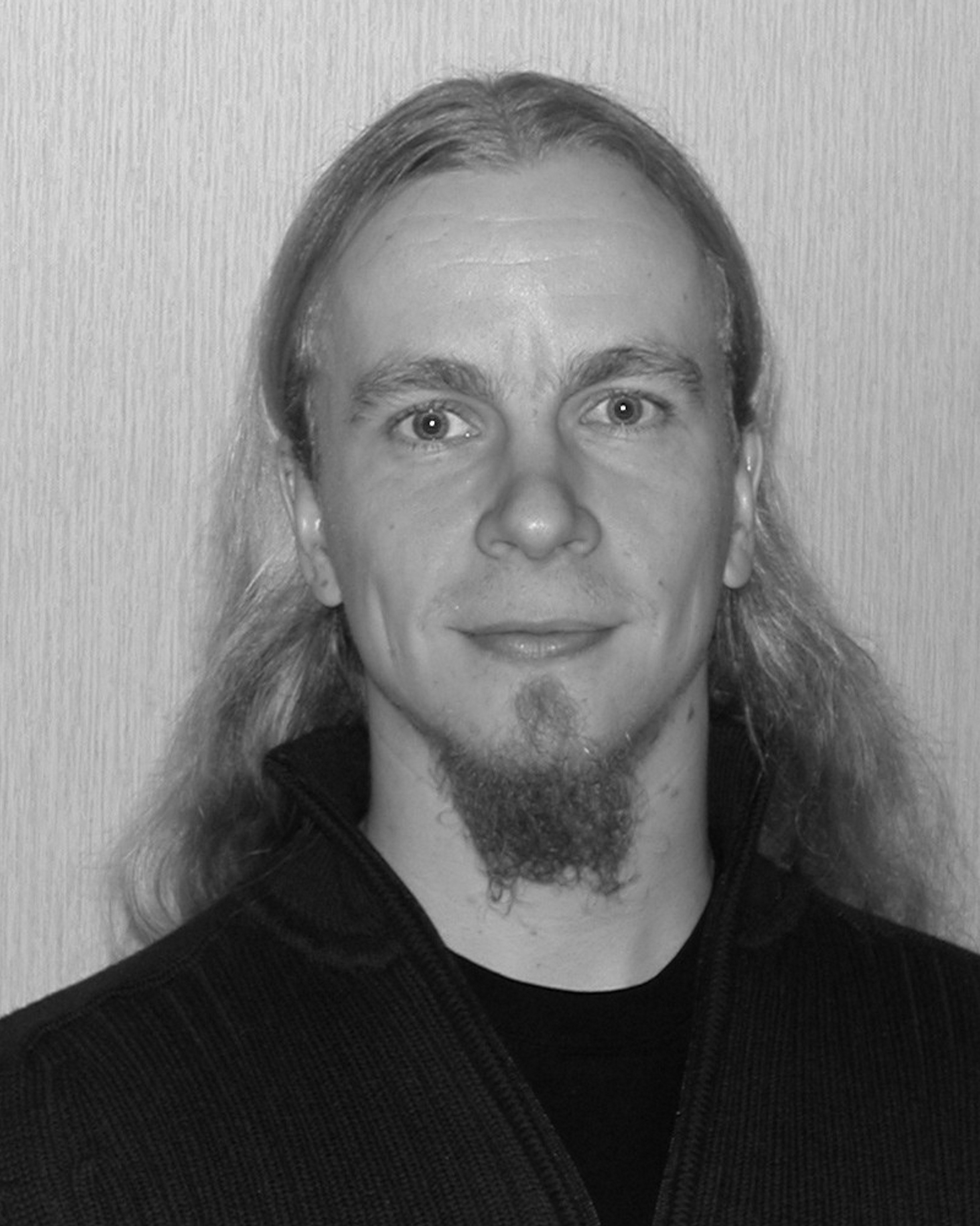}}]{Mikko Valkama} received the D.Sc. (Tech.) degree (with honors) in electrical engineering from Tampere University of Technology (TUT), Finland, in 2001. In 2003, he was a visiting post-doc research fellow with SDSU, San Diego, CA. Currently, he is a Full Professor and Department Head of Electrical Engineering at the newly formed Tampere University (TAU), Finland. His research interests include radio communications, radio localization, and radio-based sensing, with particular emphasis on 5G and beyond mobile radio networks.
\end{IEEEbiography}

% % Biographies

% \begin{IEEEbiography}{Pablo Pascual Campo}
% Pablo Pascual Campo received his B.Sc. and M.Sc. degrees in Telecommunications and Electrical Engineering in 2012 and 2014, respectively, from Universidad Politécnica de Madrid, Madrid, Spain.
% He is currently pursuing his D.Sc. degree at Tampere University’s Department of Electrical Engineering, Tampere, Finland. His research interests include digital predistortion, full-duplex systems and applications, and signal processing for wireless communications in the mmWave spectrum.
% \end{IEEEbiography}

% \begin{IEEEbiography}{Lauri Anttila}
% Biography text here.
% \end{IEEEbiography}

% \begin{IEEEbiography}{Alberto Brihuega}
% Biography text here.
% \end{IEEEbiography}

% \begin{IEEEbiography}{Matias Turunen}
% Biography text here.
% \end{IEEEbiography}

% \begin{IEEEbiography}{Dani Korpi}
% Biography text here.
% \end{IEEEbiography}

% \begin{IEEEbiography}{Markus Allen}
% Biography text here.
% \end{IEEEbiography}

% \begin{IEEEbiography}{Mikko Valkama}
% Biography text here.
% \end{IEEEbiography}

% That's all mates.
\end{document}